\documentclass[fleqn]{svmult}
\usepackage{ae}
\usepackage{aecompl}
\usepackage{amsmath}
\usepackage{amssymb}
\usepackage{epsfig}
\usepackage{epsf}
\usepackage{cite}
\usepackage{svaips}      
\usepackage{makeidx}     
\usepackage{graphicx}    
\usepackage{multicol}    
\usepackage{times}

\newcommand{\gl}[1]{Eq. (\ref{#1})}
\newcommand{\gls}[2]{Eqs. (\ref{#1},\ref{#2})}
\newcommand{\glto}[2]{Eqs. (\ref{#1}) to (\ref{#2})}
\newcommand{\bsq}[1]{\begin{subequations}\label{#1}}
\newcommand{\esq}{\end{subequations}}
\newcommand{\beq}[1]{\begin{equation}\label{#1}}
\newcommand{\eeq}{\end{equation}}
\newcommand{\beqa}[1]{\begin{eqnarray}\label{#1}}
\newcommand{\eeqa}{\end{eqnarray}}

\newcommand{\gd}{\dot{\gamma}}
\newcommand{\smop}{\Omega}
\newcommand{\smopb}{\Omega^{\dagger}}

\newcommand{\kap}{\boldsymbol{\kappa}}
\newcommand{\rb}{{\bf r}}
\newcommand{\bp}{\boldsymbol{\partial}}
\newcommand{\qb}{{\bf q}}
\newcommand{\kb}{{\bf k}}
\newcommand{\pb}{{\bf p}}
\renewcommand{\rho}{\varrho}
\renewcommand{\epsilon}{\varepsilon}

\begin{document}
\title*{Nonlinear rheological properties of dense colloidal dispersions close to a glass transition
under steady shear}
\titlerunning{Nonlinear rheology}
\author{Matthias Fuchs}
\institute{Fachbereich Physik, Universit\"at Konstanz,
 78457 Konstanz, Germany \textit{matthias.fuchs@uni-konstanz.de}}

\maketitle
\begin{abstract}
The nonlinear rheological properties of dense colloidal suspensions
under steady shear are discussed within a first principles approach.
It starts from the Smoluchowski equation of interacting Brownian
particles in a given shear flow, derives generalized Green-Kubo
relations, which contain the transients dynamics formally exactly,
and closes the equations using mode coupling approximations. Shear
thinning of colloidal fluids and  dynamical yielding of colloidal
glasses arise from a competition between a slowing down of
structural relaxation, because of particle interactions, and
enhanced decorrelation of fluctuations, caused by the shear
advection of density fluctuations. The integration through
transients approach takes account of the dynamic competition,
translational invariance enters the concept of wavevector advection,
and the mode coupling approximation enables to quantitatively
explore the shear-induced suppression of particle caging and the
resulting speed-up of the structural relaxation. Extended
comparisons with shear stress data in the linear response and in the
nonlinear regime measured in model thermo-sensitive core-shell
latices are discussed. Additionally, the single particle motion
under shear observed by confocal microscopy and in computer
simulations is reviewed and analysed theoretically.

\end{abstract}
\dominitoc
\begin{abbrsymblist}

\item[$G_\infty$] Shear modulus of a solid (transverse elastic constant or
Lame-coefficient)
\item[$\eta_0$] Newtonian viscosity of a fluid
\item[$\sigma$] Shear stress
\item[$\gd$] Shear rate
\item[$g(t)$] Time dependent shear modulus; in the linear response
regime denoted as $g^{\rm lr}(t)$ of the quiescent system;
generalized one if including dependence on shear rate
\item[$\tau$] Maxwell (final or $\alpha$- relaxation) time of structural relaxation
\item[$G'(\omega)$] Storage modulus in linear response
\item[$G''(\omega)$] Loss modulus in linear response
\item[$\eta$] Shear viscosity; defined via $\eta=\sigma(\gd)/\gd$
\item[$S_q$] Equilibrium structure factor
\item[$R_H$] Hydrodynamic radius  of a colloidal particle (radius $a=R_H$ assumed)
\item[$d$] Colloid diameter ($d=2a=2R_H$  assumed throughout)
\item[$k_B T$] Thermal energy
\item[$\eta_s$] Solvent viscosity
\item[$D_0$] Stokes Einstein Sutherland diffusion coefficient at infinite dilution
\item[Pe$_0$] Bare Peclet number
\item[Pe] Dressed Peclet or Weissenberg number
\item[$\phi$] Packing fraction $\phi= \frac{4\pi}{3} R_H^3 n$ of spheres of radius
$R_H$ at number density $n$
\item[$\epsilon$] Separation parameter in MCT giving the relative
distance in a thermodynamic control parameter to its value at the
glass transition
\item[$\lambda$] MCT exponent parameter
\item[$G'_\infty$] Instantaneous isothermal shear modulus
\item[$\eta_\infty$] High frequency viscosity
\item[$\sigma^+$] Dynamic yield stress of a shear molten glass
\item[HI] Hydrodynamic/  solvent induced interactions
\item[MCT]  Mode coupling theory
\item[ITT] Integrations through transients approach
\item[SO] Smoluchowski operator $\smop$
\item[PY] Percus-Yevick theory giving the approximate PY $S_q$
of a hard sphere fluid
\item[ISHSM] Isotropically sheared hard spheres model
\item[F$_{12}^{(\gd)}$] Schematic model without spatial resolution considering a single
correlator
\end{abbrsymblist}

%
\section{Introduction}

Rheological and elastic properties under flow and deformations are
highly characteristic for many soft materials like complex fluids,
pastes, sands and gels, viz.~soft (often metastable) solids of
dissolved macromolecular constituents \cite{Larson}. Shear
deformations, which conserve volume but stretch material elements,
often provide the simplest experimental route to investigate the
materials. Moreover, solids and fluids respond in a
characteristically different way to shear, the former elastically,
the latter by flow. The former are characterized by a  shear modulus
$G_\infty$, corresponding to a Hookian spring constant, the latter
by a Newtonian viscosity $\eta_0$, which quantifies the dissipation.

Viscoelastic materials exhibit both, elastic and dissipative,
phenomena depending on external control parameters like temperature
and/ or density, and depending on frequency or time-scale of
experimental observation. Viscoelastic fluids differ from pastes and
sands in the importance of thermal fluctuations causing Brownian
motion, which enables them to explore their phase space without
external drive like shaking, that would be required to fluidize
granular systems. The change between fluid and solid like behavior
in viscoelastic materials can have diverse origins, including phase
transitions of various kinds, like freezing and micro-phase
separation, and/or molecular mechanisms like entanglement formation
in polymer melts. One mechanism existent quite universally in dense
particulate systems is the glass transition, that structural
rearrangements of particles become progressively slower
\cite{Goe:92} because of interactions/ collisions, and that the
structural relaxation time grows dramatically.

Maxwell was the first to describe the viscoelastic response at the
fluid-to-glass transition phenomenologically. He introduced a
time-dependent shear modulus $g(t)$ describing the response of a
viscoelastic fluid to a time-dependent shear deformation, \beq{a1}
\sigma(t) = \int_0^t dt'\; g(t-t')\; \gd(t')\;. \eeq
 Here $\sigma$ is
the (transverse) stress, the thermodynamic average of an
off-diagonal element of the microscopic  stress tensor, and $\gd(t)$
is the time-dependent shear rate impressed on the material starting
at time $t=0$. Maxwell chose the Ansatz $g(t) = G_\infty \,
\exp{\{-(t/\tau)\}}$, which interpolates inbetween elastic behavior
$\sigma(t\to0)\approx G_\infty\, \gamma(t)$ for short times
$t\ll\tau$ and dissipative behavior, $\sigma(t) \approx \eta_0\,
\gd(t)$, for long times, $t\gg\tau$; the strain $\gamma(t)$ is
obtained from integrating up the strain rate, $\gamma(t)=\int_0^t
dt' \gd(t')$. Maxwell found the relation $\eta_0= G_\infty\, \tau$
which connects the structural relaxation time and the glass modulus
$G_\infty$ to the Newtonian viscosity. He thus explained the
increase of the viscosity at the glass transition by the slowing
down of the structural dynamics (viz.~the increase of $\tau$), and
provided a definition of an idealized glass state, where
$\tau=\infty$. It responds purely elastically.

Above relation (\ref{a1}) between $\sigma$ and $\gd$ is exact in
linear response, where non-linear contributions in $\gd$ are
neglected in the stress. The linear response modulus (to be denoted
as $g^{\rm lr}(t)$) itself is defined in the quiescent system and
describes the small shear-stress fluctuations always present in
thermal equilibrium \cite{Larson,russel}. Often, oscillatory
deformations at fixed frequency $\omega$ are applied and the
frequency dependent storage- ($G'(\omega)$) and loss-
($G''(\omega)$) shear moduli are measured in or out of phase,
respectively. The former captures elastic while the latter captures
dissipative contributions. Both moduli result from
Fourier-transformations of  the linear response shear modulus
$g^{\rm lr}(t)$, and are thus connected via Kramers-Kronig
relations.

The stationary, nonlinear rheological behavior  under steady
shearing provides additional insight into the physics of dense
colloidal dispersions \cite{Larson,russel}. Here, the shear rate is
constant, $\gd(t)\equiv\gd$, and the stress in the stationary state
achieved after waiting sufficiently long (taking $t\to\infty$ in
\gl{a1}) is of interest. Equation (\ref{a1}) may be interpreted
under flow to state that the non-linearity in the stress versus
shear rate curve (the relation $\sigma(\gd)$ is termed 'flow curve')
results from the dependence of the (generalized) time-dependent
shear modulus $g(t,\gd)$ on shear rate. The (often) very strong
decrease  of the viscosity, defined via $\eta(\gd)=\sigma(\gd)/\gd$,
with increasing
 flow rate is called 'shear thinning', and indicates that the particle
system is strongly affected by the solvent flow. One may thus wonder
whether the particles' non-affine, random motion relative to the
solvent differs qualitatively from the Brownian motion in the
quiescent solution. Taylor showed that this is the case for dilute
solutions. A single colloidal particle moves super-diffusively at
long times along the direction of the flow. Its mean squared
non-affine displacement grows with the third power of time, much
faster than the linear in time growth familiar from diffusion in the
quiescent system \footnote{This effect that flow speeds up the
irreversible mixing is one mechanism active when stirring a
solution. The non-affine motion even in laminar flow prevents that
stirring backwards would reverse the motion of the dissolved
constituents.}. A priori it is thus not clear, whether the
mechanisms relevant during glass formation in the quiescent system
also dominate the nonlinear rheology. Solvent mediated interactions
(hydrodynamic interactions, HI), which do not affect the equilibrium
phase diagram, may become crucially important. Also, shear may cause
ordering or layering of the particles leading to heterogeneities of
various kinds \cite{Lau:92}.

Within a number of theoretical approaches a connection between
steady state rheology and the glass transition has been suggested.
Brady worked out a scaling description of the rheology based on the
concept that the structural relaxation arrests at random close
packing \cite{Bra:93}. In the  soft glassy rheology model, the trap
model of glassy relaxation by Bouchaud was generalized by Cates and
Sollich and coworkers to describe mechanical deformations and ageing
\cite{Sol:97,Sol:98,Fie:00}. The mean field approach to  spin
glasses was generalized to systems with broken detailed balance in
order to model flow curves of glasses under shear
\cite{Ber:00,Ber:02}. The application of these novel approaches to
colloidal dispersions has lead to numerous insights, but has been
hindered by the use of unknown parameters in the approaches.

Dispersions consisting of colloidal, slightly polydisperse (near)
hard spheres arguably constitute one of the most simple viscoelastic
systems, where a glass transition has been identified. It has been
studied in detail by dynamic light scattering measurements
\cite{Pus:87,Meg:91,Meg:93,Meg:94,Meg:98,Heb:97,Bec:99,Bar:02,Eck:03},
confocal microscopy \cite{Wee:00}, linear \cite{Mas:95,Zac:06}, and
non-linear rheology
\cite{Sen:99,Sen:99b,Pet:99,Pet:02,Pet:02b,Pha:06,Bes:07,Cra:06,Cra:08,Sie:09}.
Computer simulations are available also \cite{Phu:96,Str:99,Dol:00}.
Mode coupling theory (MCT) has provided a semi-quantitative
explanation of the observed glass transition phenomena, albeit
neglecting ageing effects \cite{Pur:06} and decay processes at
ultra-long times that may cause (any) colloidal glass to flow
ultimately \cite{Goe:91,Goe:92,Goe:99}. It has thus provided a
microscopic approach recovering Maxwell's phenomenological picture
of the glass transition; $G_\infty$ and $\tau$ could be calculated
starting from the particle interactions as functions of the
thermodynamic control parameters. MCT was also generalized to
include effects of shear on time dependent fluctuations
\cite{Miy:02,miyazaki,Kob:05}, and, within the {\it integrations
through transients} (ITT) approach, to quantitatively describe all
aspects of stationary states under steady shearing
\cite{Fuc:02,Fuc:05c,Fuc:09}.

The MCT-ITT approach thus provides a microscopic route to calculate
the generalized shear modulus $g(t,\gd)$ and  other quantities
characteristic of the quiescent and the stationary state under shear
flow. While MCT has been reviewed thoroughly, see
e.g.~\cite{Goe:91,Goe:92,Goe:99}, the MCT-ITT approach shall be
reviewed here, including its recent tests by experiments in model
colloidal dispersions and by computer simulations. The recent
developments of microscopy techniques to study the motion of
individual particles under flow and the improvements in rheometry
and preparation of model systems, provide detailed information to
scrutinize the theoretical description, and to discover the
molecular origins of viscoelasticity in dense colloidal dispersions
even far away from thermal equilibrium.

The outline of the  review is as follows: At first, the microscopic
starting points, the formally exact manipulations, and the central
approximations of MCT-ITT are described in detail. Section 3
summarizes the predictions for the viscoelasticity in the linear
response regime and their recent experimental tests. These tests are
the quantitatively most stringent ones, because the theory can be
evaluated without technical approximations in the linear limit;
important parameters are introduced here, also. Section 4 is central
to the review, as it discusses the universal scenario of a glass
transition under shear. The shear melting of colloidal glasses and
the key physical mechanisms behind the structural relaxation in flow
are described. Section 5 builds on the insights in the universal
aspects and formulates successively simpler models which are
amenable to complete quantitative analysis. In the next Section,
those models are compared to experimental data on the microscopic
particle motion obtained by confocal microscopy, to data on the
macroscopic stresses in dispersions of novel model core-shell
particles close to equilibrium and under steady flow, and to
simulations providing the same information for binary supercooled
mixtures.  In the last Section, recent generalizations and open
questions are addressed.

\section{Microscopic approach}

MCT considers interacting Brownian particles, predicts a purely
kinetic glass transition and describes it using only equilibrium
structural input, namely the equilibrium structure factor $S_q$
\cite{russel,dhont} measuring thermal density fluctuations. MCT-ITT
extends this Statistical Mechanics, particle based many-body
approach to dispersions in steady flow assuming a linear solvent
velocity profile, but neglecting the solvent otherwise.

\subsection{Interacting Brownian particles}

$N$ spherical particles with radius $R_H$ are considered, which are
dispersed in a volume $V$ of solvent (viscosity $\eta_s$).
Homogeneous shear  is imposed corresponding to a constant linear
solvent velocity profile. The flow velocity points along the
$x$-axis and its gradient along the $y$-axis. The motion of  the
particles (with positions $\rb_i(t)$ for $i=1,\ldots,N$) is
described by  $N$ coupled Langevin equations \cite{dhont} \beq{b0}
\zeta \left( \frac{d \rb_i  }{d t}  - {\bf v}^{\rm solv}(\rb_i)
\right)
 = {\bf F}_i + {\bf f}_i\; .
\eeq Solvent friction is measured by the Stokes friction coefficient
$\zeta=6\pi\eta_s R_H$. The interparticle forces ${\bf
F}_i=-\partial/\partial \rb_i\, U(\left\{\rb_j\right\})$  derive
from potential interactions of particle $i$ with all other colloidal
particles; $U$ is the total potential energy. The solvent shear-flow
is given by ${\bf v}^{\rm solv}(\rb)=\gd\, y\, \hat{\bf x}$ , and
the Gaussian white noise force satisfies (with $\alpha,\beta$
denoting directions)
$$\langle f^\alpha_i(t)\; f^\beta_j(t') \rangle = 2 \zeta \, k_BT\,
\delta_{\alpha \beta}\, \delta_{ij}\, \delta(t-t')\; ,$$ where
$k_BT$ is the thermal energy. Each particle experiences
interparticle forces, solvent friction, and random kicks from the
solvent. Interaction and friction forces on each particle balance on
average, so that the particles are at rest in the solvent on
average; giving for their affine motion: $\langle \rb_i(t) \rangle
=\rb_i(0) + \gd\,t\; y_i(0)\, \hat{\bf x}$ . The Stokesian friction
is proportional to the particle's motion {\em relative to} the
solvent flow at its position; the latter varies linearly along the
$y$-direction. The random force on the level of each particle
satisfies the fluctuation dissipation relation. The interaction
forces ${\bf F}_i$ need

Even though \gl{b0} thus has been obtained under the assumption,
that solvent fluctuations are close to equilibrium, the Brownian
particle system described by it may reach macroscopic states far
from thermal equilibrium. Moreover, under (finite) shear, this holds
generally because the friction force from the solvent in \gl{b0} can
not be derived from a conservative force field. It has non-vanishing
curl, and thus the stationary distribution function $\Psi$
describing the probability of the particle positions $\rb_i$ can not
be of Boltzmann-Gibbs type \cite{Risken}.

Already the microscopic starting equation (\ref{b0}) of MCT-ITT
carries two important approximations. The first  is the neglect of
hydrodynamic interactions (HI), which would arise from the proper
treatment of the solvent flow around moving particles
\cite{russel,dhont}. As vitrification is observed in molecular
systems without HI, MCT-ITT assumes that HI are not central to the
glass formation of colloidal dispersions. The interparticle forces
are assumed to dominate and to hinder and/or prevent structural
rearrangements  close to arrest into an amorphous, metastable solid.
MCT-ITT assumes that pushing the solvent away only provides some
additional instantaneous friction, and thus lets short-time
transport properties (like single and collective short-time
diffusion coefficients, high frequency viscoelastic response, etc.)
depend on HI. The second important approximation in \gl{b0} is the
assumption of an homogeneous  shear rate $\gd$. This assumption may
be considered as a first step, before heterogeneities and
confinement effects are taken into account. The interesting
phenomena of shear localization and shear banding and shear driven
clustering \cite{dhontrev,Ben:96,Var:04,Gan:06,Bal:08} therefore are
not addressed. All difficulties in \gl{b0} thus are connected to the
many-body interactions given by the forces ${\bf F}_i$, which couple
the $N$ Langevin equations. In the absence of interactions, ${\bf
F}_i\equiv0$, \gl{b0} immediately leads to the super-diffusive
particle motion mentioned in the introduction, which often is termed
'Taylor dispersion' \cite{dhont}.

As is well known, the  considered microscopic Langevin equations,
are equivalent to  the reformulation of \gl{b0} as Smoluchowski
equation; it is a variant of a Fokker-Planck equation \cite{Risken}.
It describes the temporal evolution of the distribution function
$\Psi(\left\{\rb_i\right\} ,t)$ of the particle positions \bsq {b1}
\beq{b1a}
\partial_t \Psi(\left\{\rb_i\right\} ,t)  = \smop \; \Psi(\left\{\rb_i\right\} ,t)\; ,
\eeq employing the Smoluchowski operator (SO) \cite{russel,dhont},
\beq{b1b} \smop =  \sum_{j=1}^{N} \left[ D_0\;
\frac{\partial}{\partial \rb_j} \cdot \left(
\frac{\partial}{\partial \rb_j}  - \frac{1}{k_BT}\, {\bf F}_j
\right) - \gd\, \frac{\partial}{\partial x_j}\, y_j \right] \;,
\eeq\esq built with the Stokes-Einstein-Sutherland diffusion
coefficient $D_0=k_BT/\zeta$ of a single particle. Averages
performed with the distribution function $\Psi$ agree with the ones
obtained from the explicit Lagevin equations.

The Smoluchowski equation is a conservation law for the probability
distribution in coordinate space, $$\partial_t \, \Psi + \nabla
\cdot {\bf j} =
\partial_t \, \Psi + \sum_{i=1}^N  \frac{\partial}{\partial \rb_i} \cdot {\bf j}_i = 0\; ,$$
formed with probability current $\bf j$. Stationary distributions,
which clearly obey $\partial_t \, \Psi_s=0$,  which are not of
equilibrium type, are characterised by a non-vanishing probability
flux ${\bf j}^s_i\ne 0$, where
 $${\bf j}^s_i = D_0
\left[-  \frac{\partial}{\partial \rb_i} + \frac{1}{k_BT}\, {\bf
F}_i + \gd\, y_i \hat{\bf x} \right]\; \Psi_s \; .
$$ Under shear, ${\bf j}_s$ can not vanish, as this would require
the gradient term to balance the term proportional to $\gd$ which,
however, has a non-vanishing curl; the 'potential conditions' for an
equilibrium stationary state are violated under shear \cite{Risken}.

 The ITT approach formally exactly solves the Smoluchowski
equation, following the transients dynamics into the stationary
state. In this way the kinetic competition between Brownian motion
and shearing, which arises from the stationary flux, is taken into
account in the stationary distribution function. To explicitly, but
approximatively compute it, using ideas based on MCT, MCT-ITT
approximates the obtained averages by following the transient
structural changes encoded in the transient density correlator.

\subsection{Integration through transients (ITT) approach}

\subsubsection{Generalized Green-Kubo relations}

Formally, the H-theorem valid for general Fokker-Planck equations
states that the solution of \gl{b1} becomes unique at long times
\cite{Risken}. Yet, because colloidal particles have a
non-penetrable core and exhibit excluded volume interactions,
corresponding to regions where the potential is infinite, and the
proof of the H-theorem requires fluctuations to overcome all
barriers, the formal H-theorem may not hold for non-dilute colloidal
dispersions. Nevertheless, we assume that the system relaxes into a
unique stationary state at long times, so that
$\Psi(t\to\infty)=\Psi_s$ holds. This assumption is self-consistent,
because later on MCT-ITT finds that under shear all systems are
'ergodic' and relax into the stationary state. In cases where phase
space decomposes into disjoint pockets ('nonmixing dynamics'), the
distribution function calculated in \gl{b2} averages over all
compartments, and can thus not be used.

 As alread stated, homogeneous, amorphous systems are
assumed so that the stationary distribution function $\Psi_s$ is
translationally invariant but anisotropic. The formal solution of
the Smoluchowski equation for the time-dependent distribution
function \bsq{b2}\beq{b2a} \Psi(t) = e^{\smop\, t}\; \Psi_e\eeq can,
by taking a derivative and integrating it up to $t=\infty$, be
brought into the form \cite{Fuc:02,Fuc:05c} \beq{b2b} \Psi_s =\Psi_e
+ \frac{\gd}{k_BT} \; \int_0^\infty\!\!\!\! dt \;\Psi_e \;
\sigma_{xy} \; e^{\smopb t }\; , \eeq\esq where the adjoint
Smoluchowski $\smopb$ operator arises from partial integrations over
the particle positions (anticipating that averages built later on
with $\Psi$ are done by integrating out the particle positions). It
acts on the quantities to be averaged with $\Psi_s$. The assumption
of spatial homogeneity rules out the considerations of thermodynamic
states where the equilibrium system would e.g.~be crystalline. The
equilibrium state is described by
 $\Psi_e$, which denotes the equilibrum canonical distribution function,
$\Psi_e\propto e^{-U/(k_BT)}$, which is the time-independent
solution of \gl{b1a} for $\gd=0$; in \gl{b2b}, it gives the initial
distribution at the start of shearing (at $t=0$). The potential part
of the stress tensor \mbox{$\sigma_{xy}=-\sum_{i=1}^N F^x_i\,y_i$}
entered via $\smop \Psi_e = \gd\, \sigma_{xy}\, \Psi_e$. The simple,
exact result \gl{b2b} is central to the ITT approach as it connects
steady state properties to time integrals formed with the
shear-dependent dynamics. Advantageously, the problem to perform
steady state averages, denoted by $\langle \ldots \rangle^{(\gd)}$,
has been simplified to performing equilibrium averages, which will
be denoted as $\langle \ldots \rangle$ in the following, and contain
the familiar $\Psi_e$.  The transient dynamics integrated up in the
second term of \gl{b2b} contains slow intrinsic particle motion,
whose handling is central to the MCT-ITT approach. Generalized
Green-Kubo relations, formally valid for arbitrary $\gd$, can be
derived from \gl{b2b}.

The adjoint Smoluchowski operator was obtained using in the partial
integrations over the particle positions the incompressibility
condition, ${\rm Trace}\{\kap\}=0$, which should always holds for
the solvents of interest in this review. It takes the explicit form
(where boundary contributions are neglected throughout, simplifying
the partial integrations): $$ \smopb = \sum_i ( \bp_i + {\bf F}_i +
\rb_i\cdot \kap^T) \cdot \bp_i \; .
$$ This formula already uses a handy notation\footnote{The simplified notation with dimensionless
quantities is used in the Sections containing formal mainpulations,
and in a number of original publications.}  employing the shear rate
tensor $\kap =\gd\ \hat{\bf x} \hat{\bf y}$ (that is,
$\kappa_{\alpha\beta}=\gd \delta_{\alpha x}\delta_{\beta y}$), and
dimensionless quantities. They are introduced by using the particle
diameter $d$ as unit of length (throughout we convert $d=2R_H$), the
combination $d^2/D_0$ as unit of time, and $k_BT$ as unit of energy,
whereupon the shear rate turns into the bare Peclet number Pe$_0=\gd
d^2/D_0$. It measures the effect of affine motion with the shear
flow compared to the time it takes a single Brownian particle to
diffuse its diameter $d$. One of the central questions of the
nonlinear rheology of dense dispersions concerns the origin of very
strong shear-dependences in the viscoelasticity already for
(vanishingly) small bare Pe$_0$ numbers. Thus we will simplify by
assuming Pe$_0\ll1$, and search for another dimensionless number
characterizing the effect of shear.

The formally exact general result for $\Psi_s$ in \gl{b2b} can be
applied to compute the thermodynamic transverse stress,
$\sigma(\gd)=\langle\sigma_{xy}\rangle /V$. Equation (\ref{b2b})
leads to  an exact non-linear Green-Kubo relation:\bsq{b3}
\beq{b3a}\sigma(\gd)= \gd \int_0^\infty dt\; g(t,\gd)\; , \eeq where
the generalized shear modulus $g(t,\gd)$ depends on shear rate via
the Smoluchowski operator from \gl{b1b} \beq{b3b} g(t,\gd)=
\frac{1}{k_BT V}\; \langle\, \sigma_{xy} \; e^{\smopb t } \;
\sigma_{xy}\, \rangle^{(\gd=0)} \; . \eeq This relation is nonlinear
in shear rate, because $\gd$ appears in the time evolution operator
$\smopb$, the adjoint of \gl{b1b}.  In MCT-ITT, the slow stress
fluctuations in $g(t,\gd)$ will be approximated by following the
slow structural rearrangements, encoded in the transient density
correlators.

But, before discussing approximations, it's worthwhile to point out
that formally exact explicit expressions for arbitrary steady-state
averages can be obtained from \gl{b2b}. Using  the definition
$f(\gd) \equiv \langle f_{\bf q=0} \rangle^{(\gd)} / V$, where $\qb$
is the wavevector, and $f_{\bf q=0}=\int d\rb f(\rb)$ denotes the
integral over an arbitrary density $f(\rb)$, one finds the general
generalized Green-Kubo relation: \beqa{b3c} f(\gd) & = & \langle
f_{\bf q=0} \rangle / V + \frac{\gd}{V}  \;
\int_0^{\infty}\!\!\!\!\! dt \; \langle \sigma_{xy}  \; e^{ \smopb
\, t  }\;  \Delta f_{\bf q=0} \, \rangle \; , \eeqa\esq where the
symbol $\Delta X$ for the fluctuation in $X$ was introduced, $
\Delta X_{\bf } = X_{\bf } - \langle X_{\bf } \rangle$, because all
mean values (which are constants, for these purposes) drop out of
the ITT integrals, leaving only the fluctuating parts to contribute.
Generalizations of \gl{b3c} valid for structure functions (see
e.g.~\gl{b4b}) and  stationary correlation functions (see \gl{b6})
are presented in Ref.~\cite{Fuc:05c}. Note  that all the averages,
denoted $\langle\ldots\rangle,$ are evaluated within the (Boltzmann)
equilibrium distribution $\Psi_e$. Why only $\qb=0$ appears in
\gl{b3c} is discussed in Sect.~2.2.2 .

 It is these generalized Green-Kubo relations \gl{b3c} which are
formally exact even for arbitrary strong flows, and which form the
basis for approximations in the MCT-ITT approach. These
approximations are guided by the evident aspect that slow  dynamics
strongly affects the time integral in \gl{b3c}. Therefore, in
MCT-ITT approximations are employed that aim at capturing the slow
structural dynamics close to a glass transition. It would be
interesting to employ the generalized Green-Kubo relations also in
other contexts, where e.g.~entanglements lead to slow dynamics in
polymer melts.

\subsubsection{Aspects of translational invariance}

The generalized Green-Kubo relations contain quantities integrated/
averaged over the whole sample volume. Thus, the aspect of
translational invariance/ homogeneity does not become an issue in
\gl{b3} yet. A system is translational invariant, if the correlation
between two points $\rb$ and $\rb'$ depends on the distance
$\rb-\rb'$ between the two points only. The correlation must not
change if both points are shifted by the same amount. (Additionally,
any quantity depending on one space point $\rb$ only, must be
constant.) A system would be isotropic, if additionally, the
correlation only depended on the length of the distance vector,
$|\rb-\rb'|$; but this obviously can not be expected, because shear
flow breaks rotational symmetry of the SO in \gl{b1b}. Shear flow
also breaks translational symmetry in the SO of \gl{b1b}, therefore
it is a priori surprising, that translational invariance holds under
shear. Moreover, discussion of translational invariance introduces
the concept of an advected wavevector, which will become important
later on.

 The time-dependent distribution function $\Psi(t)$ from \gl{b2a}
can be used to show that a translationally invariant equilibrium
distribution function $\Psi_e$ leads to a translationally invariant
steady state distribution $\Psi_s$, even though the SO in \gl{b1b}
is not translationally invariant itself. To show this, a point in
coordinate space $(\rb_1,\ldots,\rb_N)$ shall be denoted by
$\Gamma$, and shall be shifted, $\Gamma \to \Gamma'$, with ${\bf
r'}_i={\bf r}_i + {\bf a}$ for all $i$; $\bf a$ is an arbitrary
constant vector.  This gives
$$ \smopb(\Gamma) =
\smopb(\Gamma') - {\bf a} \; \kap^T \; {\bf P } \; , \quad \mbox{
with }\, {\bf P} = \sum_i \bp_i \; ,$$ explicitly stating that the
SO is not translationally invariant. From \gl{b2a} follows
 $$ \Psi(\Gamma',t) = e^{\smop(\Gamma)
\, t - {\bf P} \, \kap \, {\bf a} \, t} \; \Psi_e(\Gamma) \; ,
$$
where $\Psi_e(\Gamma')=\Psi_e(\Gamma)$ was used. The SO $\smop$ and
the operator ${\bf P}\,\kap\,{\bf a}$ commute, because the shear
rate tensor satisfies $\kap\cdot \kap = 0$, and because the sum of
all internal forces vanishes due to Newton's third law: $$ \left(
{\bf P} \kap {\bf a} \right) \; \smop - \smop \; \left( {\bf P} \kap
{\bf a} \right)  = $$ $$ \sum_{ij} \left\{  \left[ \bp_i ( \bp_j
\cdot\frac{\partial U}{\partial {\bf r}_j} ) -
 ( \bp_j \cdot\frac{\partial U}{\partial {\bf r}_j} ) \bp_i \right]\kap{\bf a} -
\left[ ({\bf a}\kap^T\bp_i) \, ({\bf r}_j\kap^T\bp_j) -
(\bp_j\kap{\bf r}_j) (\bp_i\kap{\bf a})\right] \right\}  $$ $$ =
\sum_{j} \left\{ \left[ \bp_j ( \cdot \frac{\partial}{\partial {\bf
r}_j} \left( \sum_i \frac{\partial U}{\partial {\bf r}_i}\kap{\bf a}
\right) ) \right] - \left[ ({\bf a}\kap^T  \cdot \kap^T\bp_j)
\right] \right\} = 0 \; . $$
 Therefore, the Baker-Hausdorff theorem \cite{vanKampen}
simplifies \gl{b2a} to $$ \Psi(\Gamma',t) = e^{\smop(\Gamma) \, t
}\; e^{- {\bf P} \, \kap \, {\bf a} \, t} \; \Psi_e(\Gamma)  =
e^{\smop(\Gamma) \, t }\; e^{- \left( \sum_i {\bf F}_i \right) \,
\kap \, {\bf a} \, t} \; \Psi_e(\Gamma) = e^{\smop(\Gamma) \, t } \;
\Psi_e(\Gamma) \; , $$ where the last equality again holds because
the sum of all internal forces vanishes. Therefore,
$$\Psi(\Gamma',t) = \Psi(\Gamma,t) $$ holds, proving that the
time-dependent and consequently the stationary distribution function
$\Psi_s(\Gamma)=\lim_{t\to\infty} \Psi(\Gamma,t)$ are
translationally invariant. This applies, at least, in cases without
spontaneous symmetry breaking. Formally, the role of such symmetry
breaking is to discard some parts of the steady state distribution
function and keep others (with the choice dependent on initial
conditions). The distributions developed here discard nothing, and
would therefore average over the disjoint symmetry-related states of
a symmetry-broken system.

Appreciable simplifications follow from translational invariance for
steady-state quantities of
 wavevector-dependent fluctuations:
$$ f_{\bf q}(\Gamma,t) = e^{\smopb \, t} \; \sum_i\; X^f_i(\Gamma) \;
e^{i\qb \cdot {\bf r}_i}\; , $$ where e.g. $X^\rho_i=1$ describes
density fluctuations $\rho_{\bf q}(t)$, while $X^\sigma_i
=\delta_{\alpha\beta} + (1/2) \sum_j' (r^\alpha_i-r^\alpha_j)
du(|{\bf r}_i-{\bf r}_j|)/dr^\beta_i$ gives the stress tensor
element $\sigma_{\alpha\beta}({\bf q})$ for interactions described
by the pair-potential $u$. Translational invariance in an infinite
sheared system dictates that averages are independent of identical
shifts of all particle positions. As the integral over phase space
must agree for either integration variables $\Gamma$  or $\Gamma'$,
 steady-state averages can be non-vanishing for zero wavevector only:
$$\frac 1V\; \langle f_{\qb}(t)
\rangle^{(\gd)} = f_{0}(\gd)\;  \delta_{\bf q , 0} \; .$$
 The average density $n=N/V$ and the shear
stress $\sigma(\gd) = \langle \sigma_{xy}\rangle^{(\gd)}/V$ are
important examples. Wavevector-dependent steady-state structure
functions under shear become anisotropic but remain translationally
invariant, so that introduction of a single wavevector suffices. The
structure factor built with density fluctuations shall be
abbreviated by \bsq{b4}\beq{b4a} S_{\bf q}(\gd)= \frac 1N \; \langle
\delta \rho^*_\qb \; \delta \rho_\qb \rangle^{(\gd)}\; .\eeq It
needs to be kept apart from the equilibrium structure factor,
denoted by \beq{b4b} S_q = \frac 1N \; \langle \delta  \rho^*_\qb \;
\delta \rho_\qb \rangle\; ,\eeq\esq which is obtained by averaging
over the particle positions using the equilibrium  distribution
function $\Psi_e$. It will be one of the hallmarks of a shear
molten, yielding glass state, that even in the limit of vanishing
shear rate both structure factors do not agree: $S_{\bf
q}(\gd\to0)\ne S_q$ in a shear molten glassy state.
\begin{figure}[h]
\begin{center}
\epsfig{file=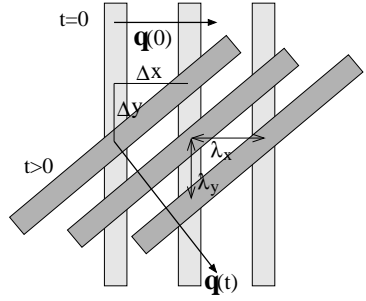,width=0.4\textwidth}
\end{center}
\caption{Shear advection of a fluctuation with initial wavevector in
$x$-direction, $\qb(t\!\!=\!\!0)=q\, (1,0,0)^T$, and advected
wavevector at later time  $\qb(t\!\!>\!\!0)= q\,  (1,-\gd t,0)^T$;
from \cite{Fuc:09}. While $\lambda_x$ is the wavelength in
$x$-direction at $t=0$, at later time $t$, the corresponding
wavelength $\lambda_y$ in (negative) $y$-direction obeys:
$\lambda_x/\lambda_y=\Delta x / \Delta y= \gd t$. At all times,
$\qb(t)$ is perpendicular to the planes of constant fluctuation
amplitude. Note that the magnitude $q(t)=q\sqrt{1+(\gd t)^2}$
increases with time. Brownian motion, neglected in this sketch,
would smear out the fluctuation.\label{Fig1}}
\end{figure}

 Translational invariance of sheared systems
takes a special form for two-time correlation functions, because a
shift of the point in coordinate space from $\Gamma$ to $\Gamma'$
gives
\[
\langle \delta f^*_{\qb}
 \;  e^{\smopb\, t }\,  \delta g_{\kb} \rangle^{(\gd)} = e^{- i \left( \kb \cdot \kap\, t + \kb -
\qb \right)\cdot
 \bf{a} }\; \langle \delta f^*_{\qb}
 \;  e^{\smopb\, t }\,  \delta g_{\kb} \rangle^{(\gd)}  \;,
\]
while obviously both averages need to agree. Therefore, a
fluctuation with wavevector $\bf q$ is correlated with a fluctuation
of ${\bf k}={\bf q}(t)$ with the {\em advected} wavevector \beq{b5}
{\bf q}(t) = {\bf q} - \qb \cdot \kap \, t\eeq at the later time
$t$; only then the exponential in the last equation becomes unity;
fluctuations with other wavevector combinations are decorrelated.
The advected wavevector's $y$-component decreases with time as
$q_y(t)=q_y-\gd\, t\; q_x$, corresponding to an (asymptotically)
decreasing wavelength, which the shear-advected fluctuation exhibits
along the $y$-direction; see Fig.~\ref{Fig1}. Taking into account
this time-dependence of the wavelength of fluctuations, a stationary
time-dependent correlation function characterized by a single
wavevector can be defined: \beq{b6} C_{fg ; {\bf q}}(t) = \frac 1N
\, \langle \delta f^*_{\qb}
 \;  e^{\smopb\, t }\,  \delta g_{\qb(t)} \rangle^{(\gd)}\;  . \eeq

Application of \gls{b2}{b3} is potentially obstructed by the
existence of conservation laws, which may cause a zero eigenvalue of
the (adjoint) SO, $\smopb$. The time integration in \gls{b2b}{b3}
would then not converge at long times. This possible obstacle when
performing memory function integrals, and how to overcome it,  is
familiar from equilibrium Green-Kubo relations \cite{Forster75}. For
Brownian particles, only the density $\rho$ is conserved. Yet,
density fluctuations do not couple in linear order to the
shear-induced change of the distribution function \cite{Fuc:05c}.
The (equilibrium) average $$\langle \sigma_{xy}\, e^{\smopb t }\,
\rho_\qb \rangle =0$$ vanishes for all $\qb$; at finite $\qb$
because of translational invariance, and at $\qb=0$ because of
inversion symmetry. Thus, the projector $Q$ can  be introduced
\beq{b7} Q = 1 - P\; ,\quad\mbox{with }\; P= \sum_\qb \delta
\rho_\qb \, \rangle \;\frac{1}{N S_q} \; \langle \, \delta
\rho^*_\qb\; . \eeq It projects any variable into the space
perpendicular to linear density fluctuations. Introducing it into
\gl{b3c} is straight forward, because couplings to linear density
can not arise in it anyway. One obtains \addtocounter{equation}{-5}
\begin{subequations}
\begin{eqnarray}\setcounter{equation}{4}\setcounter{equation}{4} f(\gd) & = & \langle
f_{\bf q=0} \rangle / V + \frac{\gd}{V}  \;
\int_0^{\infty}\!\!\!\!\! dt \; \langle \sigma_{xy}\, Q  \; e^{Q\,
\smopb \, Q\, t  }\; Q \, \Delta f_{\bf q=0} \, \rangle \;
,\label{b3d}
\end{eqnarray}\end{subequations}\addtocounter{equation}{4}
 The projection step is exact, and also formally redundant
at this stage; but it will prove useful later on, when
approximations are performed.

\subsubsection{Coupling to structural relaxation}

The generalized Green-Kubo relations, leave us with the problem of
how to approximate time-dependent correlation functions in \gl{b3}.
Their physical meaning is that at time zero, an equilibrium stress
fluctuation arises; the system then evolves under internal and
shear-driven motion until time $t$, when its correlation with a
fluctuation $\Delta f_{\bf q=0}$ is determined. Integrating up these
contributions for all times since the start of shearing gives the
difference of the shear-dependent quantities to the equilibrium
ones. During the considered time evolution, the projector $Q$
prevents linear couplings to the conserved particle density.

The time dependence and magnitudes of the correlations in \gl{b3}
shall now be approximated by using the overlaps of both the stress
and $\Delta f_{\bf q=0}$ fluctuations with appropriately chosen {\em
`relevant slow fluctuations}'. For the dense colloidal dispersions
of interest, the relevant structural rearrangements are assumed to
be {\em density fluctuations}. Because of the projector $Q$ in
\gl{b3d}, the lowest nonzero order in fluctuation amplitudes, which
we presume dominant, must then involve pair-products of density
fluctuations, $\rho_\kb\,\rho_\pb$.

The mode coupling approximation may be summarized as a rule that
applies to all fluctuation products that exhibit slow structural
relaxations but whose variables cannot couple linearly to the
density. Their time-dependence is approximated as: \bsq{b8}\beq{b8a}
Q\; e^{Q \smopb Q t}\; Q \approx \sum_{ \kb
> \pb}\; Q\, \varrho_{{\kb(-t)}}\, \varrho_{{\pb(-t)}}\; \rangle \;
\frac{\Phi_{\kb(-t)}(t) \; \Phi_{\pb(-t)}(t)}{N^2 S_{{ k}}\,  S_{{
p}}}\; \langle \;\varrho^*_{\kb} \, \varrho^*_{\pb}\, Q \eeq The
fluctuating variables are thereby projected onto pair-density
fluctuations, whose time-dependence follows from that of the
transient density correlators $\Phi_{\qb(t)}(t)$, defined in
\gl{b10} below. These describe the relaxation (caused by shear,
interactions and Brownian motion) of density fluctuations with
equilibrium amplitudes.  Higher order density averages are
factorized into products of these correlators, and the reduced
dynamics containing the projector $Q$ is replaced by the full
dynamics. The entire procedure is written in terms of {\em
equilibrium} averages, which can then be used to compute
nonequilibrium steady states via the ITT procedure. The
normalization in \gl{b8a} is given by the equilibrium structure
factors such that the pair density correlator with reduced dynamics,
which does not couple linearly to density fluctuations, becomes
approximated to: \beq{b8b}\langle \,\varrho^*_{\kb} \,
\varrho^*_{\pb}\, Q\; e^{Q \smopb Q t}\; Q\, \varrho_{{\kb(t)}}\,
\varrho_{{\pb(t)}}\, \rangle \approx N^2 S_k\, S_p\; \Phi_{\kb}(t)
\; \Phi_{\pb}(t)\;. \eeq
 \esq
This equation can be considered as central approximation of the MCT
\cite{Goe:91} and MCT-ITT approach. While the projection onto
density pairs, which is also contained/implied in \gl{b8a} may be
improved upon systematically by including higher order density or
other fluctuations, see \cite{schofield,Goe:87} for examples, no
systematic way to improve upon the breaking of averages in \gl{b8b}
has been discovered up to now, to the knowledge of the author.

 The mode
coupling approximations introduced above can now be applied to the
exact generalized Green-Kubo relations \gl{b3d}. Steady state
expectation values are approximated by projection onto pair density
modes, giving \bsq{b9} \beq{b9a} f(\gd) \approx \langle f_{\bf 0}
\rangle / V + \frac{\gd}{2V} \int_0^\infty\!\!\!dt\; \sum_{\kb}
\frac{k_xk_y(-t)S'_{k(-t)}}{k(-t)\; S^2_{k}}\; V^{f}_{\kb} \;
 \Phi^2_{\kb(-t)}(t)\, \; ,
\eeq with $t$ the time since switch-on of shear. To derive this, the
property  $\Phi^*_\kb=\Phi_{-\kb}=\Phi_{\kb}$ was used; also the
restriction $\kb>\pb$ when summing over wavevectors was dropped, and
a factor $\frac 12$ introduced, in order to have unrestricted sums
over $\kb$. Within \gl{b9a} we have already substituted the
following explicit result for the equal-time correlator of the shear
stress with density products: \beq{b9ab} \langle \sigma_{xy} \; Q \;
\rho_{\kb(-t)} \; \rho_{\pb(-t)} \rangle = N \frac{k_x \,
k_y(-t)}{k(-t)}\; S'_{k(-t)} \; \delta_{\kb(-t),-\pb(-t)} =
\frac{N}{\gd} \;
\partial_t S_{q(-t)} \; \delta_{\kb,-\pb}\; . \eeq It's an exact
equality using the equilibrium distribution function and Eq.(6)
$$\langle \sigma_{xy} \; Q \;
\rho_{\kb} \; \rho^*_{\kb} \rangle = \langle \sigma_{xy} \;
\rho_{\kb} \; \rho^*_{\kb} \rangle = \int d\Gamma \Psi_e (-\sum_i
F^x_i y_i) \; \rho_{\kb} \; \rho^*_{\kb}$$ $$ = \int d\Gamma \Psi_e
(\sum_i \partial^x_i y_i)\; \rho_{\kb} \; \rho^*_{\kb} = i k_x
\sum_{ij} \langle y_i( e^{i\kb(\rb_i-\rb_j)}-e^{-i\kb(\rb_i-\rb_j)})
\rangle
$$

Equation (\ref{b9a}), as derived via the mode-coupling rule detailed
above, contains a `vertex function' $V^f_{\kb}$, describing the
coupling of the desired variable $f$ to density pairs. This denotes
the following quantity, computed using familiar thermodynamic
equalities \beq{b9b} V^{f}_{\kb}\equiv \langle
\rho_{\kb}^*\rho_{\kb} \; Q \; \Delta f_{\bf 0} \rangle / N =
\langle \rho_{\kb}^*\rho_{\kb} \;\Delta f_{\bf 0} \rangle  / N - S_0
\left( S_k + n \, \frac{\partial S_k}{\partial n} \right) \frac 1V
\left.\frac{\partial \langle f_{\bf 0} \rangle }{\partial n}
\right)_T\; . \eeq

In ITT, the slow stress fluctuations in $g(t,\gd)$ are approximated
by following the slow structural rearrangements, encoded in the
transient density correlators. The generalized modulus becomes,
using the approximation \gl{b8a} and the vertex \gl{b9ab}: \beq{b9c}
g(t,\gd) = \frac{k_BT}{2} \, \int\!\!\frac{d^3k}{(2\pi)^3}\;
\frac{k_x^2k_yk_y(-t)}{k\, k(-t)}\; \frac{S'_kS'_{k(-t)}}{S^2_{k}}\;
\Phi^2_{\kb(-t)}(t)\; , \eeq \esq Summation over wavevectors has
been turned into integration in \gl{b9c} considering an infinite
system.

The familiar shear modulus of linear response theory describes
thermodynamic stress fluctuations in equilibrium, and is obtained
from \gls{b3b}{b9c} by setting $\gd=0$ \cite{Larson,russel,Nae:98}.
While \gl{b3b} then gives the exact Green-Kubo relation, the
approximation \gl{b9c} turns into the well-studied MCT formula (see
\gl{cc2} below).  For finite shear rates, \gl{b9c} describes how
affine particle motion causes stress fluctuations to explore shorter
and shorter length scales. There the effective forces, as measured
by the gradient of the direct correlation function, $S'_k/S_k^2 = n
c'_k=n\partial c_k/\partial k$, become smaller, and vanish
asympotically, $c'_{k\to\infty} \to 0$; the direct correlation
function $c_{k}$ is connected to the structure factor via the
Ornstein-Zernicke equation $S_k=1/(1-n\, c_k)$, where $n=N/V$ is the
particle density. Note, that the equilibrium structure suffices to
quantify the effective interactions, while shear just pushes the
fluctuations around on the 'equilibrium energy landscape'.

While, in the linear response regime, modulus and density correlator
are measurable quantities, outside the linear regime, both
quantities serve as tools in the ITT approach only. The transient
correlator and shear modulus provide a route to the stationary
averages, because they describe the decay of equilibrium
fluctuations under external shear, and their time integral provides
an approximation for the stationary distribution function.
Determination of the frequency dependent moduli under large
amplitude oscillatory shear has become possible recently only
\cite{Miy:06}, and requires an extension of the present approach to
time dependent shear rates in \gl{b1} \cite{Bra:07}.

\subsubsection{Transient density correlator}

 In ITT, the evolution towards the stationary
distribution at infinite times is approximated by following the slow
structural rearrangements, encoded in the transient density
correlator $\Phi_\qb(t)$. It is defined by \cite{Fuc:02,Fuc:05c}
\beq{b10} \Phi_\qb(t) = \frac{1}{NS_q}\; \langle\,
\delta\varrho^*_{\qb} \; e^{\smopb t } \; \delta\varrho_{\qb(t)}\,
\rangle^{(\gd=0)}\; . \eeq It describes the fate of an equilibrium
density fluctuation with wavevector $\qb$, where $\varrho_\qb =
\sum_{j=1}^N e^{i \qb \cdot \rb_j}$, under the combined effect of
internal forces, Brownian motion and shearing. Note that because of
the appearance of $\Psi_e$ in \gl{b2}, the average in \gl{b10} can
be evaluated with the equilibrium canonical distribution function,
while the dynamical evolution contains Brownian motion and shear
advection. The normalization is given by $S_q$ the equilibrium
structure factor \cite{russel,dhont} for wavevector modulus $q=|{\bf
q}|$. The {\em advected} wavevector from \gl{b5} enters in \gl{b10}.
The time-dependence in ${\bf q}(t)$ results from the affine particle
motion with the shear flow of the solvent. Again, irrespective of
the use of $\Psi_e$ in \gl{b10}, or $\Psi_s$ in \gl{b6}, in both
cases translational invariance under shear dictates that at a time
$t$ later, the  density fluctuation $\delta\varrho^*_\qb$ has a
nonvanishing overlap only with the advected fluctuation
$\delta\varrho_{\qb(t)}$.  Figure \ref{Fig1} again applies, where a
non-decorrelating fluctuation is sketched under shear. In the case
of vanishing Brownian motion, viz.~$D_0=0$ in \gl{b1b}, we find
$\Phi_{\qb}(t)\equiv1$, because the advected wavevector takes
account of simple affine particle motion. The relaxation of
$\Phi_{\qb}(t)$ thus heralds decay of structural correlations by
Brownian motion, affected by shear.

\subsubsection{Zwanzig-Mori equations of motion}

Structural rearrangements of the dispersion affected by Brownian
motion is encoded in the transient density correlator. Shear induced
affine motion, viz.~ the case $D_0=0$, is not sufficient to cause
$\Phi_\kb(t)$ to decay. Brownian motion of the quiescent correlator
$\Phi^{(\gd=0)}_{k}(t)$ leads at high densities to a slow structural
process which arrests at long times in (metastable) glass states.
Thus the combination of structural relaxation and shear is
interesting. The interplay between intrinsic structural motion and
shearing in $\Phi_\kb(t)$ is  captured by $(i)$ first a  formally
exact Zwanzig-Mori type equation of motion, and $(ii)$ second a mode
coupling factorisation in the memory function built with
longitudinal stress fluctuations \cite{Fuc:02,Fuc:05c,Fuc:09}.  The
equation of motion for the transient density correlators is
\beq{b11}
\partial_t \Phi_\qb(t) + \Gamma_\qb(t) \; \left\{
\Phi_\qb(t) + \int_0^t dt'\; m_\qb(t,t') \; \partial_{t'}\,
\Phi_\qb(t') \right\} = 0 \; , \eeq where the initial decay rate
$\Gamma_\qb(t)= D_0\, q^2(t)/S_{q(t)}$ generalizes the familiar
result from linear response theory to advected wavevectors; it
contains Taylor dispersion mentioned in the introduction, and
describes the short time behavior, $\Phi_{\qb}(t\to0)\to
1-\Gamma_{\qb}(0)\, t+\ldots$ .

\subsubsection{Mode-coupling closure}

The memory equation contains fluctuating stresses and similarly like
$g(t,\gd)$ in \gl{b9c}, is calculated in mode coupling approximation
using \gl{b8a} giving: \bsq{b12} \beq{b12a} m_\qb(t,t') =
\frac{1}{2N} \sum_{\kb} V_{\qb\kb\pb}(t,t')\;
 \Phi_{\kb(t')}(t-t')\;  \Phi_{\pb(t')}(t-t')\; ,
\eeq where we abbreviated $\pb=\qb-\kb$. The vertex generalizes the
expression in the quiescent case, see \gl{cc3c} below, and depends
on two times capturing that shearing decorrelates stress
fluctuations \cite{Fuc:02,Fuc:05c,Fuc:09} \beqa{b12b}
V_{\qb\kb\pb}(t,t') &=& \frac{S_{\qb(t)} \, S_{\kb(t')}\,
S_{\pb(t')}}{q^2(t)\, q^2(t')}\;
{\cal V}_{\qb\kb\pb}(t)\; {\cal V}_{\qb\kb\pb}(t')\, , \nonumber \\
{\cal V}_{\qb\kb\pb}(t) &=& \qb(t)\cdot \left( \, \kb(t)\; n
c_{k(t)} + \pb(t)\; n c_{p(t)}\, \right)\;.  \eeqa\esq With shear,
wavevectors in \gl{b12b} are advected according to \gl{b5}.

The summarized MCT-ITT equations  form a closed set of equations
determining  rheological properties of a sheared dispersion from
equilibrium structural input \cite{Fuc:02,Fuc:05c,Fuc:09}. Only the
static structure factor $S_q$ is required to predict $(i)$ the time
dependent shear modulus within linear response, $g^{\rm lr}(t)=
g(t,\gd=0)$, and $(ii)$ the stationary stress $\sigma(\gd)$ from
\gl{b3a}.

\subsection{A microscopic model: Brownian hard spheres}

In the microscopic ITT approach, the rheology is determined from the
equilibrium structure factor $S_q$ alone. This holds at low enough
frequencies and shear rates, and excludes a single time scale, to be
denoted by the  parameter $t_0$ in \gl{c3b}, which needs to be found
by matching to the short time dynamics. This prediction has as
consequence that the moduli and flow curves should be a function
only of the thermodynamic parameters characterizing the present
system, viz.~its structure factor. Because the structure factor for
simple fluids far away from demixing and other phase separation
regions can be mapped onto the one of hard spheres, the system of
hard spheres plays a special role in the MCT-ITT approach. It
provides the most simple microscopic model where slow strucutral
dynamics can be studied. Moreover, other experimental systems can be
mapped onto it by chosing an effective packing fraction $\phi_{\rm
eff}=(4\pi/3) n R_H^3$ and particle radius so that the structure
factors agree.

The claim  that the rheology follows from $S_q$ is supported if the
rheological properties of a dispersion only depend on the effective
packing fraction, if particle size is taken account of properly.
Obviously, appropriate scales for frequency, shear rate and stress
magnitudes need to be chosen to observe this; see Sect.~6.2. The
dependence of the rheology (via the vertices) on $S_q$
 suggests that $k_BT$ sets the energy scale as long as repulsive
interactions dominate the local packing. The length scale is set by
the average particle separation, which can be taken to scale with
$R_H$. The time scale of the glassy rheology within ITT is given by
$t_0$, which should scale with the measured dilute diffusion
coefficient $D_0$. Thus the rescaling of the rheological data can be
done with measured parameters alone.

Because the hard sphere system thus provides the most simple system
to test and explore MCT-ITT, numerical calculations only for this
model will be reviewed in the present overview. Input for the
structure factor is required, which, for simplicity, will be taken
from the analytical Percus-Yervick (PY) approximation
\cite{russel,Goe:92}. Straightforward discretization of the
wavevector integrals will be performed as discussed below, and in
detail in the quoted original papers.

\subsection{Accounting for hydrodynamic interactions}

Solvent-particle interactions (viz. the HI) act instantaneously if
the particle microstructure differs from the equilibrium one, but do
not themselves determine the equilibrium structure
\cite{dhont,russel}. If one assumes that glassy arrest is connected
with the ability of the system to explore its configuration space
and to approach its equilibrium structure, then
 it appears natural to assume that the solvent particle interactions
are characterized by a finite  time scale $\tau_{\rm HI}$. And that
they do not shift the glass transition nor affect the frozen glassy
structure.  HI would thus only lead to an increase of the high
frequency viscosity above the solvent value; this value shall be
denoted as $\eta_{\infty}$: \bsq{b13}\beq{b13a} g(t,\gd) \to
g(t,\gd) + \eta_\infty \delta( t - O+) \;.\eeq
 The parameter $\eta_\infty$ would thus characterize
a short-time, high frequency viscosity and model viscous processes
which require no structural relaxation. It can be measured from the
high frequency dissipation \beq{b13b} G''(\omega\to\infty) =
\eta_\infty\; \omega \; . \eeq For identical reasoning, also the
short time diffusion in the collective (and single particle) motion
will be affected by HI. The most simple approximation is to adjust
the initial decay rate \beq{b13c} \Gamma_\qb(t)= D_s\,
q^2(t)/S_{q(t)}\; ,\eeq where the collective short time diffusion
coefficient $D_s$ accounts for HI and other (almost) instantaneous
effects which affect the short time motion, and which are not
explicitly included in the MCT-ITT approach. \esq

The naive picture sketched here, is not correct for a number of
reasons. It is well known that for hard spheres without HI the
quiescent shear modulus diverges for short times, $g^{\rm lr, HS no
HI}(t\to0)\sim t^{-1/2}$. Lubrication forces, which keep the
particles apart,
  eliminate this divergence and render
$g^{\rm lr, HI}(t\to0)$ finite \cite{Lio:94}.
 Thus, the simple separation of the modulus into HI and potential part
 is not possible for short times, at
least for particles with a hard core. Moreover, comparison of
simulations without and with HI has shown that the increase of
$(\eta_{0}-\eta_{\infty})/\eta_{\infty}$ depends somewhat
 on HI, and thus not just on the potential interactions as implied.

  Nevertheless the sketched picture provides the most basic
view of a glass transition in colloidal suspensions, connecting it
with the increase of the structural relaxation time $\tau$.
Increased density or interactions cause a slowing down of particle
rearrangements which leave the HI relatively unaffected, as these
solvent mediated forces act on all time scales. Potential forces
dominate the slowest particle rearrangements because vitrification
corresponds to the limit where they actually prevent the final
relaxation of the microstructure. The structural relaxation time
$\tau$ diverges at the glass transition, while $\tau^{\rm HI}$ stays
finite. Thus close to arrest a time scale separation is possible,
$\tau\gg\tau^{{\rm HI}}$.

\subsection{Comparison with other MCT inspired approaches
to sheared fluids}

The MCT-ITT approach aims at describing the steady state properties
of a concentrated dispersion under shear. Stationary averages are
its major output, obtained via the integration through transients
procedure from (approximate) transient fluctuation functions, whose
strength is the equilibrium one, and whose dynamics originates from
the competition between Brownian motion and shear induced
decorrelation. In this respect, the MCT-ITT approach differs from
the interesting recent generalization of MCT to sheared systems by
Miyazaki, Reichman and coworkers \cite{Miy:02,miyazaki}. These
authors considered the stationary but time-dependent fluctuations
around the steady state, whose amplitude is the stationary
correlation function, e.g.~in the case of density fluctuations, it
is the distorted structure factor $S_\qb(\gd)$ \gl{b4a}. In the
approach by Miyazaki et al.~this structure factor is an
input-quantity required to calculate the dynamics, while it is an
output quantity, calculated in MCT-ITT from the equilibrium $S_q$ of
\gl{b4b}. Likewise, the stationary stress as function of shear rate,
viz.~the flow curve $\sigma(\gd)$, is a quantity calculated in
MCT-ITT, albeit using mode coupling approximations, while in the
approach of Refs.~\cite{Miy:02,miyazaki} additional ad-hoc
approximations beyond the mode coupling approximation are required
to access $\sigma(\gd)$. Thus, while the scenario of an
non-equilibrium transition between a shear-thinning fluid and a
shear molten glass, characterized by universal aspects in
e.g.~$\sigma(\gd)$ --- see the discussion in Sect. 4 --- forms the
core of the MCT-ITT results, this scenario can not be directly
addressed based on Refs.~\cite{Miy:02,miyazaki}.

Because the recent experiments and simulations reviewed here
concentrated on the universal aspects of the novel non-equilibrium
transition, focus will be laid on the MCT-ITT approach.
Reassuringly, however, many similarities between the MCT-ITT
equations and the results by Miyazaki and Reichman exist, even
though these authors used a different, field theoretic approach to
derive their results. This supports the robustness of the mechanism
of shear-advection in \gl{b5} entering the MCT vertices in
\gls{b9c}{b12}, which were derived independently in
Refs.~\cite{Miy:02,miyazaki} and Refs.~\cite{Fuc:02,Fuc:05c,Fuc:09}
from quite different theoretical routes. This mechanism had been
known from earlier work on the dynamics of critical fluctuations in
sheared systems close to phase transition points \cite{Onu:79}, on
current fluctuations in simple liquids \cite{Kaw:73}, and on
incoherent density fluctuations in dilute solutions \cite{Ind:95}.
Different possibilities also exist to include shear into
MCT-inspired approaches, especially the one worked out by Schweizer
and coworkers including strain into an effective free energy
\cite{Kob:05}. This approach does not recover the (idealized) MCT
results reviewed below but starts from the extended MCT where no
true glass transition exists and describes a crossover scenario
without e.g.~a true dynamic yield stress as discussed below.

\section{Microscopic results in linear response regime}

Before turning to the properties of the stationary non-equilibrium
states under shear, it is useful to investigate the quiescent
dispersion close to vitrification. Consensus on the ultimate
mechanism causing glassy arrest may yet be absent, yet, the
so-called 'cage effect' has lead to a number of fruitful insights
into glass formation in dense colloidal dispersions. For example, it
was extended to particles with  a short ranged attraction leading to
at first surprising predictions
\cite{sciortino,bergen,dawson,pham1}.
\begin{figure}[h]
\centering
\includegraphics[width=0.5\textwidth]{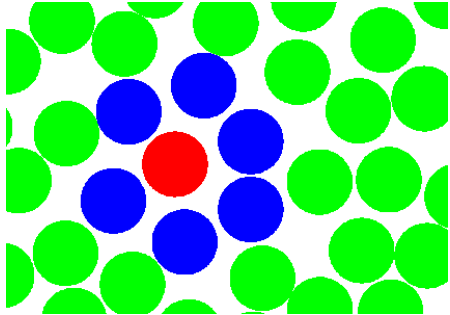}
\caption{\label{kaefig} Positions of hard disks in two dimensions
from a Monte Carlo simulation at a density close to freezing;
courtesy of Th. Franosch. A particle and its shell of neighbours is
highlighted by different colors/shadings.}
\end{figure}

Figure \ref{kaefig} shows a section of the cell of a Monte Carlo
simulation of hard disks moving in two dimensions (for simplicity of
visualization). The density is just below freezing and the sample
was carefully equilibrated. Only 100 disks were simulated, so that
finite size effects cannot be ruled out. Picking out a disk, it is
surrounded by a shell of on average 6 neighbours (in two dimension,
of 12 neighbours in three dimension), which hinder its free motion.
In order for the central particle to diffuse at long times, it needs
to escape the shell of neighbours. In order for a gap in this shell
to open at higher concentrations, the neighbours have to be able to
move somewhat themselves.  Yet, each neighour is hindered by its own
shell of neighbours, to which the originally picked particle
belongs. Thus one can expect a cooperative feedback mechanism that
with increasing density or particle interactions particle
rearrangements take more and more time. It appears natural, that
consequently stress fluctuations also slow down and the system
becomes viscoelastic.

MCT appears to capture the cage-effect in supercooled liquids and
predicts that it dominates the slow relaxation of structural
correlations close to the glass transition \cite{Goe:91,Goe:92}.
Density fluctuations play an important role because they are well
suited to describe the structure of the particle system and its
relaxation. Moreover, stresses that decay slowly because of the slow
particle rearrangements, MCT argues, also can be approximated by
density fluctuations
 using effective potentials. Density fluctuations not at large
wavelengths, but for wavelengths corresponding to the average
particle distance turn out to be the dominant ones. In agreement
with the picture of the caging of particles by structural
correlations, the MCT glass transition is independent on whether the
particles move ballistically in between interactions with their
neighbors (say collisions for hard spheres) or by diffusion.
Structural arrest happens whenever the static density correlations
for wavelengths around the average particle distance are strong
enough. The arrest of structural correlations entails an increase in
the viscosity of the dispersion connected to the existence of a slow
Maxwell-process in the shear moduli. While the MCT solutions for
density fluctuations have been thoroughly reviewed, the viscoelastic
spectra have not been presented in such detail.

\subsection{Shear moduli close to the glass transition}

\subsubsection{MCT equations and results for hard spheres}

The loss and storage moduli of small amplitude oscillatory shear
measurements \cite{Larson,russel} follow from \gl{b3b} in the linear
response case at $\gd=0$: \bsq{cc1}\beq{cc1a} G'(\omega) + i\,
G''(\omega) = i \omega\; \int_0^\infty dt\; e^{-i\, \omega\, t}\;
g^\text{lr}(t) \; . \eeq Here, the shear modulus in the linear
response regime is, again like the transient one in \gl{b3b},
obtained from equilibrium averaging: \beq{cc1b} g^\text{lr}(t)=
\frac{1}{k_BT V}\; \langle\, \sigma_{xy} \; e^{\smopb_e t } \;
\sigma_{xy}\, \rangle^{(\gd=0)} \; , \eeq yet, differently from the
transient one, the equilibrium one contains the equilibrium SO
$\smop_e$, which characterizes the quiescent system: \beq{cc1c}
\smop_e = \sum_{j=1}^{N} D_0\; \frac{\partial}{\partial \rb_j} \cdot
\left( \frac{\partial}{\partial \rb_j}  - \frac{1}{k_BT}\, {\bf F}_j
\right)  \;, \eeq\esq The linear response modulus thus quantifies
the small stress fluctuations, which are excited by thermally, and
relax because of Brownian motion.
\begin{figure}[h] \centering
\includegraphics[width=0.8\textwidth]{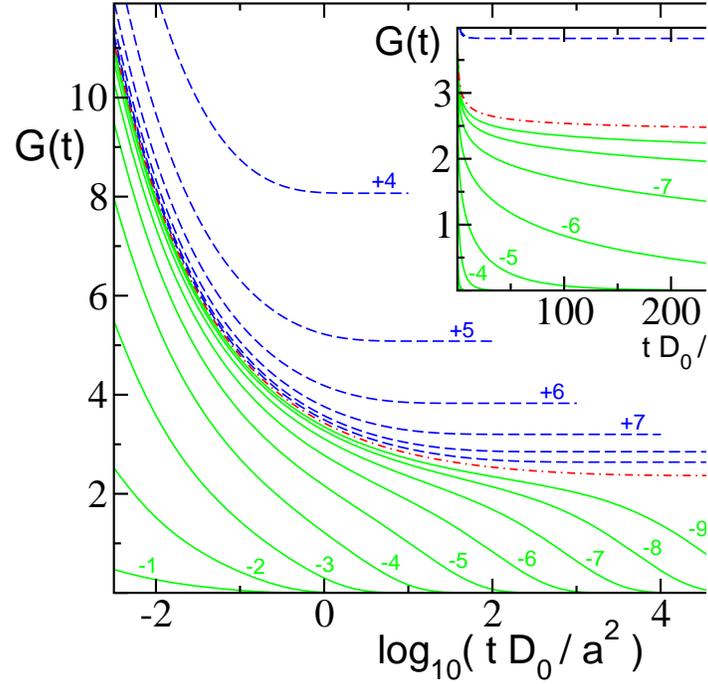}
\caption{\label{bild2} Equilibrium potential or linear response
shear modulus $G(t)=g^\text{lr}(t)$ (in units of $k_{B}T/R_H^{3}$)
for  Brownian
  hard spheres  with radius $a=R_H$ for packing fractions close to
  vitrification at $\phi_{c}$; from Ref.~\cite{Fuc:05}. Densities are measured by the separation
  parameter $\epsilon=(\phi-\phi_{c})/\phi_{c}=\pm 10^{-|n|/3}$, and labels
  denote the value $n$. Positive values belong to glass ($\epsilon>0$),
  negative to   fluid states ($\epsilon<0$);
 the label $c$ gives the transition. The inset shows a
  subset of the curves on a linear time axis; the increase of
  $g^\text{lr}(t)$
  for short times cannot be resolved. }
\end{figure}

Predictions of the (idealized) MCT equations for the potential part
of the equilibrium, time-dependent shear modulus $g^\text{lr}(t)$ of
hard spheres for various packing fractions $\phi$ are shown in Fig.
\ref{bild2} and calculated from the  limit of \gl{b9c} for vanishing
shear rate: \beq{cc2}  g^\text{lr}(t) \approx \frac{k_BT}{60\pi^2}
\; \int_0^\infty\!\!dk\; k^4\; \left( \frac{\partial
\ln{S_k}}{\partial k} \right)^2\; \Phi^2_{{ k}}(t)\; , \eeq The
normalized density fluctuation functions  are calculated
self-consistently within MCT from the \gls{b11}{b12} at vanishing
shear rate \cite{Goe:91,Goe:92}, which turn into the quiescent MCT
equations: \bsq{cc3}\beq{cc3a}
\partial_t \Phi_q(t) + \Gamma_q \; \left\{
\Phi_q(t) + \int_0^t dt'\; m_q(t-t') \; \partial_{t'}\, \Phi_q(t')
\right\} = 0 \; , \eeq where the initial decay rate $\Gamma_q= D_s\,
q^2/S_{q}$ describes diffusion with a short-time diffusion
coefficient $D_s$ differing from the $D_0$ because of HI; $D_s=D_0$
will be taken for exemplary calculations, while $D_s\ne D_0$ is
required for analyzing experimental data. The memory kernel becomes
(again with abbreviation $\pb=\qb-\kb$) \beq{cc3b} m_q(t) =
\frac{1}{2N} \sum_{\kb} V_{qkp}\;
 \Phi_{k}(t)\;  \Phi_{p}(t)\;  ,
\eeq  \beq{cc3c} V_{qkp} = \frac{S_q \, S_k\, S_{p}}{q^4}\;\left(
\qb\cdot \left[ \, \kb\; n c_{k} + \pb\; n c_{p}\, \right]
\right)^2\;.  \eeq\esq
\begin{figure}[h]
\centering
\includegraphics[width=0.8\textwidth]{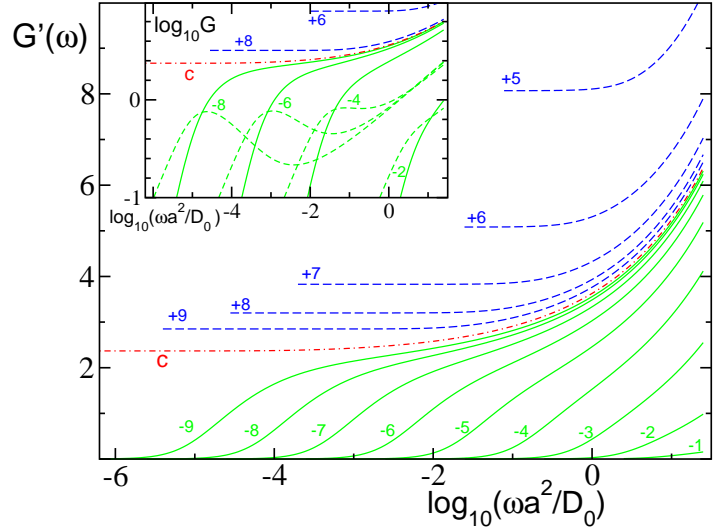}
\caption{\label{bild3} Storage part of the shear modulus
$G'(\omega)$ corresponding to Fig. \protect\ref{bild2}. The inset
shows storage and loss moduli (only for fluid states) for a number
of densities. }
\end{figure}
\begin{figure}[h]
\centering
\includegraphics[width=0.8\textwidth]{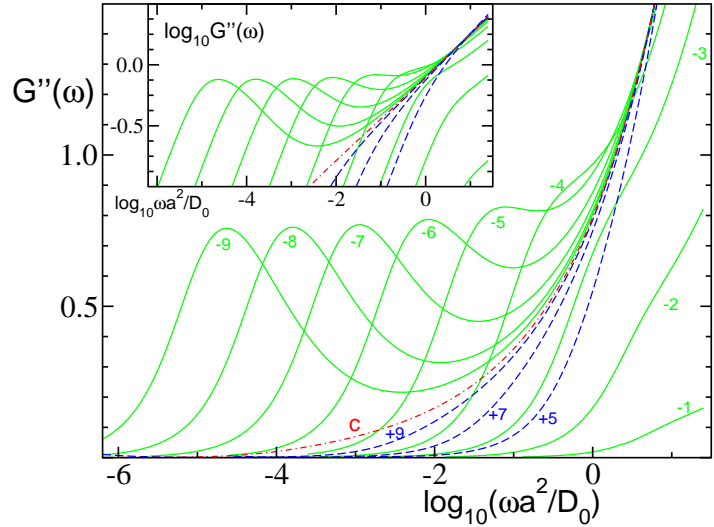}
\caption{\label{bild4} Loss part of the shear modulus $G''(\omega)$
corresponding to Fig. \protect\ref{bild2}. The inset shows the same
data in a double logarithmic representation.}
\end{figure}
 Packing fractions are conveniently  measured in relative
separations $\epsilon=(\phi-\phi_{c})/\phi_{c}$ to the glass
transition point, which for this model of hard spheres lies at
$\phi_{c}=0.516$ \cite{Goe:91,Fra:97}.
 Note that this result depends on the
static structure factor $S(q)$ , which is taken from Percus-Yevick
theory, and that the experimentally determined value $\phi_{c}^{\rm
expt.}=0.58$ lies somewhat higher \cite{Meg:93,Meg:94}. The
wavevector integrals  were discretized using $M=100$ wavevectors
chosen from $k_{\rm min}=0.1/R_H$ up to $k_{\rm max}=19.9/R_H$ with
separation $\Delta k = 0.2/R_H$ for Figs.~\ref{bild2} to
\ref{bild4}, or using $M=600$ wavevectors chosen from $k_{\rm
min}=0.05/R_H$ up to $k_{\rm max}=59.95/R_H$ with separation $\Delta
k = 0.1/R_H $ for Figs.~\ref{Fig5} and \ref{Fig6}, and in Sect. 6.2
in Fig. \ref{Fig:5}. Time was discretized with initial step-width
$dt=2\,10^{-7} R_H^2/D_s$, which was doubled each time after 400
steps. Slightly different discretizations in time and wavevector of
the MCT equations  were used in Sects. 3.2.1 and 4, causing only
small quantitative differences whose discussion goes beyond the
present review. The quiescent density correlators $\Phi_q(t)$
corresponding to the following linear response moduli have
thoroughly been discussed in Ref.~\cite{Fra:97}.

 For low packing fractions, or large negative
separations, the modulus decays quickly on a time-scale set by the
short-time diffusion of well separated particles. The strength of
the modulus increases strongly at these low densities, and its
behavior at short times presumably depends sensitively on the
details of hydrodynamic and potential interactions; thus Fig.
\ref{bild2} is not continued to small times, where the employed
model (taken from Ref. \cite{Fra:97,Fuc:99}) is too crude
\footnote{The MCT shear modulus at short times depends sensitively
on the large cut-off $k_{\rm max}$ for hard spheres \cite{Nae:98},
\mbox{$g(t,\gd=0)=(n^2k_BT/60\pi^2) \int_{k_{\rm min}}^{k_{\rm max}}
dk\, k^4 (c'_k)^2 \; S_k^2\, \Phi_k^2(t)$} gives the qualitatively
correct \cite{Lio:94,Ver:97} short time $g^\text{lr}(t\to0) \sim
t^{-1/2}$, or high frequency divergence \mbox{$G'(\omega \gg
D_0/R_H^2) \sim \sqrt{\omega}$} only for $k_{\rm max}\to\infty$. }.
Approaching the glass transition from below, $\epsilon\nearrow0$,
little changes in $g^\text{lr}(t)$ at short times, because the
absolute change in density becomes small. Yet, at long times a
process in $g^\text{lr}(t)$ becomes progressively slower upon taking
$\epsilon$ to zero. It can be considered the MCT analog of the
phenomenological Maxwell-process. MCT finds that it depends on the
equilibrium structural correlations only, while HI and other short
time effects only shift its overall time scale. Importantly, this
overall time scale applies to the slow process in coherent and
incoherent density fluctuations as well as in the stress
fluctuations \cite{Fra:98}. This holds even though e.g. HI are known
to affect short time diffusion coefficients and high frequency
viscosities differently. Upon crossing the glass transition, a part
of the relaxation freezes out and the amplitude $G_{\infty}$ of the
Maxwell-process does not decay; the modulus for long times
approaches the elastic constant of the glass
$g^\text{lr}(t\to\infty)\to G_{\infty}>0$. Entering deeper into the
glassy phase the elastic constant increase quickly with packing
fraction.

 The corresponding storage $G'(\omega)$ and loss
$G''(\omega)$ moduli are shown as functions of frequency in Figs.
\ref{bild3} and \ref{bild4}, respectively. The slow Maxwell-process
appears as a shoulder in $G'$ which extends down to lower and lower
frequencies when approaching glassy arrest, and reaches to zero
frequency  in the glass, $G'(\omega=0)=G_{\infty}$. The slow process
shows up as a peak in $G''$ which in parallel motion (see inset of
Fig. \ref{bild3}) shifts to lower frequencies when
$\epsilon\nearrow0$. Including
 hydrodynamic interactions into the calculation by adjusting $\eta_\infty$
 would affect
the frequency dependent moduli at higher frequencies only. For the
range of smaller frequencies which is of interest here, only a small
correction would arise.

\subsubsection{Comparison with experiments}

Recently, it has been demonstrated that suspensions of
thermosensitive particles present excellent model systems for
measuring  the viscoelasticity of dense concentrations. The
particles consist of a solid core of polystyrene onto which a
thermosensitive network of poly(N-isopropylacrylamide) (PNIPAM) is
attached \cite{Cra:06,Cra:08}.  The PNiPAM shell of these particles
swells when immersed in cold water (10 - 15$^o$C). Water gets
expelled at higher temperatures leading to a considerable shrinking.
Thus, for a given number density the effective volume fraction
$\phi_\text{eff}$ can be adjusted within wide limits by adjusting
the temperature. Senff \textit{et al.} (1999) were the first to
demonstrate the use of these particles as model system for studying
the dynamics in concentrated suspensions \cite{Sen:99,Sen:99b}. The
advantage of these systems over the classical hard sphere systems
are that dense suspensions can be generated \textit{in situ} without
shear and mechanical deformation. The previous history of the sample
can be erased by raising the temperature and thus lowering the
volume fraction to the fluid regime.

\begin{figure}[h]
\centering
\includegraphics[ width=0.8\columnwidth]{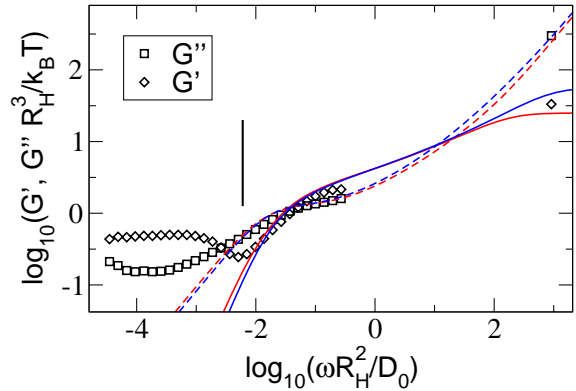}
\caption{The reduced storage (diamonds and solid lines) and loss
(squares and broken lines) modulus for a fluid state at effective
packing fraction $\phi_{\rm eff}=0.540$; from Ref.~\cite{Cra:08}.
The vertical bars mark the minimal rescaled frequency above which
the influence of crystallisation can be neglected. Parameters in the
MCT calculation given as blue lines: $\epsilon=-0.01$,
$\frac{D_{S}}{D_{0}}=0.15$, and $\eta_{\infty}=0.3\,
k_{B}T/(D_{0}R_{H}) $; moduli scale factor $c_{y}=1.4 $. For the
other lines see Ref.~\cite{Cra:08}. \label{Fig5} }
\end{figure}
\begin{figure}[h]
\centering
\includegraphics[ width=0.8\columnwidth]{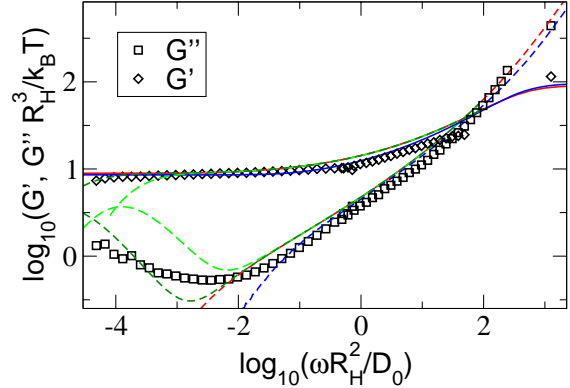}
\caption{The moduli for a glass state at effective packing fraction
$\phi_{\rm eff}=0.622$; storage (diamonds and solid lines) and loss
(squares and broken lines) moduli from \cite{Cra:08}. MCT fits are
shown as blue lines with parameters: $\epsilon=0.03$,
$\frac{D_{S}}{D_{0}}=0.08$, and $\eta_{\infty}=0.3\,
k_{B}T/(D_{0}R_{H}) $; moduli scale factor $c_{y}=1.4 $ .  For the
other lines see Ref.~\cite{Cra:08}. \label{Fig6} }
\end{figure}

Frequency dependent moduli were measured spanning a wide density and
frequency range by combining different techniques. The moduli
exhibit a qualitative change when increasing the effective packing
fraction from around 50\% to above 60\%. For lower densities (see
Fig.~\ref{Fig5}), the spectra $G''(\omega)$ exhibit a broad peak or
shoulder, which corresponds to the final or $\alpha$-relaxation. Its
peak position (or alternatively the crossing of the moduli,
$G'=G''$) is roughly given by $\omega \tau=1$ . These properties
characterize a viscoelastic fluid. For higher density, see
Fig.~\ref{Fig6}, the storage modulus exhibits an elastic plateau at
low frequencies. The loss modulus drops far below the elastic one.
These observations characterize a soft solid \footnote{The loss
modulus rises again at very low frequencies, which may indicate that
the colloidal glass at this density is metastable and may have a
finite lifetime (an ultra-slow process is discussed in Ref.
\cite{Cra:08}).}.

The linear response moduli are   affected by the presence of small
crystallites. At low frequencies, $G'(\omega)$ and $G''(\omega)$
increase above the behavior expected for a solution
($G'(\omega\to0)\to\eta_0\, \omega$ and $G''(\omega\to0)\to c\,
\omega^2$) even at low density, and exhibit elastic contributions
(apparent from $G'(\omega)>G''(\omega)$); see Fig.~\ref{Fig5}. This
effect tracks the crystallisation of the system during the
measurement after a strong preshearing at $\dot{\gamma}=100$
$s^{-1}$. Only data can be considered which were collected before
the crystallisation time; they lie to the right of the vertical bar
in Fig.~\ref{Fig5}. While this experimental restriction limits more
detailed studies of the shapes of the spectra close to the glass
transition, the use of a system with a rather narrow size
distribution provides the quantitatively closest comparison with MCT
calculations for a monodisperse hard sphere system. Especially the
magnitude of the stresses and the effective densities can be
investigated quantitatively.

Included in figures \ref{Fig5} and \ref{Fig6} are calculations using
the microscopic MCT given by \glto{cc1}{cc3} evaluated for hard
spheres in PY approximation. The only a priori unknown, adjustable
parameter is a frequency or time scale, which was adjusted by
varying the short time diffusion coefficient $D_s$ appearing in the
initial decay rate in \gl{cc3a}. Values for $D_s/D_0$ are reported
in the captions. The viscous contribution to the stress is mimicked
by including $\eta_\infty$ like in \gl{b13}; it can directly be
measured at the highest frequencies. Gratifyingly, the stress values
computed from the microscopic approach are close to the measured
ones; they are too small by 40\% only, which may arise from the
approximate structure factors entering the MCT calculation; the
Percus-Yevick approximation was used here \cite{russel}. In order to
compare the shapes of the moduli the MCT calculations were scaled up
by a factor $c_y=1.4$ in Figs.~\ref{Fig5} and \ref{Fig6}.
Microscopic MCT also does not hit the correct value for the glass
transition point \cite{Goe:92,Goe:91}. It finds $\phi^{\rm
MCT}_c=0.516$, while experiments give $\phi^{\rm exp}_c\approx0.58$.
Thus, when comparing, the relative separation from the respective
transition point needs to be adjusted as, obviously, the spectra
depend sensitively on the distance to the glass transition; the
fitted values of the separation parameter $\epsilon$ are included in
the captions.

Overall, the semi-quantitative agreement between the linear
viscoelastic spectra and first-principles MCT calculations is very
promising. Yet, crystallization effects in the data prevent a closer
look, which will be given in Section 6.2, where data from a more
polydisperse sample are discussed.

\subsection{Distorted structure factor}

\subsubsection{Linear order in $\gd$}

The stationary structural correlations of a dense fluid of spherical
particles undergoing Brownian motion, neglecting hydrodynamic
interactions, change with shear rate $\gd$ in response to a steady
shear flow. In linear order, the structure is distorted only in the
plane of the flow, while already in second order in $\gd$, the
structure factor  changes under shear also for wavevectors lying in
the plane perpendicular to the flow. Consistent with previous
theories, MCT-ITT finds regular expansion coefficients in linear and
quadratic order in $\gd$ for fluid (ergodic) suspensions
\cite{Hen:07}.  For the steady state structure factor $S_{\bf
q}{(\dot{\gamma})}$ of density fluctuations under shearing in plain
Couette flow defined in \gl{b4a}, the change from the equilibrium
one in linear order in shear rate is given by the following
ITT-approximation: \beq{cc4} S_q{(\dot{\gamma})}=S_q +\dot{\gamma}\,
\left\{\frac{q_x q_y}{q}S'_q\,\int_0^\infty\!\!\!\!\! dt\;
\Phi_q^2(t)\right\} + {\cal O}(\text{Pe}^2)\;. \eeq This relation
follows from \gl{b9a} in the limit of small $\gd$, where the
quiescent density correlators can be taken from quiescent MCT in
\gl{cc3}.

\subsubsection{Comparison with simulations}

\begin{figure}[h]
\begin{center}
\includegraphics[width=0.8\columnwidth]{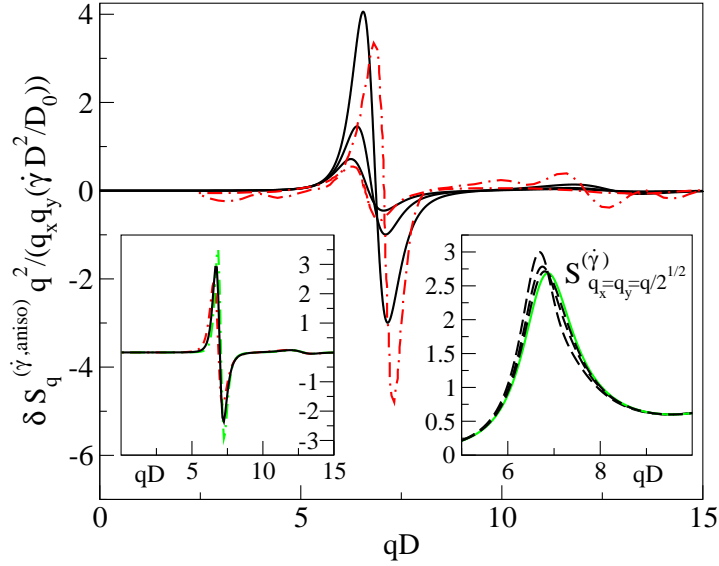}
\end{center}
\caption{ Contribution $\delta S_\qb^{\gd}$ linear in shear rate to
the distorted structure factor $S_\qb(\gd)$ normalized to
$\dot{\gamma}q_x q_y/q^2$ (black solid lines); from
Ref.~\cite{Hen:07}, here $D$ denotes the hard sphere diameter.
Decreasing the relative separations from the critical point as
$\varepsilon=-0.2, -0.15, -0.1$, the magnitude of $\delta S_{\bf
q}^{(\dot{\gamma})}$ increases. The dashed-dotted red curves are
Brownian dynamics simulation data from Ref.~\cite{Szamel01} using
the same normalization at $\varepsilon=-0.259, -0.138$. The right
inset shows the unnormalized $S_{\bf q}^{(\gd)}$ along the
extensional axis $q_x=q_y=q/\sqrt2$ at $\varepsilon=-0.1$ and, from
bottom to top, Pe $=\gd\tau= 0$ (green-solid), $1/8$, $1/4$ and
$1/2$ (all black-dashed), where $\Phi_{q_p}(t=\tau)=0.1$ defines
$\tau$. Pe$/$Pe$_0=1.66$ holds at this $\varepsilon$. The left inset
shows the data of the main figure rescaled with the dressed Peclet
number, $\delta S_{\bf q}^{(\dot{\gamma},{\rm aniso})}/({\rm Pe }\,
q_x q_y/q^2$); for $\varepsilon=-0.1, -0.05$, and $-0.01$ (with
increasing peak height) the values Pe/Pe$_0=1.66$, Pe/Pe$_0=8.06$,
and Pe/Pe$_0=419$ are used.}\label{Fig7}
\end{figure}

Figure \ref{Fig7} shows the  contribution $\delta S_{\bf
q}^{(\dot{\gamma}}$ to the distorted structure factor in leading
linear order in $\gd$ for packing fractions $\phi=0.36, 0.44$ and
$0.46$. Data taken from Brownian dynamics simulations at $\phi=0.43$
and $0.5$ from Ref.~\cite{Szamel01} are also included. In both cases
the data was divided by a factor $\dot{\gamma}q_x q_y/q^2$ which is
the origin of the trivial anisotropy in the leading linear order.
The distortion $\delta  S_{\bf q}^{(\dot{\gamma})}$ of the
microstructure grows strongly with $\phi$, because of the approach
to the glass transition.  The $\delta  S_{\bf q}^{(\dot{\gamma})}$
is proportional to the $\alpha$-relaxation time $\tau$, as proven in
the left inset of Fig.~\ref{Fig7}. Here, $\tau$ is estimated from
$\Phi_{q_p}(t=\tau)=0.1$, where $q_p$ denotes the position of the
primary peak in $S_q$. Rescaling the data with Pe, collapses the
curves at different distances to the glass transition. The strongest
shear-dependence occurs for the direction of the extensional
component of the flow, $q_x=q_y$. Here, the mesoscale order of the
dispersion grows; the peak in $\delta S_{\bf q}^{(\dot{\gamma})}$
increases and sharpens. The ITT results qualitatively agree with the
simulations in these aspects \cite{Szamel01}.

The most important finding of Fig.~\ref{Fig7} concerns the magnitude
of the distortion of the microstructure, and the dimensionless
parameter measuring the effect of shear relative to the intrinsic
particle motion. This topic can  already be discussed using the
linear order result, and is not affected by considerations of
hydrodynamic interactions, as can be glanced from comparing Brownian
dynamics simulations \cite{Szamel01} and experiments on dissolved
particles \cite{Johnson88}. In previous theories, shear rate effects
enter when the bare Peclet number Pe$_0$ becomes non-negligible. In
the present ITT approach the dressed Peclet/ Weissenberg number
Pe$=\gd\tau$ governs shear effects; here, $\tau$ is the (final)
structural relaxation time. Shear flow competes with structural
rearrangements that become arbitrarily slow compared to diffusion of
dilute particles when approaching the glass transition. The
distorted microstructure results from the competition between shear
flow and cooperative structural rearrangements. It is thus no
surprise that previous theories using Pe$_0$, which is
characteristic for dilute fluids or strong flows, had severely
underestimated the magnitude of shear distortions in hard sphere
suspensions for higher packing fractions;
Refs.~\cite{Szamel01,lionberger} report an underestimate by roughly
a decade at $\phi=0.50$. The ITT approach actually predicts a
divergence of $\lim_{\gd\to0} (S_{\bf q}^{(\dot{\gamma})}-S_q)/\gd$
for density approaching the glass transition at $\phi_c\approx0.58$.
And for (idealized) glass states, where $\tau=\infty$ holds in MCT
following Maxwell's phenomenology, the stationary structure factor
becomes non-analytic, and differs from the equilibrium one even for
$\gd\to0$. The distortion $\delta S^{(\gd)}_{\qb}$ thus
qualitatively behaves like the stress, which goes to zero linear in
$\gd$ in the fluid, but approaches a yield stress $\sigma^+$ in the
glass for $\gd\to0$.

The reassuring agreement of ITT results on $S_{\bf q}^{(\gd)}$ with
the data from simulations and experiments shows that in the ITT
approach the correct expansion parameter Pe has been identified.
This can be taken as support for the ITT-strategy  to connect the
non-linear rheology of dense dispersions with the structural
relaxation studied at the glass transition.

\section{Universal aspects of the glass transition in steady shear}

\begin{figure}[htp]
\centering
\includegraphics[ width=0.8\columnwidth]{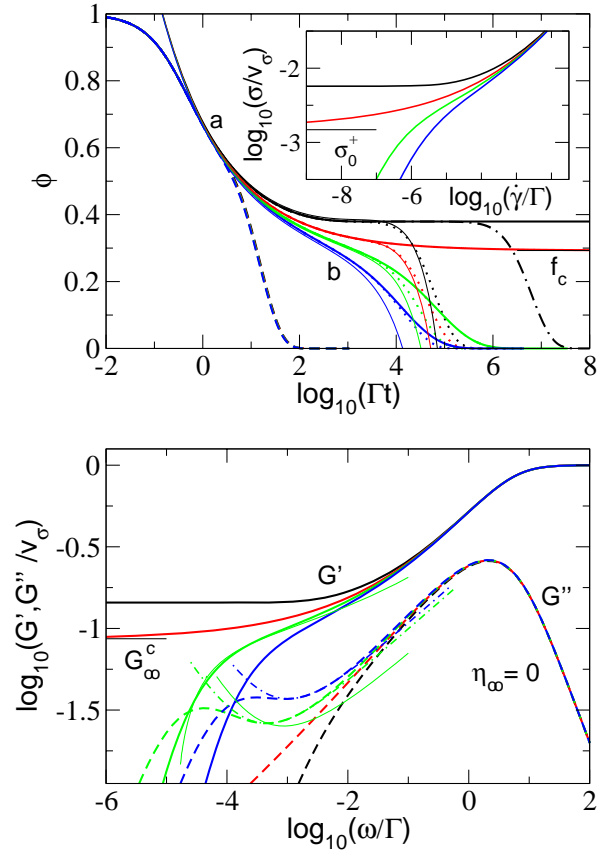}
\caption{Overview of the properties of the F$_{12}^{(\gd)}$-model
characteristic for the transition between  fluid and yielding glass;
from Ref.~\cite{Cra:08}. The upper panel shows numerically obtained
transient correlators $\Phi(t)$ for $\epsilon=0.01$ (black curves),
$\epsilon=0$ (red), $\epsilon=-0.005$  (green), and $\epsilon=-0.01$
(blue). The shear rates are $\left|\dot{\gamma}/\Gamma\right|=0$
(thick solid lines), $\left|\dot{\gamma}/\Gamma\right|=10^{-6}$
(dotted lines), and $\left|\dot{\gamma}/\Gamma\right|=10^{-2}$
(dashed lines). For the glass state at $\epsilon=0.01$ (black),
$\left|\dot{\gamma}/\Gamma\right|=10^{-8}$ (dashed-dotted-line) is
also included. All curves were calculated with  $\gamma_{c}=0.1$ and
$\eta_{\infty}=0$. The thin solid lines give the factorization
result \gl{c1} with scaling functions $\cal G$  for
$\left|\dot{\gamma}/\Gamma\right|=10^{-6}$;  label $a$ marks the
critical law (\ref{c3b}), and label $b$ marks the von Schweidler-law
(\ref{c4}).  The critical glass form factor $f_c$ is indicated. The
inset shows the flow curves for the same values for $\epsilon$. The
thin black bar shows the yield stress $\sigma^+_c$ for $\epsilon=0$.
The lower panel shows the viscoelastic storage (solid line) and loss
(broken line)  modulus  for the same values of $\epsilon$. The thin
green lines are the Fourier-transformed factorization result \gl{c1}
with scaling function $\cal G$ taken from the upper panel for
$\epsilon=-0.005$. The dashed-dotted lines show the  fit formula
\gl{cneu} for the spectrum in the minimum-region with $G_{\rm
min}/v_{\sigma}=0.0262$, $\omega_{\rm min}/\Gamma=0.000457$ at
$\epsilon=-0.005$ (green) and $G_{\rm min}/v_{\sigma}=0.0370$,
$\omega_{\rm min}/\Gamma=0.00105$ at $\epsilon=-0.01$ (blue). The
elastic constant at the transition $G_\infty^c$ is marked also,
while the high frequency asymptote $G'_\infty=G'(\omega\to\infty)$
is not labeled explicitly.} \label{Fig8}
\end{figure}

The summarized microscopic MCT-ITT equations contain  a
non-equilibrium transition between a shear thinning fluid and a
shear-molten glassy state; it is the central novel transition found
in MCT-ITT \cite{Fuc:02}. Close to the transition, (rather)
universal predictions can be made about the non-linear dispersion
rheology and the steady state properties. Following
Refs.~\cite{Fuc:03,Cra:08}, the central predictions are introduced
in this section and summarized in the overview figure \ref{Fig8};
the following results sections contain more examples. Figure
\ref{Fig8} is obtained from the schematic model which is also often
used to analyse  data, and which is introduced in the following
section 5.2 .

A dimensionless separation parameter $\epsilon$ measures the
distance to the transition which lies at $\epsilon=0$. A fluid state
($\epsilon < 0$) possesses a (Newtonian) viscosity,
$\eta_0(\epsilon<0) =\lim_{\gd\to0} \sigma(\gd)/\gd$, and shows
shear-thinning upon increasing $\gd$. Via the relation
$\eta_0=\lim_{\omega\to0}\, G''(\omega)/\omega$, the Newtonian
viscosity can also be taken from the linear response loss modulus at
low frequencies, where $G''(\omega)$ dominates over the storage
modulus. The latter varies like $G'(\omega\to0)\sim \omega^2$. A
glass ($\epsilon\ge0$), in the absence of flow, possesses an elastic
constant $G_\infty$, which can be measured in the elastic shear
modulus $G'(\omega)$ in the limit of low frequencies,
$G'(\omega\to0,\epsilon\ge0)\to G_\infty(\epsilon)$. Here the
storage modulus dominates over the loss one, which drops like
$G''(\omega\to0)\sim \omega$. The high frequency modulus
$G'_\infty=G'(\omega\to\infty)$ is characteristic of the particle
interactions, see Footnote 4, and exists, except for the case of
hard sphere interactions without HI, in fluid and solid states. The
dissipation at high frequencies $G''(\omega\to\infty)\to\eta_\infty
\omega$ also shows no anomaly at the glass transition and depends
strongly on HI and solvent friction.

Enforcing steady shear flow melts the glass. The stationary stress
of the shear-molten glass always exceeds a (dynamic) yield stress.
For decreasing shear rate, the viscosity increases like $1/\gd$, and
the stress levels off onto the yield-stress plateau,
$\sigma(\gd\to0,\epsilon\ge0)\to\sigma^+(\epsilon)$.

Close to the transition, the zero-shear viscosity $\eta_0$, the
elastic constant $G_\infty$, and the yield stress $\sigma^+$ show
universal  anomalies as functions of the distance to the transition.
\begin{figure}[htp]
\centering
\includegraphics[ width=0.6\columnwidth]{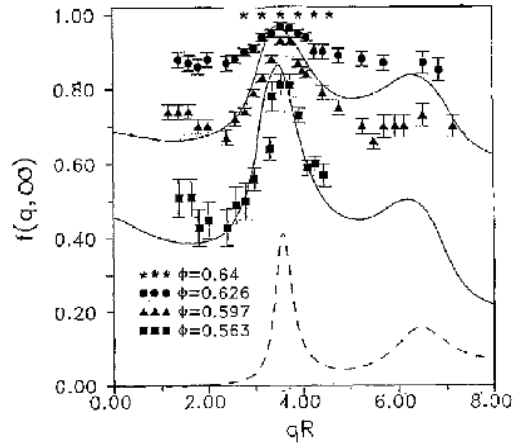}
\caption{Glass form factors $f_q$ as function of wavevector $q$ in a
colloidal glass of hard spheres for packing fractions $\phi$ as
labeled. Data obtained by van Megen and coworkers by dynamic light
scattering are qualitatively compared to MCT computations using the
PY-$S_q$ at $\phi$ values chosen ad hoc to match the experimental
data; from \cite{Meg:91}. The PY structure factor at the glass
transition density $\phi=0.58$ is shown as broken line, rescaled by
a factor $1/10$ .} \label{Figneu2}
\end{figure}

The described results follow from the stability analysis of
\gls{b11}{b12} around an arrested, glassy structure $f_q$ of the
transient correlator \cite{Fuc:02,Fuc:03}. Considering the time
window where $\Phi_\qb(t)$ is close to arrest at $f_q$, and taking
all control parameters like density, temperature, etc. to be close
to the values at the transition, the stability analysis yields the
'factorization' between spatial and temporal dependences \beq{c1}
\Phi_\qb(t) = f^c_q + h_q \; {\cal G}(\,t/t_0,\epsilon,\gd t_0\,) +
\ldots\; , \eeq where the (isotropic) glass form factor $f^c_q$ and
critical amplitude $h_q$ describe the spatial properties of the
metastable glassy state. The critical glass form factor $f^c_q$
gives the long-lived component of density fluctuations right at the
transition $\epsilon=0$, while
$\Phi_\qb(t\to\infty,\epsilon\ge0,\gd=0)=f_q>0$ characterizes states
even deep in the glass with $f_q$ obeying \cite{Goe:91}: \beq{c2}
\frac{f_q}{1-f_q} = \frac{1}{2N} \sum_\kb\; V_{qkp}\;
 f_{k}\;  f_{p}\; ,
\eeq with $V_{qkp}$ from \gl{cc3c} as follows from \gls{b11}{b12}
asymptotically in the limit of vanishing shear rate \cite{Fuc:03}.
 Figure \ref{Figneu2} shows dynamic light
scattering data for the glass form factors at a number of densities
in PMMA hard sphere colloids, comparing them to solutions of \gl{c2}
evaluated for hard spheres using the PY structure factor; it is
included in Fig.~\ref{Figneu2} for the packing fraction
$\phi_c=0.58$ of the experimental glass transition. The glass form
factor varies with the average particle separation and in phase with
the structure factor. Both, $f_q$ and $h_q$ thus describe local
correlations, the so-called 'cage-effect', and can be taken as
constants independent on shear rate and density, as they are
evaluated from the vertices in \gl{b12} at the transition point
$(\epsilon=0,\gd=0)$.

All time-dependence and (sensitive) dependence on the external
control parameters is contained in the function $\cal G$, which
often is called '$\beta$-correlator' and obeys the non-linear
stability equation \cite{Fuc:02,Fuc:03,Fuc:09} \bsq{c3}\beq{c3a}
\tilde \epsilon - c^{(\gd)}\; (\gd t)^2 + \lambda\; {\cal G}^2(t) =
\frac{d}{dt} \int_0^t\!\!\!dt'\; {\cal G}(t-t')\; {\cal G}(t')\; ,
\eeq with initial condition \beq{c3b} {\cal G}(t \to 0) \to
(t/t_0)^{-a} \; . \eeq\esq The two parameters $\lambda$ and
$c^{(\gd)}$ in \gl{c3a} are determined by the static structure
factor at the transition point, and take values around
$\lambda\approx 0.73$ and $c^{(\gd)}\approx 0.7$ for the PY $S_q$
for hard spheres. The transition point then lies at packing fraction
$\phi_c=\frac{4\pi}{3} n_c R_H^3\approx 0.52$ (index $c$ for
critical), and the separation parameter measures the relative
distance,  $\tilde \epsilon=C\,\epsilon$ with
$\epsilon=(\phi-\phi_c)/\phi_c$ and $C \approx 1.3$. The 'critical'
exponent $a$ is given by the exponent parameter $\lambda$ via
$\lambda=\Gamma(1-a)^2/\Gamma(1-2a)$, as had been found  in
quiescent MCT \cite{Goe:91,Goe:92}.

The time scale $t_0$ in \gl{c3b} provides the means to match the
function ${\cal G}(t)$ to the microscopic, short-time dynamics. The
\gls{b11}{b12} contain a simplified description of the short time
dynamics in colloidal dispersions via the initial decay rate
$\Gamma_\qb(t)$.  From this model for the short-time dynamics, the
time scale $t_0\approx 1.6\, 10^{-2} R_H^2/D_0$  is obtained.
Solvent mediated effects on the short time dynamics are well known
and are neglected in $\Gamma_\qb(t)$ in \gl{b11}. The most simple
minded approximation to account for HI is given in \gl{b13}. It only
shifts the value of $t_0$. Within the ITT approach, this finding
holds more generally. Even if HI lead to more substantial changes of
\gl{b11}, all of the mentioned universal predictions
 would remain true, as long as HI do
not affect the mode coupling vertex in \gl{b12}. Like in the
quiescent MCT \cite{Fra:98}, in MCT-ITT hydrodynamic interactions
can thus  be incorporated into the theory of the glass transition
under shear, and amount to a rescaling of the matching time $t_0$,
only.

The parameters $\epsilon$, $\lambda$ and $c^{(\gd)}$ in \gl{c3a} can
be determined from the equilibrium structure factor $S_q$ at or
close to the transition, and, together with $t_0$ and the shear rate
$\gd$ they capture the essence of the rheological anomalies in dense
dispersions. A divergent viscosity follows from the prediction of a
strongly increasing final relaxation time in $\cal G$ in the
quiescent fluid phase \cite{Goe:92,Goe:91}: \beq{c4} {\cal G}(t \to
\infty , \epsilon < 0, \gd = 0 ) \to - \left(t / \tau \right)^{b}
\quad ,\;\; \mbox{with }\; \frac{t_0}{\tau} \propto
(-\epsilon)^\gamma\; . \eeq The entailed temporal power law, termed
von Schweidler law, initiates the final decay of the correlators,
which has a density and temperature independent shape $\tilde
\Phi_q(\tilde t)$.  In MCT, the (full) correlator thus takes the
characteristic form of a two-step relaxation. The final decay, often
termed $\alpha$-relaxation, depends on $\epsilon$ only via the time
scale $\tau(\epsilon)$ which rescales the time, $\tilde t=t/\tau$.
Equation (\ref{c3}) establishes the crucial time scale separation
between $t_0$ and $\tau$, the divergence of $\tau$, and the
stretching (non-exponentiality) of the final decay; it also gives
the values of the exponents via
$\lambda=\Gamma(1+b)^2/\Gamma(1+2b)$, and $\gamma=(1/a+1/b)/2$.
Using \gl{cc2}, the MCT-prediction for the divergence of the
Newtonian viscosity follows \cite{Goe:92,Goe:91}. During the final
decay the quiescent shear modulus also becomes a function of
rescaled time, $\tilde t=t/\tau$, leading to $\eta_0 \propto
\tau(\epsilon)$; its initial value is given by the elastic constant
at the transition, $G_\infty^c$.

The two asymptotic temporal power-laws of MCT also affect the
frequency dependence of $G''$ in the minimum region. The scaling
function $\cal G$  describes the minimum as crossover between two
power laws in frequency. The approximation for the modulus around
the minimum in the quiescent fluid becomes \cite{Goe:91}: \beq{cneu}
G''(\omega) \approx \frac{G_{\rm min}}{a+b} \; \left[\, b\, \left(
\frac{\omega}{\omega_{\rm min}}\right)^a + a\, \left(
\frac{\omega_{\rm min}}{\omega}\right)^{b} \, \right]\; . \eeq
 The
parameters in this approximation follow from \gls{c3}{c4} which give
$G_{\rm min}\propto\sqrt{-\epsilon}$ and $\omega_{\rm min}\propto
(-\epsilon)^{1/2a}$. Observation of this handy expression requires
that the relaxation time $\tau$ is (very) large, viz.~that time
scale separation holds (extremely well) for (very) small
$-\epsilon$; even in the exemplary Fig.~\ref{Fig8}, the chosen
distances to the glass transition are too large in order for the
\gl{cneu} to agree with the true $\beta$-correlator $\cal G$, which
is also included in Fig.~\ref{Fig8}. The reason for this difficulty
is the aspect that the expansion in \gl{c1} is an expansion in
$\sqrt{\epsilon}$, which requires very small separation parameters
$\epsilon$ for corrections to be negligible. For packing fractions
too far below the glass transition, the final relaxation process is
not clearly separated from the high frequency relaxation. This holds
in the experimental data shown in Fig.~\ref{Fig5}, where the final
structural decay process only forms a shoulder.

On the glassy side of the transition, $\epsilon\ge 0$, the transient
density fluctuations stays close to a plateau value for intermediate
times which increases when going deeper into the glass, \beq{c5}
{\cal G}(t_0 \ll t \ll 1/|\gd| , \epsilon \ge 0 ) \to
\sqrt{\frac{\tilde \epsilon}{1-\lambda}} + {\cal O}(\epsilon)\; .
\eeq Entered into \gl{cc2}, the square-root dependence of the
plateau value translates into the square-root anomaly of the elastic
constant $G_\infty$, and causes the increase of the yield stress
close to the glass transition.

Only, for vanishing shear rate, $\gd=0$, an ideal glass state exists
in the ITT approach for steady shearing. All density correlators
arrest at a long time limit, which from \gl{c5} close to the
transition is given by
$\Phi_\qb(t\to\infty,\epsilon\ge0,\gd=0)=f_q=f^c_q + h_q
\sqrt{\tilde \epsilon/(1-\lambda)} +{\cal O}(\epsilon)$.
Consequently the modulus remains elastic at long times,
$g(t\to\infty,\epsilon\ge0,\gd=0)=G_\infty>0$. Any (infinitesimal)
shear rate, however, melts the glass and causes a final decay of the
transient correlators. The function $\cal G$ initiates the decay
around the critical plateau of the transient correlators and sets
the common time scale for the final decay under shear \beq{c6} {\cal
G}(t \to \infty  ) \to - \sqrt{\frac{c^{(\gd)}}{\lambda-\frac 12}}
\; |\gd t| \equiv - \frac{t}{\tau_{\gd}}\; ,\quad \text{ with } \;
\tau_{\gd} = \sqrt{\frac{\lambda-\frac 12}{c^{(\gd)}}} \;
\frac{1}{|\gd|} \; . \eeq Under shear all correlators decay from the
plateau as function of $|\gd t|$; see e.g.~Figs.~\ref{Fig10},
\ref{Fig11},  \ref{Fig20} and \ref{Fig21}. Steady shearing thus
prevents non-ergodic arrest and restores ergodicity. Shear melts a
glass and produces a unique steady state at long times. This
conclusion is restricted by the already discussed assumption to
neglect aging of glassy states. It could remain because of
non-ergodicity in the initial quiescent state, which needs to be
shear-molten before ITT holds. Ergodicity of the sheared state,
however, suggests aging to be unimportant under shear, and that it
should be possible to melt initial non-ergodic contributions
\cite{Ber:00,Fie:00}. The experiments in model colloidal dispersions
reported in Sect. 3.1.2 and Sect. 6.2 support this notion, as
history independent steady states could be achieved at all
densities\footnote{An ultra-slow process causing the metastability
of glassy states even without shear may have contributed to restore
ergodicity in Refs.~\cite{Cra:08,Sie:09}.}.

The described universal scenario of shear-molten glass and
shear-thinnig fluid makes up the core of the MCT-ITT predictions
derived from \glto{b9}{b12}. Their consequences for the nonlinear
rheology will be discussed in more detail in the following sections,
while the MCT results for the linear viscoelasticity were reviewed
in Sect.~3. Yet, the anisotropy of the equations has up to now
prevented more complete solutions of the MCT-ITT equations of
Sect.~2. Therefore, simplified MCT-ITT equations become important,
which can be analysed in more detail and recover the central
stability equations \gls{c1}{c3}. The two most important ones will
be reviewed next, before the theoretical picture is tested in
comparison with experimental and simulations' data.

\section{Simplified models}

Two progressively more simplified  models  provide insights into the
generic scenario of non-Newtonian flow, shear melting and solid
yielding which emerge from the ITT approach.

\subsection{Isotropically sheared hard sphere model}

On the fully microscopic level of description of a sheared
 colloidal suspension, affine motion of the particles with
 the solvent leads to anisotropic dynamics. Yet, recent
simulation data of steady state structure factors indicate  a rather
isotropic distortion of the structure for Pe$_0\ll1$, even though
the Weissenberg number Pe is already large  \cite{Ber:02,Var:06}.
Confocal microscopy data on concentrated solutions support this
observation \cite{Bes:07}. The shift of the advected wavevector in
\gl{b5} with time to higher values, intially is anisotropic, but
becomes isotropic at longer times, when the magnitude of $q(t)$
increases along all directions. As the effective potentials felt by
density fluctuations evolve with increasing wavevector, this leads
to a decrease of friction functions, speed-up of structural
rearrangements and shear-fluidization. Therefore, one may hope that
an 'isotropically sheared hard spheres model' (ISHSM), which  for
$\gd=0$ exhibits the nonlinear coupling of density correlators with
wavelength equal to the average particle distance (viz. the
``cage-effect''), and which, for $\gd\ne0$, incorporates
shear-advection, captures some spatial aspects of shear driven
decorrelation.

\subsubsection{Definition of the ISHSM}

Thus, in the ISHSM, the equation of motion for the density
fluctuations at time $t$ after starting the shear is approximated by
the  one of the quiescent system, namely \gl{cc3a} (with
$\Gamma_q=q^2 D_s / S_q$). The memory function also is taken as
isotropic and modeled close to the unsheared situation
\cite{Fuc:02,Fuc:03} \bsq{dd1}\beq{dd1a}
 m_q(t) \approx
\frac{1}{2N} \sum_{\kb}  V^{(\gd)}_{qkp}(t)\;
 \Phi_k(t) \, \Phi_{p}(t) \; ,
\eeq with \beq{dd1b} V^{(\gd)}_{qkp}(t) = \frac{n^2S_qS_kS_p}{q^4}
\left[ \qb\cdot\kb\; c_{\bar k(t)} + \qb\cdot\pb\; c_{\bar p(t)}
\right] \left[ \qb\cdot\kb\; c_{k} + \qb\cdot\pb\; c_{p} \right]
\eeq where $\pb=\qb\!-\!\kb$, and the length of the advected
wavevectors is approximated by $\bar k(t)=k(1+(t\gd/\gamma_c)^2
)^{1/2}$ and  equivalently for $\bar p(t)$ . Note, that the memory
function thus only depends on one time, and that shear advection
leads to a dephasing of the two terms in the vertex \gl{dd1b}, which
form a perfect square in the quiescent vertex of \gl{cc3c} without
shear. This (presumably) also is the dominant effect of shear in the
full microscopic MCT-ITT memory kernel in \gl{b12}. The fudge factor
$\gamma_c$ is introduced in order to correct for the underestimate
of the effect of shearing in the ISHSM \footnote{Except for the
introduction of the parameter $\gamma_c$, further quantitatively
small, but qualitatively irrelevant differences exist between the
ISHSM defined here and used in Sect. 6.1 according to
Ref.~\cite{Fuc:09}, and the one originally defined in
Ref.\cite{Fuc:02} and shown in Sect. 5.1; see Ref.~\cite{Fuc:09} for
a discussion.}.

The expression  for the potential part of the transverse stress may
be simplified to \beq{dd1c} \sigma=  \gd\; \int_0^\infty\!\! dt\;
g(t,\gd)\; , \;\text{with }\; g(t,\gd) \approx\frac{k_BT}{60\pi^2}\;
\int\!\! dk\;
  \frac{k^5\, c'_k\; S'_{\breve k(t)}}{\breve k(t) S_k^2}\; \Phi_{\breve k(t)}^2(t) \; ,
\eeq \esq where, in the last equation \gl{dd1c}, the advected
wavevector is chosen as $\breve k(t)=k(1+(t\gd)^2/3 )^{1/2}$ , as
follows from straight forward isotropic averaging of $\kb(t)$ . For
the numerical solution of the ISHSM for hard spheres using $S_q$ in
PY approximation, the wavevector integrals were discretized as
discussed in Sect. 3.1.1 and following Ref.~\cite{Fra:97}, using
$M=100$ wavevectors from $k_{\rm min}=0.1/R$ up to $k_{\rm
max}=19.9/R$ with separation $\Delta k = 0.2/R $. Again, time was
discretized with initial step-width $dt=2\,10^{-7} R^2/D_s$, which
was doubled each time after 400 steps \cite{Fuc:91}. The model's
glass transition lies at $\phi_c=0.51591$, with exponent parameter
$\lambda=0.735$ and $c^{(\gd)}\approx 0.45 / \gamma_c^2$ ; note that
these values still change somewhat if the discretization is made
finer. The separation parameter  $\epsilon=(\phi-\phi_c)/\phi_c$),
and $\gd$ are the two relevant control parameters determining the
rheology.

\subsubsection{Transient correlators}

\begin{centering}
\begin{figure}[htp]
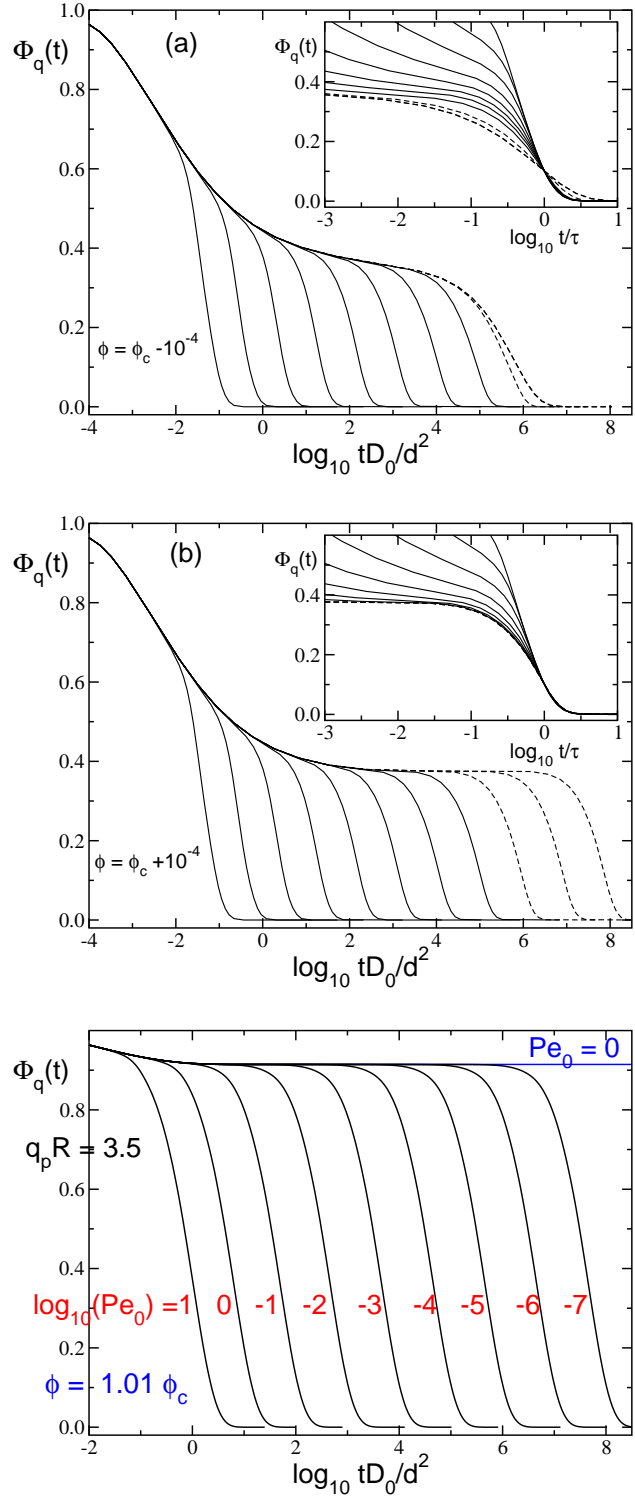

\centerline{\psfig{file=figures/ishsmphivontfluidk.eps,width=0.7\columnwidth}}
\vspace*{0.5cm}
\centerline{\psfig{file=figures/ishsmphivontglassk.eps,width=0.7\columnwidth}}
\vspace*{0.5cm}
\centerline{\psfig{file=figures/ishsmphivontglassneu.eps,width=0.7\columnwidth}}
\caption{Normalized transient density correlators $\Phi_q(t)$ of the
ISHSM at wavevector $q=3.4/d$ below (panel (a) at
$\phi=\phi_c-10^{-4}$) and above  (panel (b) at
$\phi=\phi_c+10^{-4}$) the transition for increasing shear rates
Pe$_0=9^n*10^{-8}$ with $n=0,\ldots,10$ from right to left; the
distances correspond to $\epsilon=\pm 10^{-3.53}$. Curves for $n=$
9, 10 carry short and for $n=$ 8 long dashes; note the collapse of
the two short dashed curves in (a). The insets show the data
rescaled so as to coincide at $\Phi(t=\tau)=0.1$. Panel (c) shows
glass correlators at another wavevector $q=7/d$ for parameters as
labeled; from Ref.~\cite{Fuc:03}. } \label{Fig10}
\end{figure}
\end{centering}
The shapes of the transient density fluctuation functions can be
studied with spatial resolution in the ISHSM. Figure \ref{Fig10}
displays density correlators at two densities, just below (panel
(a)) and just above (panels (b,c)) the transition, for varying shear
rates. Panel (b) and (c) compare correlators at different
wavevectors to exemplify the spatial variation. In almost all cases
the shear rate is so small that the bare Peclet number Pe$_0$ is
negligibly small and the short-time dynamics is not affected.
\begin{centering}
\begin{figure}[htp]
\centerline{\psfig{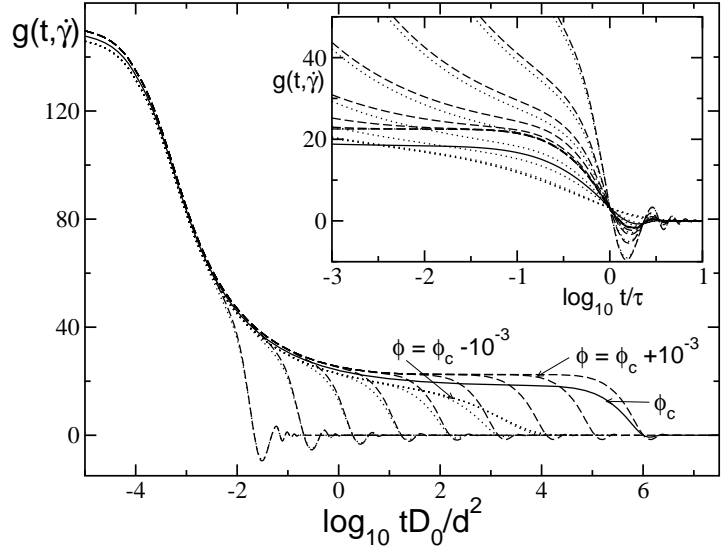}}
\caption{Transient non-Newtonian shear modulus $g(t,\gd)$ of the
ISHSM in units of  $k_BT/d^3$ for the packing fractions
 $\phi=\phi_c\pm10^{-3}$ ($\epsilon=\pm
10^{-2.53}$;  dashed/ dotted lines, respectively)
 for increasing shear rates Pe$_0=9^n*10^{-6}$ with $n=0,\ldots,8$
from right to left; note the collapse of fluid lines for the
smallest Pe$_0$; from Ref.~\cite{Fuc:03}.  The solid line gives
$g(t,\gd)$ for $\phi=\phi_c$ and  Pe$_0=10^{-6}$. The inset shows
the data rescaled so as to coincide at $g(t=\tau,\gd)=5$; note the
collapse of the $n=$ 6,7 \& 8 curves for both $\phi>\phi_c$ and
$\phi<\phi_c$. } \label{Fig11}
\end{figure}
\end{centering}

In the fluid case, the final or $\alpha$-relaxation is also not
affected for the two smallest dressed Peclet Pe values, but for
larger Pe it becomes faster and less stretched; see the inset of
fig. \ref{Fig10}(a).

Above the transition, the quiescent system forms an (idealized)
glass \cite{Goe:91,Goe:92}, whose density correlators arrest at the
glass form factors $f_q$ from Fig.~\ref{Figneu2}, and which exhibits
a finite elastic constant $G_\infty$, that describes the
(zero-frequency) Hookian response of the amorphous solid to a small
applied shear strain $\gamma$: $\sigma = G_\infty \gamma$ for
$\gamma\to0$; the plateau $G_\infty$ can be seen in Fig.~\ref{bild2}
and, for intermediate times in Fig.~\ref{Fig11}. If steady flow is
imposed on the system, however, the glass melts for any arbitrarily
small shear rate. Particles are freed from their cages and diffusion
perpendicular to the shear plane also becomes possible. Any finite
shear rate, however small, sets a finite longest relaxation time,
beyond which ergodicity is restored; see Figs.~\ref{Fig10}(b,c) and
\ref{Fig11}.

 The glassy curves at $\epsilon>0$, panels
(b,c), exhibit a  shift of the final relaxation with $\tau_{\gd}$
from \gl{c6} and asymptotically approach a scaling function
$\Phi_q^+(t/\tau_{\gd})$. The master equation for the ``yielding''
scaling functions $\Phi_q^+$ in the ISHSM can be obtained from
eliminating the short-time dynamics in \gl{cc3a}. After a partial
integration, the equation with $\partial_t \Phi_q(t) = 0$ is solved
by the scaling functions: \bsq{dd2} \beq{dd2a} \Phi_q^+(\tilde t) =
m^+_q(\tilde t) - \frac{d}{d\tilde t}\, \int_0^{\tilde t}d\tilde
t'\; m^+_q(\tilde t-\tilde t')\; \Phi_q^+(\tilde t')\; , \eeq where
$\tilde t=t/\tau_{\gd}$, and the memory kernel is given by
\beq{dd2b}
 m^+_q(\tilde t) =
\frac{1}{2N} \sum_{\kb} V^{(\tilde{\gd})}_{q,\kb}(\tilde t)\;
 \Phi^+_k(\tilde t) \, \Phi^+_{|\qb-\kb|}(\tilde t) \; .
\eeq  While the vertex is evaluated at fixed shear rate,
$\tilde{\gd}=\sqrt{\frac{\lambda-\frac 12}{c^{(\gd)}}}$, it depends
on the equilibrium parameters. The initialization for the correlator
is given by \beq{dd2c}\Phi_q^+(\tilde t \to 0) = f_q\; ,\eeq with
glass form factor taken from \gl{c2}. The two-step relaxation
 and the shift of the final relaxation with
$\tau_{\gd}$ are quite apparent in Fig.~\ref{Fig10}.

Figure \ref{Fig11} shows the transient shear modulus $g(t,\gd)$ of
the ISHSM from \gl{dd1c}, which determines the viscosity via
$\eta=\int_0^\infty\!\! dt \, g(t,\gd)$. It is the time derivative
of the shear stress growth function $\eta^+(t,\gd)$  (or transient
start up viscosity; here, the $^+$ labels the shear history
\cite{Larson,Zau:08}), $g(t,\gd)=\frac{d}{dt} \eta^+(t,\gd)$, and in
the Newtonian-regime reduces to the time dependent shear modulus,
$g^\text{lr}(t)$. The $g(t,\gd)$ shows all the features exhibited by
the correlator of the density correlators in the ISHSM, and thus the
discussion based on the stability analysis in Section 4 and the
yielding scaling law carries over to it: $g(t,\gd)=G^c_\infty +
h_g\, {\cal G} + \ldots$; especially the dependence of the final
relaxation step on rescaled time $t/\tau_{\gd}$ is apparent in the
glass curves. But in contrast to the density correlators
$\Phi_q(t)$, the function $g(t,\gd)$ becomes negative (oscillatory)
in the final approach towards zero, an effect more marked at high
Pe. This behavior originates in the general expression for
$g(t,\gd)$, \gl{dd1c}, where the vertex reduces to a positive
function (complete square) only in the absence of shear advection. A
overshoot and oscillatory approach of the start up viscosity to the
steady state value, $\eta^+(t\to\infty,\gd)\to\eta(\gd)$, therefore
are  generic features predicted from our approach.

\subsubsection{Flow curves}

\begin{centering}
\begin{figure}[htp]
\centerline{\psfig{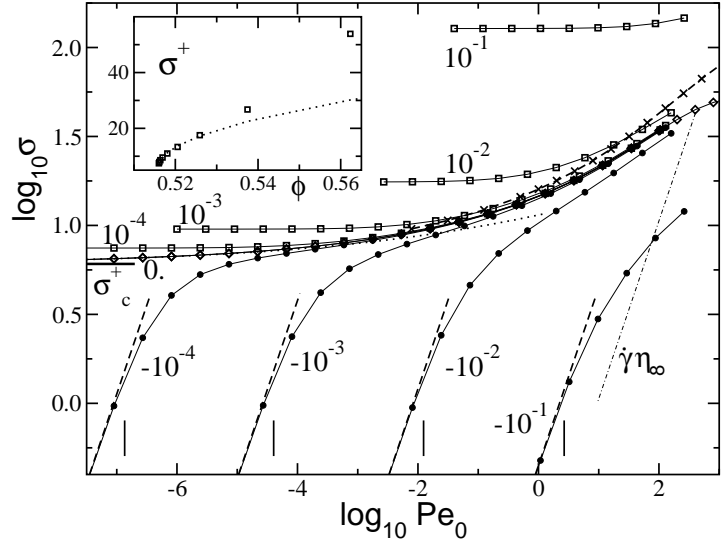}}
\caption{Steady state shear stress
 $\sigma$  in units of $k_BT/d^3$
 versus Pe$_0=\gd d^2/D_0$, for the ISHSM
at various distances from its glass transition, $\phi-\phi_c$ as
labeled; circles correspond to fluid, diamonds to the critical, and
squares to glassy densities; from Ref.~\cite{Fuc:03}, where the
additional lines are discussed. For the fluid cases, $\phi<\phi_c$,
dashed lines indicate Newtonian fluid behavior, $\sigma=\eta\gd$,
while vertical bars mark Pe$=\gd\tau=1$, with the structural
relaxation time taken from $\Phi_{q=7/d}(t=\tau)=0.1$. The
additional stress which would arise from the background solvent
viscosity, $\sigma=\gd \eta_\infty$, is marked by a dot-dashed line.
For the critical density, $\phi_c$, the critical yield stress,
$\sigma^+_c=6.04$, is shown by a horizontal bar. The inset shows the
rise of the dynamical yield stress
$\sigma^+=\sigma(\varepsilon\ge0,\gd\to0+)$ in the glass together
with a fitted power-law asymptote,
$\sigma^+=\sigma_c^++112\sqrt{\phi-\phi_c}$.} \label{Fig12}
\end{figure}
\end{centering}
As discussed in Section 4, in the fluid, MCT-ITT finds a  linear or
Newtonian regime in the limit $\gd\to0$, where it recovers the
standard MCT approximation for Newtonian viscosity $\eta_0$ of a
viscoelastic fluid \cite{Goe:91,Goe:92}. Hence $\sigma \to \gd
\,\eta_0$ holds for Pe $\ll 1$, as shown in Fig.~\ref{Fig12}, where
Pe calculated with the structural relaxation time $\tau$ is
included. As discussed, the growth of $\tau$ (asymptotically)
dominates all transport coefficients of the colloidal suspension and
causes an proportional increase in the viscosity $\eta$. For Pe
$>1$, the non-linear viscosity shear thins, and $\sigma$ increases
sublinearly with $\gd$. The stress versus strain rate plot in
Fig.~\ref{Fig12} clearly exhibits a broad crossover between the
linear Newtonian and a much weaker (asymptotically)
$\gd$-independent variation of the stress. In the fluid, the flow
curve takes a S-shape in double logarithmic representation, while in
the glass it is bent upward only.

Above the transition, the glass  melts for any shear rate.
 Nonetheless, a finite limiting
stress (yield stress) must be overcome in order to maintain the flow
of the glass: $$\sigma(\gd,\epsilon\ge0)\ge \sigma^+(\epsilon\ge 0)
= \lim_{\gd\to0} \sigma(\gd,\epsilon>0)\; .$$ For $\varepsilon\ge0$
and $\gd\to0$, the time $\tau^{(\gd)}$ for the final decay ,
\gl{c6}, can become arbitrarily slow  compared to the time
characterizing the decay onto $f_q$. Inserting the scaling functions
$\Phi^{+}$ from \gl{dd2} into the expression \gl{dd1c} for the
stress,
 the long time contribution separates out. Importantly, the integrands
containing the $\Phi^{+}$ functions depend on time only via $\tilde
t=t/\tau_{\gd}\propto \gd t$, so that nontrivial limits for the
stationary stress follow in the limit $\gd\to0$. In the ISHSM for
$\epsilon\ge0$, the yield stress is given by: \beq{dd2d}
\sigma^+=\frac{k_BT\, \tilde{\gd}}{60\pi^2}\; \int_0^\infty\!\!
d\tilde t\, \int\!\! dk\; k^5 \; \frac{S'_k S'_{\breve{k}(\tilde
t)}}{\breve k(\tilde t)\, S^2_{k}}\; \left(\Phi^{+}_{\breve k(\tilde
t)}(\tilde t)\right)^2 \; , \eeq\esq The existence of a dynamic
yield stress in the glass phase is thus  seen to arise from the
scaling law in \gl{dd2}, which is clearly borne out in the
Figs.~\ref{Fig10} and \ref{Fig11}. The yield stress arises from
those fluctuations which require the presence of shearing to prevent
their arrest. Even though $\sigma^+$ requires the solution of
dynamical equations, in MCT-ITT it is completely determined by the
equilibrium structure factor $S_q$. This may suggest a connection of
MCT to the potential energy paradigm for glasses, as recently
discussed \cite{Angelani00,Broderix00}. One might argue that
$\sigma^+$ arises because the external driving allows the system to
overcome energy barriers so that different metastable states can be
reached. This interpretation would agree with ideas from spin-glass
\cite{Ber:00} and soft-glassy rheology \cite{Sol:97,Sol:98,Fie:00}.
MCT-ITT indicates how shear achieves this in the case of colloidal
suspensions. It pushes fluctuations to shorter wavelengths where
smaller particle rearrangements cause their decorrelation.

The increase of the amplitude of the yielding master functions
$\Phi^+$ in \gl{dd2} originates in the increase of the arrested
structure in the unsheared glass \gl{c5}. In consequence the yield
stress should rapidly increase as one moves further into the glass
phase $\sigma^+ - \sigma^+_c\propto\sqrt{\epsilon}$ should
(approximately) hold; see however \cite{Haj:09} for the more
complicated rigorous expression. Indeed, the inset of
Fig.~\ref{Fig11} shows a good fit of this anomalous increase to the
numerical data. It is one of the hallmarks of the weakening of the
glass upon approaching the glass transition from the low temperature
or high density side.

\subsection{Schematic  $F_{12}^{(\dot{\gamma})}$-model}

The universal aspects described in the previous Section 4 are
contained in any ITT model that contains the central bifurcation
scenario and recovers \gls{c1}{c3}. Equation (\ref{c1}) states that
spatial and temporal dependences decouple in the intermediate time
window. Thus it is possible to investigate ITT models without proper
spatial resolution. Because of the technical difficulty to evaluate
the anisotropic functionals in \gls{b9c}{b12}, it is useful to
restrict the description to few or to a single transient correlator.
The best studied version of such a one-correlator model is the
$F_{12}^{(\dot{\gamma})}$-model.

\subsubsection{Definition and parameters}

In the schematic $F_{12}^{(\dot{\gamma})}$-model \cite{Fuc:03} a
single 'typical' density correlator $\Phi(t)$, conveniently
normalized according to $\Phi(t\to0)= 1-\Gamma t$, obeys a
Zwanzig-Mori memory equation which is modeled according to \gl{b11}
\bsq{f12} \begin{equation}\label{d1}
\partial_t \Phi(t) + \Gamma \left\{ \Phi(t) + \int_0^tdt'\; m(t-t') \;
\partial_{t'} \Phi(t') \right\}  = 0 \; .
\end{equation}
The parameter $\Gamma$ mimics the short time, microscopic dynamics,
and depends on structural and hydrodynamic correlations. The memory
function describes stress fluctuations which become more sluggish
together with density fluctuations, because slow structural
rearrangements dominate all quantities. A self consistent
approximation closing the equations of motion  is made mimicking
\gl{b12a}. In the $F_{12}^{(\dot{\gamma})}$-model one includes a
linear term (absent in \gl{b12a}) in order to sweep out the full
range of $\lambda$ values in \gl{c2}, and in order to retain
algebraic simplicity: \beq{d2} m(t)= \frac{v_1 \, \Phi(t) + v_2\,
\Phi^2(t)}{1+\left(\dot\gamma t/\gamma_c\right)^2} \eeq

This model, for the quiescent case $\dot\gamma=0$, was introduced by
G\"otze in 1984 \cite{Goe:84,Goe:91} and describes the development
of slow structural relaxation upon increasing the coupling vertices
$v_i\ge0$; they mimic the dependence of the vertices  in \gl{b12b}
at $\gd=0$ on the equilibrium structure given by $S_q$. Under shear
an explicit time dependence of the couplings in $m(t)$ captures the
decorrelation by shear  in \gl{b12b}. The parameter $\gamma_c$ sets
a scale that is required in order for the accumulated strain $\gd t$
to matter. Shearing causes the dynamics to decay for long times,
because fluctuations are advected to smaller wavelengths where small
scale Brownian motion relaxes them. Equations (\ref{d1},\ref{d2})
lead, with $\Phi(t)=f^c+(1-f^c)^2\,{\cal G}(t,\epsilon,\gd)$, and
the choice of the vertices $v_2=v_2^c=2$, and
$v_1=v_1^c+\epsilon\,(1-f^c)/f^c$, where $v_1^c=0.828$,   to the
critical glass form factor $ f^c= 0.293$ and to the stability
equation (\ref{c3}), with parameters
$$\lambda= 0.707\, ,\;  c^{(\gd)}= 0.586/\gamma_c^2\, \mbox{, and }\; t_0= 0.426 /\Gamma \;.
$$
The $F_{12}^{(\dot{\gamma})}$-model possesses a line of glass
transitions where the long time limit $f=\Phi(t\to\infty)$ jumps
discontinuously; it obeys the equivalent equation to \gl{c2}. The
glass transition line is parameterized by
$(v_1^c,v_2^c)=((2\lambda-1),1)/\lambda^2$ with $0.5\le \lambda< 1$,
and $f^c=1-\lambda$. The present choice of transition point
$(v_1^c,v_2^c)$ is  a typical one, which corresponds to the given
typical $\lambda$-value. The separation parameter $\epsilon$ is the
crucial control parameter as it takes the system through the
transition.

For simplicity, the quadratic dependence of the generalized shear
modulus on density fluctuations is retained from the microscopic
\gl{b9c}. It simplifies because only one density mode is considered,
and as, for simplicity, a dependence of the vertex (prefactor)
$v_\sigma$ on shear is neglected \beq{d4} g(t,\gd)= v_\sigma
\,\Phi^2(t) + \eta_\infty \; \delta(t-0+) \; .
 \eeq\esq
The parameter $\eta_\infty$ characterizes a short-time, high
frequency viscosity and models viscous processes which require no
structural relaxation, like in the general case \gl{b13}. Together
with $\Gamma$, it is the only model parameter affected by HI. Steady
state shear stress under constant shearing, and viscosity then
follow via integrating up the generalized modulus: \beq{d5}
 \sigma = \eta \; \dot\gamma =  \dot{\gamma}\; \int_0^\infty\!\!\!
dt\; g(t)  = \dot{\gamma}\; \int_0^\infty\!\!\!dt\;v_\sigma
\Phi^2(t) + \dot{\gamma}\; \eta_\infty \; . \eeq Also, when setting
shear rate $\gd=0$ in \gls{d1}{d2}, so that the schematic correlator
belongs to the quiescent, equilibrium system, the frequency
dependent moduli are obtained from Fourier transforming: \beq{d6}
G'(\omega) + i\, G''(\omega)  =  i\, \omega \, \int_0^\infty dt\;
e^{-i\, \omega\, t}\;  v_\sigma \left.
\Phi^2(t)\right|_{\dot\gamma=0} + i \omega\, \eta_\infty\; . \eeq
Because of the vanishing of the Fourier-integral in \gl{d6} for high
frequencies, the parameter $\eta_\infty$ can be identified as high
frequency viscosity: \beq{d7} \lim_{\omega\to\infty} G''(\omega) /
\omega =\eta^{\omega}_\infty\quad , \; \mbox{with
}\;\eta^{\omega}_\infty = \eta_{\infty}\; . \eeq At high shear, on
the other hand, \gl{d2} leads to a vanishing of $m(t)$, and \gl{d1}
gives an exponential decay of the transient correlator, $\Phi(t) \to
e^{-\Gamma\, t}$ for $\gd\to0$. The high shear viscosity thus
becomes \beq{d8} \eta^{\gd}_\infty = \lim_{\gd\to\infty}
\sigma(\gd)/ \gd = \eta_\infty + \frac{v_\sigma}{2\, \Gamma} =
\eta^\omega_\infty + \frac{v_\sigma}{2\, \Gamma} \; . \eeq

\subsubsection{Correlators and stability analysis}

Representative solutions of the F$_{12}^{(\gd)}$-model are
summarized in Fig.\ref{Fig8}, which bring out the discussed
universal aspects included in all ITT models.
\begin{figure}[htp]
\resizebox{0.8\columnwidth}{!}{\includegraphics{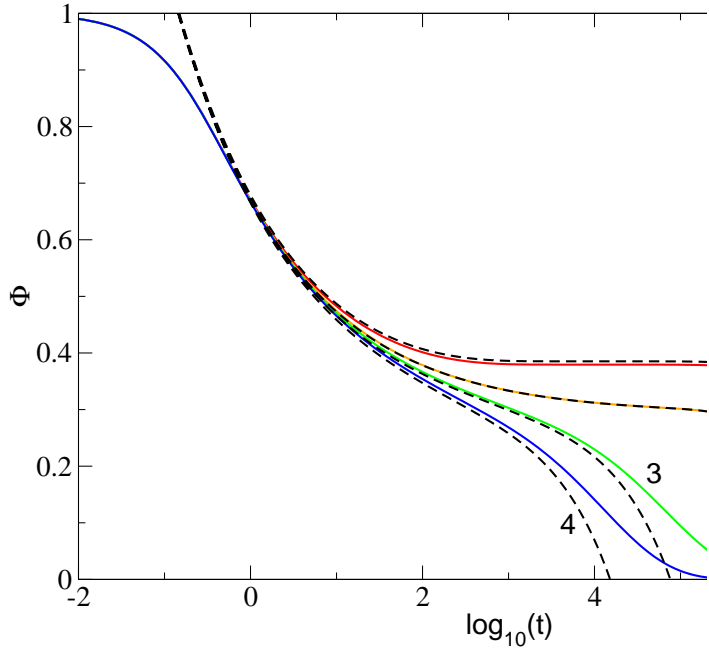}}
\caption{\label{fig:fig_1} Numerically obtained transient
correlators $\Phi\left(t\right)$ (solid lines) for
$\varepsilon=0.01$ (red, 1), $\varepsilon=0$ (orange, 2),
$\varepsilon=-0.005$ (green, 3) and $\varepsilon=-0.01$ (blue, 4)
for the   $F_{12}^{(\dot{\gamma})}$-model from Ref.~\cite{Haj:09}.
All curves were calculated with $\dot{\gamma}=10^{-7}$. The dashed
lines show the corresponding numerically obtained functions
$f_{c}+\left(1-f_{c}\right)^{2}\mathcal{G}\left(t\right)$.}
\end{figure}
For small separation parameters and shear rates the correlators
develop a stretched dynamics located around the critical plateau
value $f_{c}$, according to  \glto{c1}{c3}. The discussion of the
dynamics around this plateau was topic of Section 4. Figure
\ref{fig:fig_1}  shows these aspects in the
$F_{12}^{(\dot{\gamma})}$-model and presents typical correlators and
the corresponding $\beta$-correlators. The latter describe how the
glassy structure, which is present on intermediate times, is molten
either because the density is too low, or the temperature too high,
or, alternatively, because of the effect of shearing. For long times
$\mathcal{G}\left(t\right)$ merges into the linear asymptote
$-t/\tau_{\dot{\gamma}}$  from \gl{c6}. In the liquid region, for
short times $\mathcal{G}\left(t\right)$ follows
$\left(t/t_{0}\right)^{-a}$ from \gl{c3b} and merges into the second
power law $-\left(t/\tau_{0}\right)^{b}$ from \gl{c4} for
intermediate times with the von MCT Schweidler exponent
\cite{Goe:91}. In the transition region close to $\varepsilon=0$,
after following $\left(t/t_{0}\right)^{-a}$ the function
$\mathcal{G}\left(t\right)$ merges directly into the long time
asymptote $-t/\tau_{\dot{\gamma}}$. In the yielding glass region,
$\mathcal{G}\left(t\right)$ follows $\left(t/t_{0}\right)^{-a}$,
arrests on the plateau value
$\sqrt{\varepsilon/\left(1-\lambda\right)}$ for intermediate times
and merges into the linear asymptote $-t/\tau_{\dot{\gamma}}$ only
for long times. So we can summarize that the short- and long time
asymptotes are common for all $\varepsilon$ if $\dot{\gamma}\neq0$
is common. Fig. \ref{fig:fig_2} shows an overview of the properties
of $\mathcal{G}\left(t\right)$.
\begin{figure}[htp]
\resizebox{0.8\columnwidth}{!}{\includegraphics{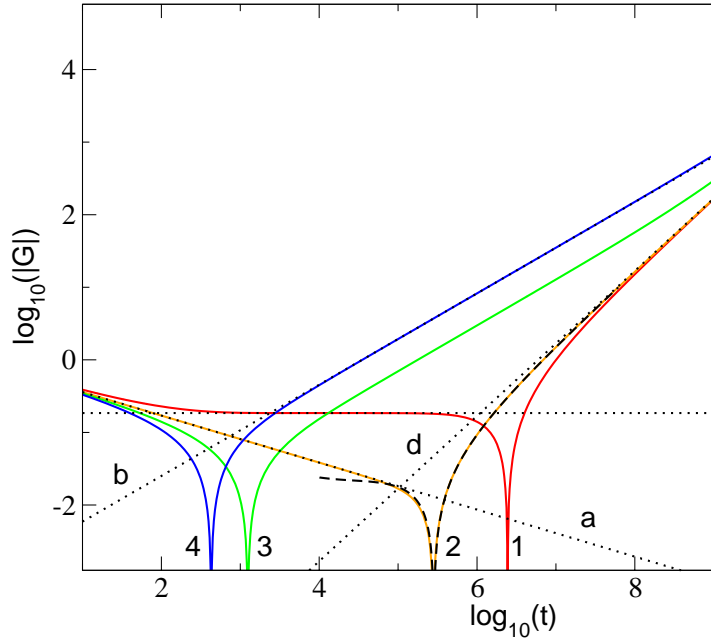}}
\caption{\label{fig:fig_2} An overview of the properties of
$\mathcal{G}\left(t\right)$ (solid lines) for the same values for
$\varepsilon$ and $\dot{\gamma}$  as in Fig. \ref{fig:fig_1}; from
Ref.~\cite{Haj:09}. The dotted lines show the leading asymptotes for
the corresponding time scales: The critical decay
$\left(t/t_{0}\right)^{-a}$ (a), the von Schweidler law
$-\left(t/\tau_{0}\right)^{b}$ (b), the arrest on the plateau value
$\sqrt{\varepsilon/\left(1-\lambda\right)}$ (c) and the
shear-induced linear asymptote $-t/\tau_{\dot{\gamma}}$ (d). The
dashed line shows a generalization of the latter law evaluated to
higher order with a fitted parameter $a_{1}=-1.39\cdot 10^{3}$ (at
$\varepsilon=0$).}
\end{figure}

\begin{figure}[htp]
\resizebox{0.8\columnwidth}{!}{\includegraphics{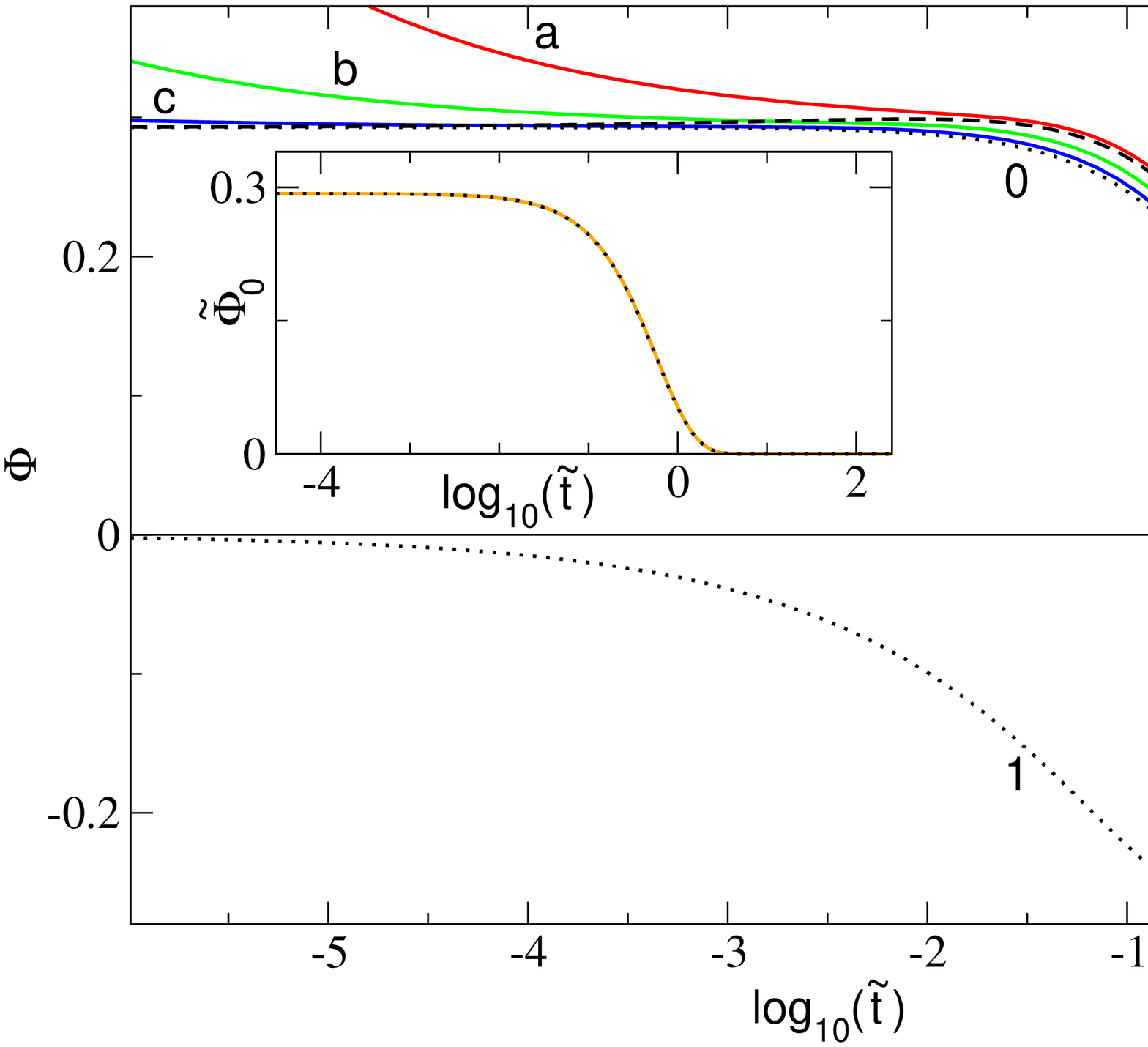}}
\caption{\label{fig:fig_a} The numerically determined master
function of the yielding-process obeying a 'time-shear-superposition
principle' (dotted line) $\Phi^+(\tilde
t)=\tilde{\Phi}_0\left(\tilde{t}\right)$ (label 0); from
Ref.~\cite{Haj:09}. The solid lines show numerically obtained
transient density correlators for $\varepsilon=0$ and
$\dot{\gamma}=10^{-7}$ (red, a), $\dot{\gamma}=10^{-9}$ (green, b)
and $\dot{\gamma}=10^{-12}$ (blue, c), plotted as functions of the
rescaled time $\tilde{t}$. The plots demonstrate that the rescaled
correlators converge to the yield master function
$\tilde{\Phi}_{0}\left(\tilde{t}\right)$ from the analog of \gl{dd2}
in the  $F_{12}^{(\dot{\gamma})}$-model for $\dot{\gamma}\rightarrow
0$, the blue curve (c) is already quite close to the master curve
(0). The dashed line shows
$\tilde{\Phi}_{0}\left(\tilde{t}\right)+a_1\left|\dot{\gamma}\right|^{c}\tilde{\Phi}_{1}\left(\tilde{t}\right)$
for $\dot{\gamma}=10^{-7}$ and the same numerical value for $a_1$ as
in Fig. \ref{fig:fig_2}, using  the leading correction
$\tilde{\Phi}_{1}\left(\tilde{t}\right)$ (1), which is shown in the
lower panel. This first order expansion already describes quite well
the shear-induced decay of the red curve (a). The inset demonstrates
that the master function $\tilde{\Phi}_{0}\left(\tilde{t}\right)$
(dotted line) can be well approximated by an exponential function
(solid line); the curves overlap completely. }
\end{figure}

The present $\beta$-scaling law bears some similarity to the one
presented by G\"otze and Sj\"ogren for the description of thermally
activated processes in   glasses \cite{Goe:87,Hil:92}. In both
cases, ideal glass states are destroyed by additional decay
mechanisms. Yet, the ITT equations and the generalised MCT equations
differ qualitatively in the mechanism melting the glass. The
similarity between both scaling laws thus underlines the
universality of the glass stability analysis, which is determined by
quite fundamental principles. In \gl{c3}, the shear rate can only be
a relevant perturbation (at long times) if it appears multiplied by
time itself. Symmetry dictates the appearance of $(\dot\gamma t)^2$,
because the sign of
 the shear rate must not matter. The aspect that shear melts the glass
  determines the negative sign of $(\dot\gamma t)^2$.

The melting of the  glassy structure during the yield process of
\gl{dd2} can be explicitly evaluated in the schematic
$F_{12}^{(\dot{\gamma})}$-model at $\epsilon=0$.  The yield master
function does not depend on $\dot{\gamma}$, and while its form is
model-dependent, its initial decay follows from the universal
stability equation \gl{c3}. Figure \ref{fig:fig_a} shows numerical
results, which can be well approximated by an exponential function.

The qualitative agreement between the transient correlators of the
ISHSM and the schematic $F_{12}^{(\dot{\gamma})}$-model support the
simplification to disregard the spatial structure. The only cost to
be paid, is the fixed plateau value $f_c$, which can be varied with
wavevector in the ISHSM, but not in the schematic model.

\subsubsection{Asymptotic laws of flow curves}

\begin{figure}[htp]
\resizebox{0.8\columnwidth}{!}{\includegraphics{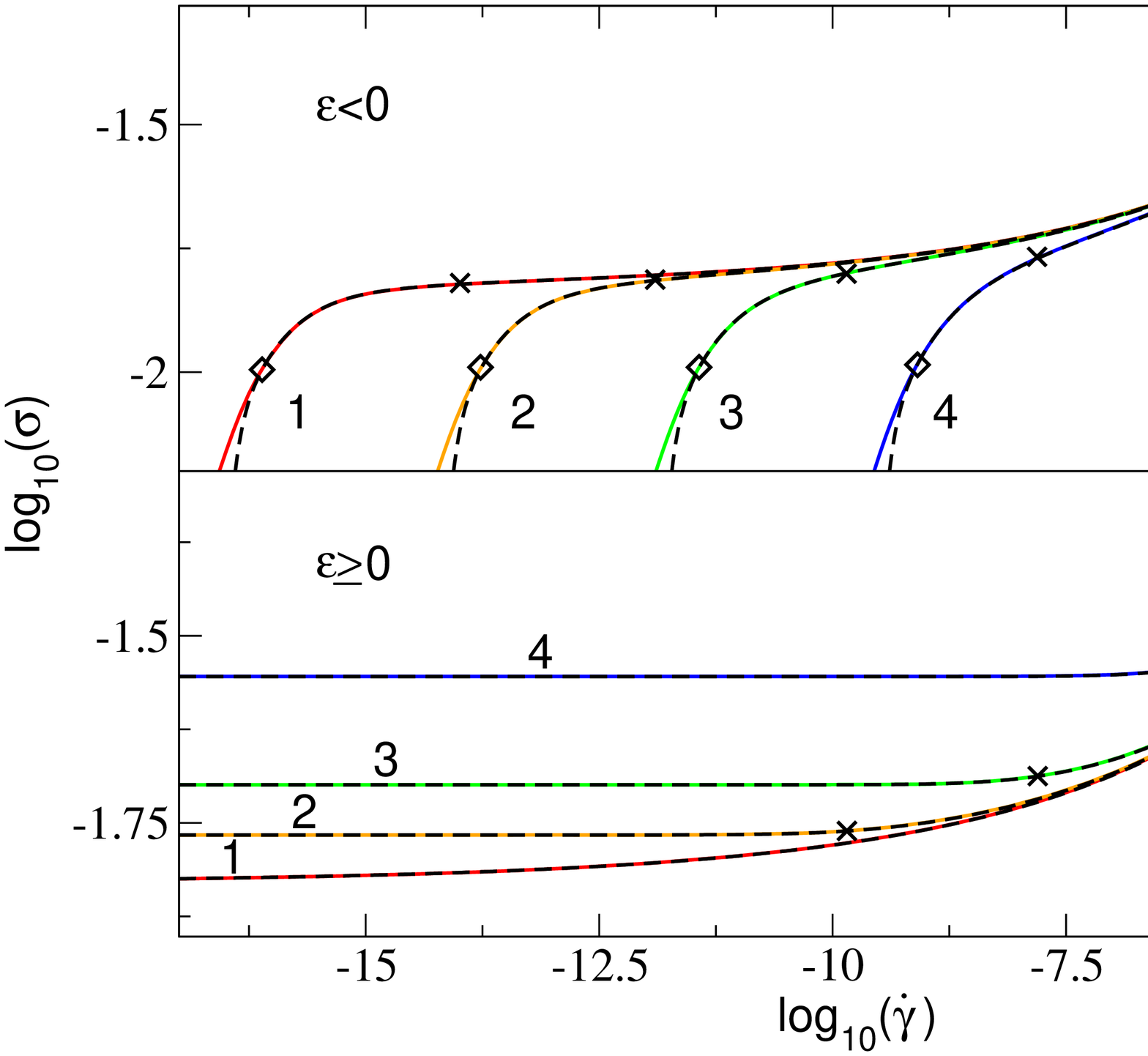}}
\caption{\label{fig:5} Overview of the numerically obtained flow
curves (solid lines) and the asymptotic $\Lambda$-formula evaluated
numerically (dashed lines); from Ref.~\cite{Haj:09}. The liquid
curves in the upper panel are shown for $\varepsilon=-10^{-7}$ (red,
1), $\varepsilon=-10^{-6}$ (orange, 2), $\varepsilon=-10^{-5}$
(green, 3) and $\varepsilon=-10^{-4}$ (blue, 4). The lower panel
shows the glassy curves for $\varepsilon=0$ (red, 1),
$\varepsilon=10^{-5}$ (orange, 2), $\varepsilon=10^{-4}$ (green, 3)
and $\varepsilon=10^{-3}$ (blue, 4). Crosses mark the points with
$\left|\varepsilon\right|=\varepsilon_{\dot{\gamma}}=\left|\dot{\gamma}t_{0}\right|^{\frac{2a}{1+a}}$.
The natural upper boundary for the shear rate, $\gd_*$, where the
range of validity of the $\Lambda$-formula is left, is also
indicated. For $\varepsilon<0$, the natural lower limits for the
shear rates, below which the $\Lambda$-formula does not describe the
flow curves, are marked by diamonds.}
\end{figure}
A major advantage of the simplified $F_{12}^{(\dot{\gamma})}$-model
is that it allows for asymptotic expansions that qualitatively
capture the flow curves. They can thus be investigated in detail
addressing such questions as e.g.~for the existence of power-law
shear thinning \cite{Larson}, or the dependence of the yield stress
on separation parameter. Fig. \ref{fig:5} shows an overview of the
numerically obtained flow curves and the corresponding asymptotic
results given by the so-called $\Lambda$-formula. While the glass
flow curves exhibit an upward curvature only, the fluid curves show
a characteristic S-shape, where the initial downward curvature
changes to an upward one for increasing  shear rate. Both behaviors
are captured by the asymptotic expansions. For positive separation
parameters the range of validity of the $\Lambda$-formula is given
by $\left|\varepsilon\right|\ll1$ and
$\left|\dot{\gamma}t_{0}\right|\ll1$. These two requirements ensure
that $\mathcal{G}\left(t\right)$ describes the dynamics of
$\Phi\left(t\right)$ with a sufficiently high accuracy; see
Figs.~\ref{fig:fig_1} and \ref{fig:fig_2}. For sufficiently small
negative separation parameters the $\Lambda$-formula is valid in
finite shear rate windows only, as it does not reproduce the linear
asymptotes for low shear rates. Precise criteria for the range of
its validity are known \cite{Haj:09}, and  Fig. \ref{fig:5} presents
an overview of the flow curves and their asymptotic laws.
\begin{figure}[htp]
\resizebox{0.8\columnwidth}{!}{\includegraphics{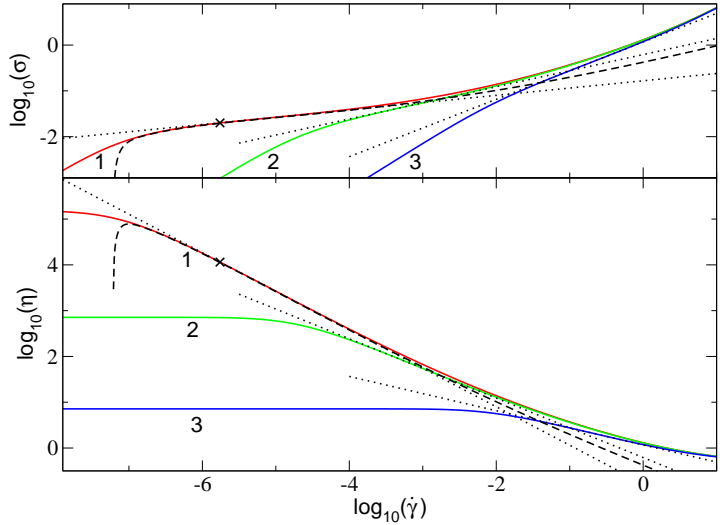}}
\caption{\label{fig:8} The upper panel shows numerically obtained
flow curves (solid lines) for $\varepsilon=-10^{-3}$ (red, 1),
$\varepsilon=-10^{-2}$ (green, 2) and $\varepsilon=-10^{-1}$ (blue,
3). The dotted lines show the corresponding inflection tangents,
with exponents $p=0.16$, $0.35$, and $0.63$ from left to right. The
dashed line shows the numerically evaluated $\Lambda$-formula for
$\varepsilon=-10^{-3}$. The shear rate with
$\varepsilon=-\varepsilon_{\dot{\gamma}}$ is marked by a cross. The
lower panel shows the corresponding results for the viscosity; from
Ref.~\cite{Haj:09}.}
\end{figure}

While the detailed discussion of the flow curves and their
asymptotics leads beyond the present review, see Ref.~\cite{Haj:09},
the important conclusions from Fig.~\ref{fig:5} in the present
context are: that the universal aspects discussed in Sect.~4 are
recovered, that qualitative agreement is obtained with the results
of the ISHSM, and that analytical expressions for the flow curves
can be obtained. For example, the critical flow curve follows a
generalized Herschel-Bulkley law:
$$
\sigma\left(\varepsilon=0,\dot{\gamma}\right)=\sigma_{c}^{+}\sum_{n=0}^{3}\;
c_{n}\left|\dot{\gamma}/\dot{\gamma}_{*}\right|^{mn}\;,
$$
where $\sigma_{c}^{+}$ is the critical dynamic yield stress and
$\dot{\gamma}_{*}$ defines a natural scale for the shear rates in
the asymptotic expansion; the upper limit $3$ of the summation is
discussed in Ref.~\cite{Haj:09}. At the transition, this law
describes the flow curve correctly for sufficiently small shear
rates, see Fig. \ref{fig:5}. This result also implies a generalized
power-law weakening of the yield stress $\sigma^+(\epsilon)$ when
approaching the glass transition for $\epsilon\searrow 0+$, which is
shown in the inset of Fig.~\ref{Fig12}.
 In fluid states for $\varepsilon<0$, the flow curves in
double logarithmic presentation, viz.~
$\log_{10}\left(\sigma\right)$ as function of
$\log_{10}\left(\dot{\gamma}\right)$, show an inflection point
defined by
$$
\frac{d^{2}\left(\log_{10}\left(\sigma\right)\right)}{d\left(\log_{10}\left(\dot{\gamma}\right)\right)^{2}}=0\;.
$$  But then in some finite shear rate windows the
flow curves can be approximated by the corresponding inflection
tangents. The slopes $p$ of the inflection tangents can be
interpreted as exponents occurring in some pseudo power laws:
$$
\sigma \propto \dot{\gamma}^{p}\;,\quad \Leftrightarrow \quad \eta
\propto \gd^{p-1} \;. $$ Figure \ref{fig:8} shows some examples. The
asymptotic formula also describes the neighborhood of the inflection
point correctly for sufficiently small $\varepsilon$, but does not
represent a real power law. In the framework of asymptotic
expansions, there is thus no real exponent $p$. The power-law shear
thinning often reported in the literature, in the ITT-flow curves
thus is a trivial artifact of the double logarithmic plot. Rather,
the flow curves on the fluid side exhibit a characteristic S-shape.
While this shape is rather apparent when plotting stress versus
shear rate, plotting the same data as viscosity as function of shear
rate hides it, because the vertical axis gets appreciably stretched.

\begin{figure}[htp]
\resizebox{0.8\columnwidth}{!}{\includegraphics{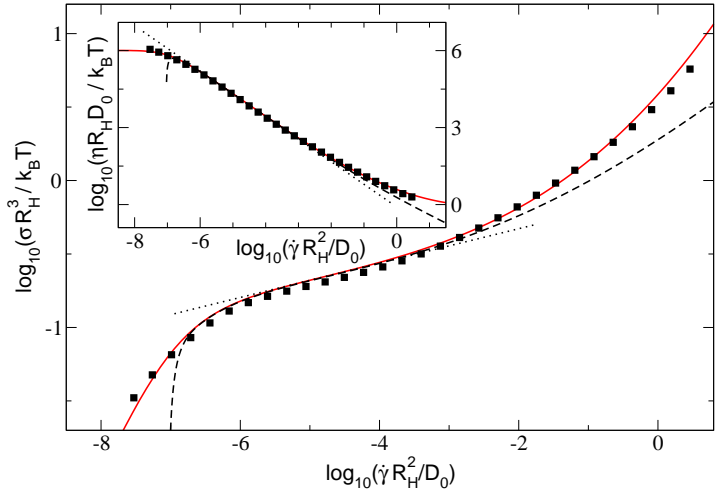}}
\caption{\label{fig:fig_exp1} Reduced flow curves for a core-shell
dispersion at an effective volume fraction of $\phi_{eff}=0.580$;
data from Ref.~\cite{Sie:09}, analysis from Ref.~\cite{Haj:09}. Here
$R_H$ denotes the hydrodynamic radius and $D_0$ the self diffusion
coefficient of the colloidal particles; $k_BT $ is the thermal
energy. The solid line (red) shows the result for the fitted
$F_{12}^{\left(\dot{\gamma}\right)}$-model with $v_2^c=2.0$. The
fitted parameters are: $\varepsilon=-0.00042$, $\gamma_c=0.14$,
$v_{\sigma}=70k_BT/R_H^3$, $\Gamma=80D_0/R_H^2$ and
$\eta_{\infty}=0.394k_BT/R_HD_0$. The dashed line shows the
corresponding result for the $\Lambda$-formula. The dotted line
shows the inflection tangent of the numerically determined flow
curve with a slope of $p=0.12$. The inset shows the corresponding
results for the viscosity. }
\end{figure}
\begin{figure}[htp]
\resizebox{0.8\columnwidth}{!}{\includegraphics{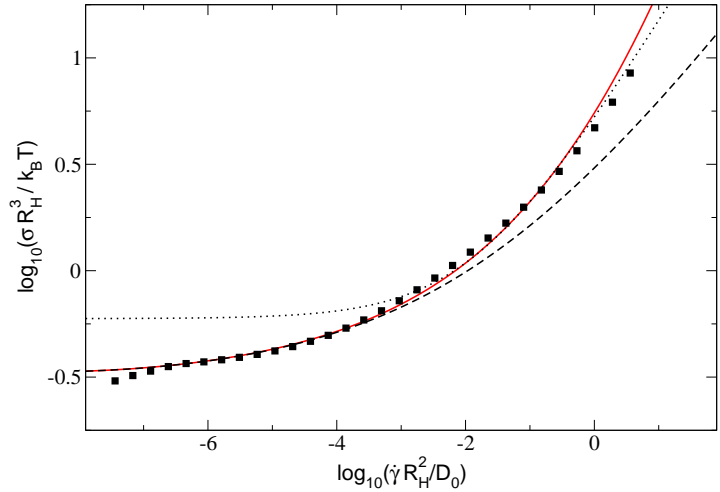}}
\caption{\label{fig:fig_exp2} Reduced flow curves for a core-shell
dispersion at an effective volume fraction of $\phi_{eff}=0.629$;
quantities as defined in the caption of Fig. \ref{fig:fig_exp1}. The
solid line (red) shows the result for the fitted
$F_{12}^{\left(\dot{\gamma}\right)}$-model with $v_2^c=2.0$. The
fitted parameters are: $\varepsilon=0.000021$, $\gamma_c=0.16$,
$v_{\sigma}=115k_BT/R_H^3$, $\Gamma=120D_0/R_H^2$ and
$\eta_{\infty}=0.431k_BT/R_HD_0$. The dashed line shows the
corresponding result for the $\Lambda$-formula. The dotted line
shows the fitted Herschel-Bulkley law given by Eq. (\ref{d9}) with
the analytically calculated exponent $\tilde m=0.489$; data from
Ref.~\cite{Sie:09}, analysis from Ref.~\cite{Haj:09}. }
\end{figure}

\subsubsection{Test of asymptotics in a polydisperse dispersion}

While the asymptotic expansions in the previous section provide an
understanding of the contents of the MCT-ITT scenario, experimental
tests of the asymptotic laws require flow curves over appreciable
windows in shear rate.

Figure \ref{fig:fig_exp1} and Fig. \ref{fig:fig_exp2}  show
experimental data recently obtained by Siebenb\"urger et al.
\cite{Sie:09} on polydisperse dispersions of the thermosensitive
core-shell particles introduced in Sect. 3.1.2 \cite{Cra:06}. In all
cases stationary states were achieved after shearing long enough,
proving that ageing could be neglected even for glassy states.
Because of the appreciable poyldispersity in particle size (standard
deviation 17 \%) crystallization could efficiently be prevented and
flow curves over extremely wide windows could be obtained.  Two flow
curves from their work can be used to test the asymptotic results.

Fig. \ref{fig:fig_exp1} shows the result for a liquid-like flow
curve where the asymptotic $\Lambda$-formula holds for approximately
four decades. The pseudo power law resulting from the inflection
tangent of the flow curve holds for approximately two decades within
the range of validity of the $\Lambda$-formula.

Fig. \ref{fig:fig_exp2} shows the result for a flow curve, where a
small positive separation parameter was necessary to fit the flow
curve and the linear viscoelastic moduli simultaneously. The data
are compatible with the (ideal) concept of a yield stress, but fall
below the fit curves for very small shear rates. This indicates the
existence of an additional decay mechanism neglected in the present
approach \cite{Cra:08,Sie:09}. Again, the $\Lambda$-formula
describes the experimental data correctly for approximately four
decades. For higher shear rates, an effective Herschel-Bulkley law
\beq{d9} \sigma(\dot \gamma_*\ll \left|\dot \gamma\right| \ll
\Gamma,\varepsilon=0 )  = \tilde \sigma_0 + \tilde \sigma_1
\left|\dot \gamma t_0 \right|^{\tilde m} \eeq with constant
amplitudes and exponent $\tilde m=0.49$ can be fitted in a window of
approximately two decades. The constant $\tilde \sigma_0$ is not the
actual yield stress, $\sigma^+$, which is obtained in the limit of
vanishing shear rate, $\sigma^+=\sigma(\dot\gamma\to0)$, but is
larger,

The experimental data of the polydisperse samples, which exhibit
structural dynamics over large windows, and their fits with the full
schematic model, will be taken up again in Sect. 6.2, where
additionally the linear response moduli are considered, as had been
done Sect. 3.1.2 for the less polydisperse samples affected by
crystallization.

\section{Comparison of theory and experiment}

As MCT-ITT contains uncontrolled approximations,  justification to
studying it, can  be obtained only from its power to rationalize
experimental observations. Because the transient density
fluctuations are the central quantity in the approach, density
correlators shall be considered first. Flow curves have been studied
in most detail experimentally and in simulations, and thus are
considered next.
\begin{figure}[htp]
\centering
\includegraphics[ width=0.8\columnwidth]{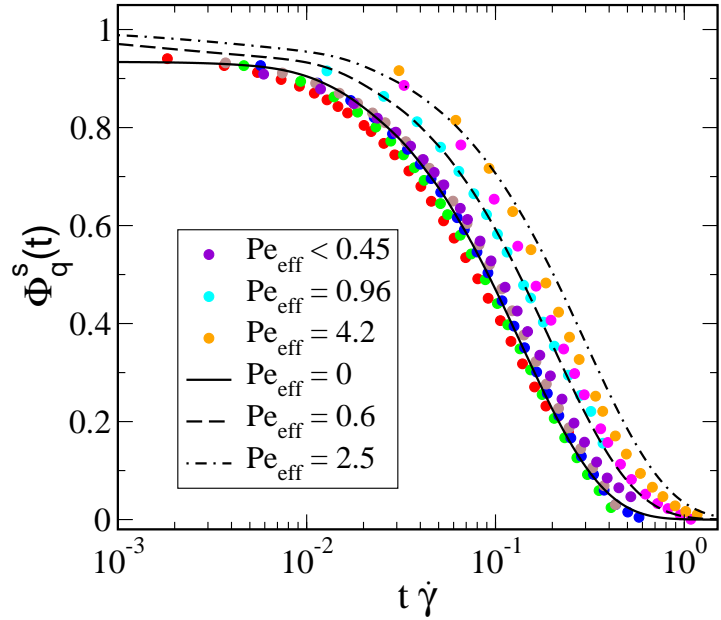}
\caption{Steady state incoherent intermediate scattering functions
$\Phi^s_q(t)$ as functions of accumulated strain $\gd t$ for various
shear rates $\gd$; the data were obtained in a colloidal hard sphere
dispersion at packing fraction $\phi=0.62$ (at $\epsilon\approx
0.07$) using confocal microscopy \cite{Bes:07}; the wavevector
points in the vorticity ($\bf \hat z$) direction and has $q=3.8/R$
(at the peak of $S_q$).  The effective Peclet numbers Pe$_{\rm
eff}=4 R^2 \gd/D_s$ are estimated with the short time self diffusion
coefficient $D_s\approx D_0/10$ at this concentration \cite{Meg:98}.
ISHSM calculations with separation parameter $\epsilon=0.066$ at
$qR=3.9$ (PY-$S_q$ peaking at $qR=3.5$), and for strain parameter
$\gamma_c=0.033$, are compared to the data for the Pe$_{\rm eff}$
values labeled. The yielding master function at Pe$_{\rm eff}=0$
lies in the  data curves which span $0.055 \le$ Pe$_{\rm eff} \le
0.45$, but discussion of the apparent systematic trend of the
experimental data would require ISHSM to better approximate the
shape of the final relaxation process; from Ref.~\cite{Fuc:09}.
\label{Fig20}}
\end{figure}

\subsection{ISHSM and single particle motion under steady shear}

Detailed measurements of the stationary dynamics under shear of a
colloidal hard sphere glass have recently been obtained by confocal
microscopy \cite{Bes:07}. Single particle motion was investigated in
a shear-molten glass at roughly the wavevector inverse to the
average particle separation. Figure \ref{Fig20}  shows
self-intermediate scattering functions measured for wavevectors
along the vorticity direction where neither affine particle motion
nor wavevector advection appears. The stationary correlators deep in
the glass, for shear rates spanning almost two decades, are shown as
function of accumulated strain $\gd t$, to test whether a simple
scaling $\tau_{\gd}\sim 1/\gd$ as predicted by \gl{dd2} holds. Small
but systematic deviations are apparent which have been interpreted
as a power law $\tau_{\gd}\sim \gd^{-0.8}$ \cite{Bes:07,Sal:08}.
ISHSM computations were performed for a nearby wavevector where
$S_q$ is around unity so that coherent and incoherent correlators
may be assumed to be similar \cite{Pus:78}. Additionally, for the
comparison it was assumed that  time dependent transient and
stationary fluctuation functions agree.  The yielding master
function from \gl{dd2} in ISHSM can be fitted to  the data measured
at small effective Peclet numbers Pe$_{\rm eff}$, by using for the
phenomenological `strain rescaling parameter' $\gamma_c=0.033$; the
smallness of the fitted value, which would be expected to be of
order unity, is not yet understood. The effective Peclet number
Pe$_{\rm eff}=4R^2 \gd / D_s$ with $D_s/D_0=0.1$ taken from
\cite{Meg:98} measures the importance of shear relative to the
Brownian diffusion time obtained from the short time self diffusion
coefficient $D_s$ at the relevant volume fraction.  At the larger
effective Peclet numbers, Pe$_{\rm eff}\ge 0.5$, for which the
short-time and final (shear-induced) relaxation processes move
closer together, the model gives quite a good account of the
$\gd$-dependence.
\begin{figure}[htp]
\centering
\includegraphics[ width=0.8\columnwidth]{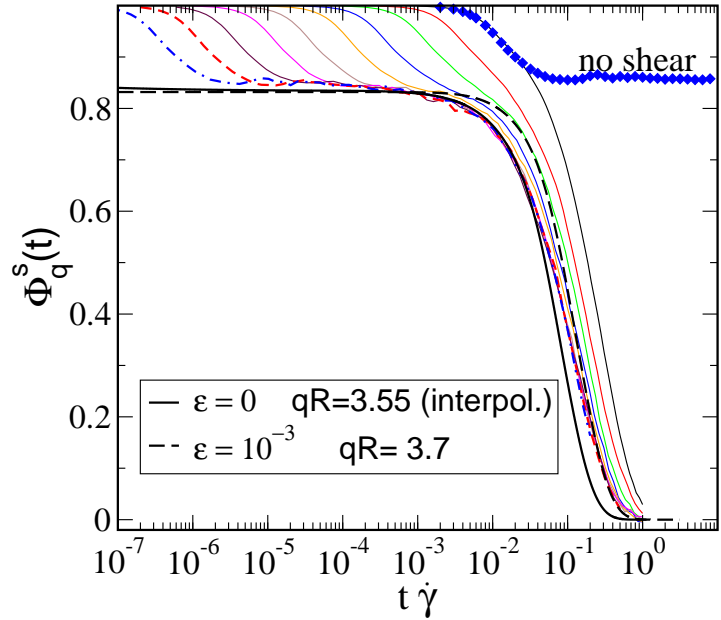}
\caption{Steady state incoherent intermediate scattering functions
$\Phi^s_q(t)$ measured in the vorticity direction as functions of
accumulated strain $\gd t$ for various shear rates $\gd$; data from
molecular dynamics simulations of a supercooled binary Lenard-Jones
mixture below the glass transition are taken from
Ref.~\cite{Var:06b}. These collapse onto a yield scaling function at
long times. The wavevector is $q=3.55/R$ (at the peak of $S_q$). The
quiescent curve, shifted to agree with the one at the highest $\gd$,
shows ageing dynamics at longer times outside the plotted window.
The apparent yielding master function from simulation is compared to
the ones calculated in ISHSM for glassy states at or close to the
transition (separation parameters $\epsilon$ as labeled) and at
nearby wave vectors (as labeled). ISHSM curves were chosen to match
the plateau value $f_q$, while strain parameters $\gamma_c=0.083$ at
$\epsilon=0$ (solid line) and $\gamma_c=0.116$ at $\epsilon=10^{-3}$
(dashed line) were used; from Ref.~\cite{Fuc:09}. \label{Fig21}}
\end{figure}

The shape of the final relaxation step in a shear-molten glass can
be studied even more closely in recent computer simulations, where a
larger separation of short and long time dynamics could be achieved
\cite{Var:06b}. In these molecular dynamics simulations of an
undercooled binary Lenard-Jones mixture, schematic ITT models give a
good account of the steady state flow curves, $\sigma(\gd)$
\cite{Var:06,Hen:05}; this will be discussed in Sect.~6.3. Figure
\ref{Fig21} shows the corresponding stationary self intermediate
scattering functions for a wavevector near the peak in $S_q$,
oriented along the vorticity direction, for shear rates spanning
more than four decades. Collapse onto a master function when plotted
as function of accumulated strain is nicely observed as predicted by
\gl{dd2}. At larger shear rates, the correlators raise above the
master function; this resembles the behaviour observed in the
confocal experiments in Fig.~\ref{Fig20}, and in the theoretical
calculations in Figs.~\ref{Fig10} and \ref{fig:fig_a}. Assuming
again that transient coherent correlators can be fitted to
stationary incoherent ones, the shape of the master function can be
fitted with the ISHSM, using again an unaccountedly small strain
parameter $\gamma_c$ . After this rescaling, modest but visible
differences in the shapes remain: the theoretical master function
decays more steeply than the one from simulations.

Overall, theory and experiment agree in finding a two step
relaxation process, where shear has a strong effect on the final
structural relaxation, while the short time diffusion is not much
affected. This supports the central MCT-ITT prediction that shearing
speeds up the structural rearrangements in a concentrated dispersion
close to vitrification. More detailed comparisons await better
theoretical calculations where the effect of shear on the stationary
density fluctuation functions is taken into account more faithfully
than in the ISHSM.

\subsection{F$_{12}^{(\gd)}$-model and shear stresses in equilibrium and under flow in
a polydisperse dispersion}

A central result of MCT-ITT concerns the close connection between
structural relaxation at the glass transition and the rheological
properties far from equilibrium. The ITT approach aims to unify the
understanding of these two phenomena, which were introduced
wrt.~experimental data in Sects. 3.1.1 and 5.2.4, respectively.
MCT-ITT requires, as sole input, information on the equilibrium
structure (namely $S_q$), and, first gives a formally exact
generalization of the shear modulus to finite shear rates,
$g(t,\gd)$, which is then approximated in a consistent way. A novel
dense colloidal dispersion serves as experimental model system,
whose linear and nonlinear rheology can be determined over very
broad windows of control parameters. The generalized modulus
$g(t,\gd)$ can thus be investigated as function of shear rate and
time (more precisely frequency), and the MCT-ITT approach can be
tested excruciatingly. Thermosensitive core-shell particles
consisting of a polystyrene core and a crosslinked
poly(N-isopropylacrylamide)(PNIPAM) shell were synthesized and their
slightly polydisperse dispersions (standard deviation 17 \%)
characterized in detail \cite{Sie:09}; see Sect. 3.1.1. While their
precise structure factor has not been measured yet, the system can
be well considered a slightly polydisperse mixture of hard spheres.
Because polydispersity prevents crystallization and the lack of
attractions prevents demixing and coagulation, this system opens a
window on structural relaxation, which can nicely be tuned by
changing the effective packing fraction by varying temperature.

\begin{figure*}
    \centering
    \includegraphics[width=0.9\columnwidth]{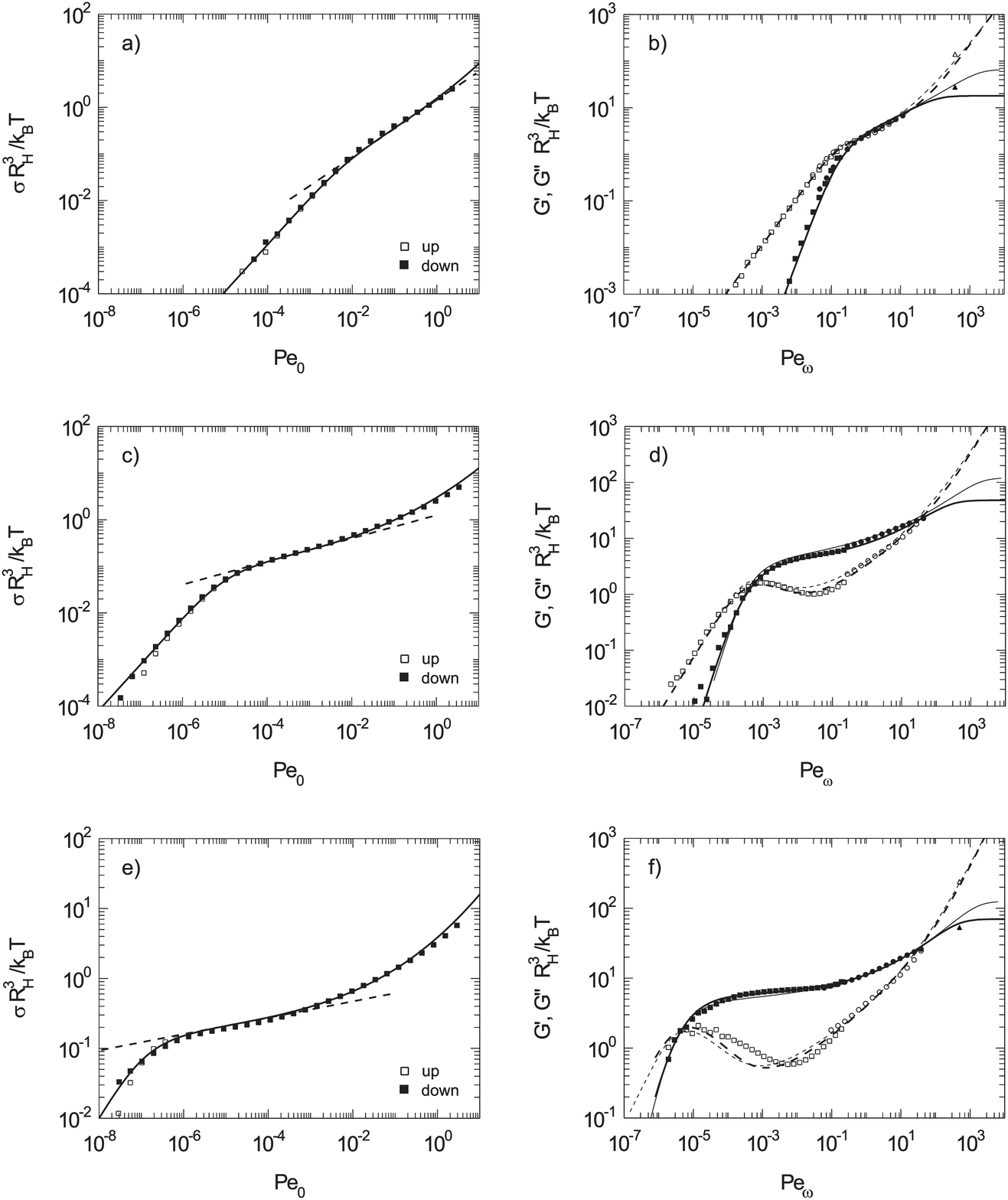}
    \includegraphics[width=0.9\columnwidth]{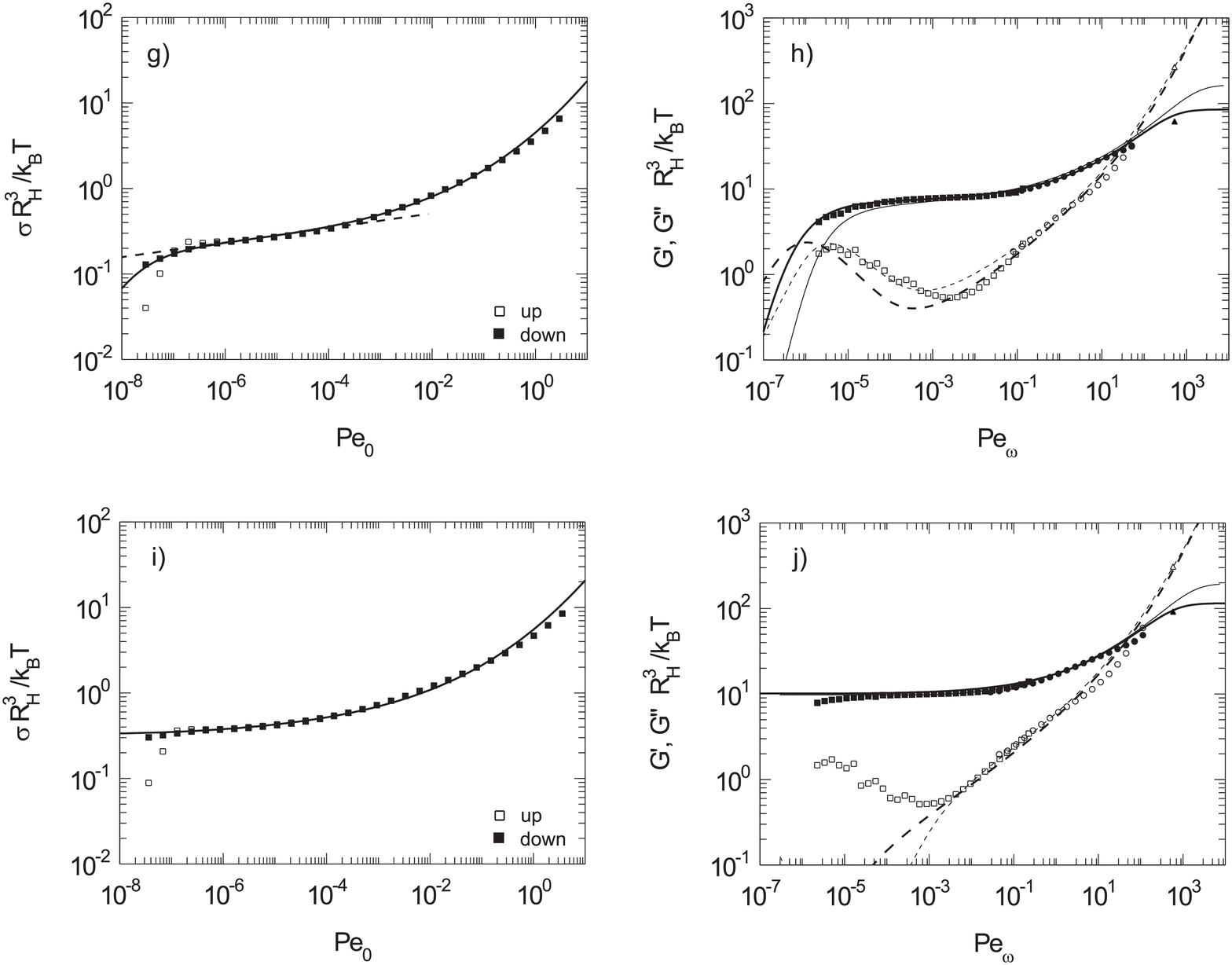}
\caption{Left column: Reduced flow curves (filled squares) for
different volume fractions. The solid lines are the results of the
schematic model, the dashed line represent the pseudo power law
behaviour; from Ref.~\cite{Sie:09}.\break
 Right column: Reduced frequency dependent moduli for different volume fractions.
  Full symbols/solid lines represent G', hollow symbols/dashed lines represent
  G''.
Thick lines are the results of the schematic model, the thin lines
the results of the microscopic MCT.\break Graphs in one row
represent the continuous and dynamic measurements at one volume
fraction. \textit{a} and \textit{b}  at $\phi_\text{eff}$  = 0.530 ,
\textit{c} and \textit{d} at  $\phi_\text{eff}$  = 0.595 ,
\textit{e} and \textit{f} at $\phi_\text{eff}$ = 0.616 ,  \textit{g}
and \textit{h} at $\phi_\text{eff}$ = 0.625  ,  and \textit{i} and
\textit{j} at $\phi_\text{eff}$ = 0.627 .\label{Fig:5} }
    \end{figure*}

Shear stresses measured in non-linear response of the dispersion
under strong steady shearing, and frequency dependent shear moduli
arising from thermal shear stress  fluctuations in the quiescent
dispersion were measured and fitted with results from the schematic
F$_{12}^{(\gd)}$-model. Some results from the microscopic MCT for
the equilibrium moduli were included also; see Fig. \ref{Fig:5}. The
fits with quiescent MCT for (monodisperse) hard spheres using the PY
$S_q$ support the finding of Sect. 3.1.2 that MCT accounts for the
magnitude of the stresses at the glass transition
semi-quantitatively. Because of the polydispersity of the samples,
which is neglected in the calculations performed according to the
presentation in Sect. 3.1.2, somewhat larger rescaling factors $c_y$
are required; they are included in Table 1. Also, the critical
packing fraction $\phi_c$ of the glass transition again is
reproduced with some small error. Because of polydispersity, the
experimental estimate $\phi^c_\text{eff}\approx 0.625$ \cite{Sie:09}
lies somewhat higher than in the (more) monodisperse case
\cite{Cra:08}, which is as expected \cite{Voi:03}.

 Figure \ref{Fig:5} gives the
comparison of the experimental flow curves and the linear response
moduli $G'$ and $G''$ with theory for  five given different
effective volume fraction $\phi_\text{eff}$ adjusted by varying
temperature. On the left-hand side the flow curves
$\sigma(\dot{\gamma})$ are presented as function of the bare Peclet
number Pe$_0=k_BT/(6\pi\eta_s R_H^3)\, \gd$, on the right-hand side
$G'$ and $G''$ are displayed as the function of the frequency-Peclet
or Deborrah number Pe$_{\omega}=k_BT/(6\pi\eta_s R_H^3)\, \omega$,
calculated with the respective frequency $\omega$. Table 1 gathers
the effective volume fractions together with the fit parameters of
the $F_{12}^{\dot{\gamma}}$-model. Note that $G'$ and $G''$ have
been obtained over nearly 7 orders of magnitude in frequency, while
the flow curves extend over more than eight decades in shear rate.

The generalized shear modulus $g(t, \dot{\gamma})$ of the
$F_{12}^{\dot{\gamma}}$-model (cf. eq. \ref{d4}) presents the
central theoretical quantity used in these fits. Within the
schematic model, the vertex prefactor $v_{\sigma}$ is kept as a
shear-independent quantity. It can easily be obtained from the
stress and modulus magnitudes. Hydrodynamic interactions enter
through $\eta_{\infty}$, which can be obtained from measurements
done at high frequencies, and through $\Gamma$, which can be
obtained via \gl{d8} from measurements done at high shear rates.
Given these three parameters, both the shapes of the flow curves as
well as the shapes of the moduli $G'$ and $G''$ may be obtained as
function of the two parameters $\epsilon$ and $\dot\gamma/\gamma_c$.
The former sets the separation to the glass transition and thus
(especially) the longest relaxation time, while the latter tunes the
effect of the shear flow on the flow curve.

 All measured
quantities, namely $\sigma$, $G'$, and $G''$ were converted to the
respective dimensionless quantities by multiplication with
$R_H^3/k_BT$ where $R_H$ is the hydrodynamic radius at the
respective temperature.  As already discussed above, the
experimental control parameters $\dot{\gamma}$ and $\omega$ also
were converted by $6\pi\eta_s R_H^3/k_BT$ to the respective Peclet
numbers.
\begin{table}
    \centering
\begin{tabular}{cccccccccc}
\hline \hline  $\;\phi_\text{eff}\;$ &    $\epsilon$ & $v_\sigma$ &
$\Gamma$ &    $\gamma_c$ & $\eta_{\infty}$ &
$\epsilon^\text{micro}$ & $D_s / D_0$ & $c_y$ & p \\

  & F$_{12}^{(\gd)}$-mod. & $\left[\frac{k_BT}{R_H^3}\right]$ &$\left[\frac{D_0}{R_H^2}\right]$&&$\left[\frac{k_BT}{D_0R_H}\right]$ &
  micro.  & & & \\

\hline

      0.530  & -0.072000  &         18 &         20 &    0.0845  &    0.2250  & -0.10 & 0.3 & 2.3   & 0.631\\

      0.595  & -0.003500  &         48 &         50 &    0.1195  &    0.2400  &   -0.008 & 0.3 & 2.3   & 0.248 \\

      0.616  & -0.000420  &         70 &         80 &    0.1414  &    0.3938  &  -0.001 & 0.3 & 2.3    & 0.117\\

      0.625  & -0.000170  &         85 &         90 &    0.1491  &    0.4250  &   -0.001 & 0.3 & 3.0   & 0.0852\\

     0.627  &  0.000021  &        115 &        120 &    0.1622  &    0.4313  &    0.002 & 0.3 & 3.5   & -\\

\hline \hline
\end{tabular}
\caption{Packing fraction  $\phi_\text{eff}$ and parameters
$v_\sigma$, $\Gamma$, $\gamma_c$ and $\eta_{\infty}$ of the fit
using the schematic $F_{12}^{\dot{\gamma}}$-model for  the
measurements shown in Fig. \ref{Fig:5}; from Ref.~\cite{Sie:09}. The
parameters $\epsilon$, short time diffusion coefficient $D_s/D_0$,
and rescaling factor $c_y$ from the microscopic linear response
calculation using MCT , and the pseudo-power law exponent $p$ are
included, also.\label{table:2}}
\end{table}
Evidently, both the reduced moduli, the Pe number, and the packing
fraction depend on the effective particle volume $R_H^3$. In the
polydisperse sample, actually a distribution of values $R_H^3$
exists, whose variance may be determined by disc centrifugation at
low concentration, and which is fixed for one given sample. Close to
the glass transition the size distribution thus is (almost) density
independent, and $R_H^3(T)$ is the single experimental control
parameter, whose small change upon varying temperature $T$ drives
the system through the glass transition.

Figure \ref{Fig:5} demonstrates that the rheological behavior of a
non-crystallizing colloidal dispersion can be modeled in a highly
satisfactory manner by five parameters that display only a weak
dependence on the effective volume fraction of the particles.
Increasing the effective packing fraction drives the system towards
the glass transition, viz.~$\epsilon$ increases with
$\phi_\text{eff}$. Stress magnitudes (measured by $v_\sigma$) also
increase with $\phi_\text{eff}$, as do high frequency and high shear
viscosities; their difference determines $\Gamma$. The strain scale
$\gamma_c$ remains around the reasonable value 10\%. In spite of the
smooth and small changes of the model parameters, the
$F_{12}^{\dot{\gamma}}$-model achieves to capture the qualitative
change of the linear and non-linear rheology. The measured Newtonian
viscosity increases by a factor around $10^5$. The elastic modulus
$G'$ at low frequencies is utterly negligible at low densities,
while it takes a rather constant value around $10 k_BT/ R^3_H$ at
high densities. An analogous observation holds for the steady state
shear stress $\sigma(\dot\gamma)$, which at high densities takes
values around $0.3 k_BT/ R^3_H$ when measured at lowest shear rates.
For lower densities, shear rates larger by a factor around $10^7$
would be required to obtain such high stress values.

At volume fractions around 0.5 the suspension is Newtonian at small
Pe$_0$.  Approaching the glass transition leads to a characteristic
S-shape of the flow curves and the Newtonian region becomes more and
more restricted to the region of smallest $Pe_0$. Concomitantly, a
pronounced minimum in $G''$ starts to develop, separating the slow
structural relaxation process from faster, rather density
independent motions, while $G'$ exhibits a more and more pronounced
plateau. At the highest density (Fig.~\ref{Fig:5} panels i and j),
the theory would conclude that a yielding glass is formed, which
exhibits a finite elastic shear modulus (elastic constant)
$G_\infty=G'(\omega\to0)$, and a finite dynamic yield stress,
$\sigma^+=\sigma(\dot\gamma\to0)$. The experiment shows, however,
that small deviations from this glass like response exist at very
small frequencies and strain rates. Description of this ultra-slow
process requires extensions of the present MCT-ITT which are
discussed in Ref.~\cite{Cra:08}.

Considering the low frequency spectra in $G'(\omega)$ and
$G''(\omega)$, microscopic MCT and schematic model provide
completely equivalent descriptions of the measured data. Differences
in the fits in Fig.~\ref{Fig:5} for Pe$_\omega \le 1$ only remain
because of slightly different choices of the fit parameters which
were not tuned to be close. These differences serve to provide some
estimate of uncertainties in the fitting procedures. Main conclusion
of the comparisons is the agreement of the moduli from microscopic
MCT, schematic ITT model, and from the measurements. This
observation strongly supports the universality of the glass
transition scenario which is a central line of reasoning in the ITT
approach to the non-linear rheology.

\subsection{F$_{12}^{(\gd)}$-model and flow curves of a simulated supercooled binary  liquid}
\begin{figure}[htp]
\centering
\includegraphics[ width=0.8\columnwidth]{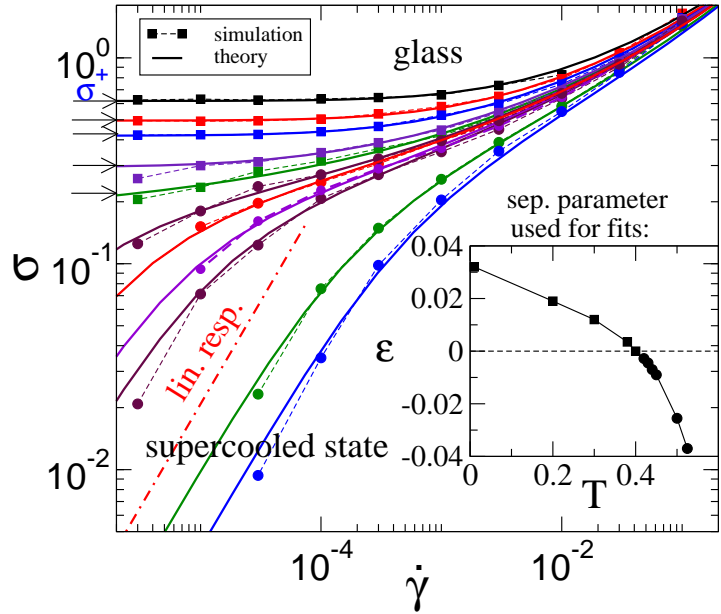}
\caption{\label{flowcurve}Flow curves $\sigma(\dot{\gamma})$
reaching from the supercooled to the glassy state of a simulated
binary LJ mixture. The data points  correspond to the temperatures
$T=0.525, 0.5, 0.45, 0.44, 0.43, 0.42, 0.4, 0.38, 0.3, 0.2$ and
$0.01$ in LJ-units (from bottom to top). $F_{12}^{(\gd)}$-model
curves fitted by eye are included as lines. The inset shows the
relation between the fitted separation parameters and temperature.
Units are converted by $\sigma=1.5 \sigma_{\rm theo}$ and
$\gd=1.3\gd_{\rm theo}\Gamma$; from \cite{Hen:05}. The arrows mark
the values of the extrapolated dynamic yield stresses
$\sigma^+(\varepsilon)$.}
\end{figure}
In  large scale molecular dynamics simulations a 80:20 binary
mixture of Lennard-Jones (LJ) particles at constant density was
supercooled under shear. This model has well known equilibrium
properties and many aspects that can be understood consistenly
within MCT \cite{Kob95}. To account for shearing, it was used
together with Lees-Edwards boundary conditions and the SLLOD
equations of motion to develop a linear velocity profile. Note that
in the simulation solvent effects are obviously lacking, and the
simulated flow curves thus provide support for the notion that shear
thinning can arise from shear-induced speed up of the structural
relaxation; it is evident in Fig.~\ref{Fig21}. Because the
microscopic motion is Newtonian, the theoretical description of this
model goes beyond the framework of Sect. 2. Yet the universality of
the structural long time dynamics, predicted by MCT \cite{Fra:98}
and also retained in MCT-ITT, supports to apply MCT-ITT to the
simulation data. Moreover, the independence of the glassy dynamics
on the employed microscopic motion was explicitly confirmed in
simulations of the mixture \cite{Gle:98}. This supercooled simple
liquid has been characterized quite extensively under shear
\cite{Ber:02,Var:04,Var:06,Hen:05,Var:06b}, and thus fits of the
flow curves provide challenging tests to  the schematic
F$_{12}^{(\gd)}$-model.

\begin{figure}[htp] \centering
\includegraphics[ width=0.6\columnwidth]{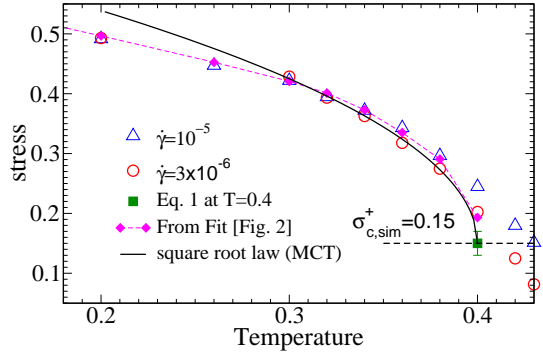}
\caption{Dynamic yield stress estimated from the simulations of a
supercooled binary LJ mixture under steady shear shown in
Fig.~\ref{flowcurve}, and its temperature dependence (in LJ units);
from Ref.~\cite{Var:06}. The estimate uses the stress values for the
two lowest simulated shear rates, namely $\gd=10^{-5}$ (triangle)
and $\gd=3 \times 10^{-6}$ (circle); the extrapolation with the
$F_{12}^{(\gd)}$-model is shown by diamonds. At temperatures below
$T=0.38$, (almost) the same shear stress is obtained for both values
of $\gd$ and the extrapolation, indicating the presence of a yield
stress plateau.} \label{fig:sigma+}
\end{figure}

Figure \ref{flowcurve} shows the stress-shear rate dependence as
flow curves, ranging from supercooled states to the glassy regime;
LJ units are used as described in \cite{Var:06}. The solid lines are
fits to the simulation data with the
$F_{12}^{(\dot{\gamma})}$-model, which reproduce the transition from
a shear-thinning fluid to a yielding glass quite well. Coming from
high shear rates, the flow curves of the supercooled state pass to
the linear response regime in the lower left corner, indicated by a
dashed-dotted line with slope 1. On approaching the transition
point, the linear response regime shifts to lower and lower shear
rates. Beyond the fluid domain, the existence of a dynamic yield
stress $\sigma^+=\lim_{\dot{\gamma}\to 0} \sigma >0$ is supported by
the simulation results, which sustain  a stress plateau over three
decades in shear rate. Best $F_{12}^{(\gd)}$-model fits are obtained
for a $T_c=0.4$ suggesting a slightly lower transition temperature
\cite{Flenner05} as determined from the simulations of the quiescent
system, where $T_c=0.435$ was found \cite{Kob95}. The reason may be
the ergodicity restoring processes which were also observed in the
colloidal experiments shown in Figs.~\ref{Fig6}, \ref{fig:fig_exp2},
and \ref{Fig:5}

 The stress plateau is best developed for temperatures deep in
the glassy phase extending over about two decades in shear rate. Its
onset is shifted toward progressively lower $\gd$ as the temperature
is increased toward $T_c$. This makes an estimate of the dynamic
yield stress, $\sigma^+(T) \equiv \sigma(T; \gd \to 0)$, a difficult
task for temperatures below but close to $T_c$. Nevertheless, an
estimate of $\sigma^+(T)$ is interesting because it highlights the
anomalous weakening of the glass when heating to $T_c$. Testing the
MCT predictions below $T_c$ has previously not been possible in
simulations because of problems to reach the equilibrated or steady
state at sufficiently low shear rates. Figure \ref{fig:sigma+}
significantly supports the notion of a glass transition under shear
as it presents the first simulations result exibiting the predicted
anomalous softening in an elastic property of the glass upon
approaching the transition from the glass side, viz.~upon heating.

\section{Summary and outlook}

The present review explored the connection between the physics of
the glass transition and the rheology of dense colloidal
dispersions, including in strong steady shear flow. A microscopic
theoretical approach for the shear-thinning of concentrated
suspensions and the yielding of colloidal glasses was presented,
which builds on the mode coupling theory (MCT) of idealized glass
transitions. The extension to strongly driven stationary states uses
the so-called ITT (integration through transients) approach, which
leads to a scenario of shear melting a glass, whose universal
aspects can be captured in simplified  schematic models. Consecutive
generalizations of ITT to arbitrary time-dependent states far from
equilibrium \cite{Bra:07} and to arbitrary flow geometries
\cite{Bra:08}, have yielded a non-Newtonian constitutive equation
applicable to concentrated dispersions in arbitrary homogeneous
flows (not reviewed here), albeit still under the approximation that
hydrodynamic interactions are neglected. Within the theory, this
approximation becomes valid close to the glass transition and for
weak but nonlinear flows, where the slow structural relaxation
dominates the system properties, and where hydrodynamic interactions
only affect the over-all time scale.

The structural dynamics under flow is predicted to result from a
competition between local particle hindrance (termed cage effect)
and the compression / stretching (i.e. advection) of the wavelength
of fluctuations induced by the affine particle motion with the flow.
Measurements of the single particle motion in the stationary state
under shear support the theoretical picture that shear speeds up the
structural dynamics, while instantaneous structural correlations
remain rather unaffected. Model dispersions made of thermo-sensitive
core-shell particles allow to investigate the close vicinity of the
transition. Measurements of the equilibrium stress fluctuations,
viz.~linear storage and loss moduli, and measurements of flow
curves, viz.~nonlinear steady state shear stress versus shear rate,
for identical external control parameters verify that the glassy
structural relaxation can be driven by shearing and in turn itself
dominates the low shear or low frequency rheology.

In the employed theoretical approach, the equilibrium structure
factor $S_q$ captures the particle interactions. Theory misses an
ultra-slow decay of all glassy states, and neglects (possible)
ageing effects.

\subsection*{Acknowledgment}
It is a great pleasure to thank all my colleagues for the enjoyable
and fruitful collaboration on this topic. I especially thank Mike
Cates for introducing me to rheology, and Matthias Ballauff for his
inspiring studies. Kind hospitality in the group of John Brady,
where part of this review was written, is gratefully acknowledged.
Financial support is acknowledged by the Deutsche
Forschungsgemeinschaft in SFB-TR6, SFB 513, IRTG 667, and via grant
Fu 309/3.

\end{document}